\documentclass[useAMS,usenatbib]{mn2e}
\usepackage{subfigure}
\usepackage{graphicx}
\usepackage{amsmath}
\usepackage[dvips,usenames]{color}
\def\aj{AJ}%
\def\araa{ARA\&A}%
\def\apj{ApJ}%
\def\apjl{ApJ}%
\def\apjs{ApJS}%
\def\aap{A\&A}%
\def\mnras{MNRAS}%
\def\pasp{PASP}%
\def\solphys{Sol.~Phys.}%
\newcommand{\dmo}{\mbox{$(m\!-\!M)_{0}$}}

\newcommand{\av}{\mbox{$A_V$}}

\newcommand{\feh}{\mbox{\rm [{\rm Fe}/{\rm H}]}}
\newcommand{\mh}{\mbox{\rm [{\rm M}/{\rm H}]}}
\newcommand{\Msun}{\mbox{$M_{\odot}$}}

\newcommand{\chisqmin}{\mbox{$\chi^2_{\rm min}$}}

\newcommand{\comment}[1]{}
\newcommand{\beq}{\begin{equation}}
\newcommand{\eeq}{\end{equation}}
\newcommand{\beqa}{\begin{eqnarray}}
\newcommand{\eeqa}{\end{eqnarray}}

\title[The SFH of NGC~1846 and NGC~1783] {The star formation history
  of the Large Magellanic Cloud star clusters NGC~1846 and
  NGC~1783\,\thanks{Based on observations with the NASA/ESA {\it
      Hubble Space Telescope}, obtained at the Space Telescope Science
    Institute, which is operated by the Association of Universities
    for Research in Astronomy, Inc., under NASA contract NAS5-26555} }

\author[Rubele et al.]{Stefano Rubele$^{1}$, L\'eo Girardi$^{1}$, Vera
  Kozhurina-Platais$^{2}$, Leandro Kerber$^{3}$, \newauthor Paul
  Goudfrooij$^{2}$, Alessandro Bressan$^{4}$, Paola Marigo$^{5}$
  \\
  $^{1}$ Osservatorio Astronomico di Padova -- INAF,
  Vicolo dell'Osservatorio 5, 35122 Padova, Italy \\
  $^{2}$ Space Telescope Science Institute, 3700 San Martin Drive,
  Baltimore, MD 21218, USA \\
  $^{3}$ Universidade Estadual de Santa Cruz, 
  Rodovia Ilh\'eus-Itabuna, km. 16 -- 
  45662-000 Ilh\'eus, Bahia, Brazil \\
  $^{4}$ SISSA, via Bonomea 265, 34136 Trieste, Italy \\
  $^{5}$ Dipartimento di Fisica e Astronomia, Universit\`a di Padova,
  Vicolo dell'Osservatorio 2, 35122 Padova, Italy \\
}

\begin{document}

\date{Accepted ... Received ...; 
}

\pagerange{\pageref{firstpage}--\pageref{lastpage}} \pubyear{2012}

\maketitle

\label{firstpage}

\begin{abstract}
  NGC~1846 and NGC~1783 are two massive star clusters in the Large
  Magellanic Cloud, hosting both an extended main sequence turn-off
  and a dual clump of red giants. They present similar masses but
  differ mainly in angular size. Starting from their high-quality ACS
  data in the F435W, F555W and F814W filters, and updated sets of
  stellar evolutionary tracks, we derive their star formation rates as
  a function of age, SFR$(t)$, by means of the classical method of CMD
  reconstruction which is usually applied to nearby galaxies. The
  method confirms the extended periods of star formation derived from
  previous analysis of the same data. When the analysis is performed
  for a finer resolution in age, we find clear evidence for a
  $\sim50$-Myr long hiatus between the oldest peak in the SFR$(t)$,
  and a second prolonged period of star formation, in both clusters.
  For the more compact cluster NGC~1846, there seems to be no
  significant difference between the SFR$(t)$ in the cluster centre
  and in an annulus with radii between 20 and 60\arcsec\ (from 4.8 to
  15.4 pc).  The same does not occur in the more extended NGC~1783
  cluster, where the outer ring (between 33 and 107\arcsec, from 8.0
  to 25.9 pc) is found to be slightly younger than the centre. We also
  explore the best-fitting slope of the present-day mass function and
  binary fraction for the different cluster regions, finding hints of
  a varying mass function between centre and outer ring in NGC~1783.
  These findings are discussed within the present scenarios for the
  formation of clusters with multiple turn-offs.
\end{abstract}


\section{Introduction}
\label{intro}

The star clusters NGC~1846 and NGC~1783 represent prototypes of
massive intermediate-age star clusters in the LMC containing multiple
main sequence turn-offs \citep[MMSTO;][]{Mackey_BrobyNielsen2007,
  Mackey_etal08, Milone_etal08, Goudfrooij_etal09}. They have masses
of about 1.5 and $1.7\times10^5$~\Msun, respectively, which locates
them among the most massive LMC clusters except for the old globulars
\citep[see e.g.\ fig. 13 in][]{Girardi_etal95}. In addition to the
MMSTOs, they also seem to present a dual clump of red giants, in
similarity to the SMC cluster NGC~419 \citep{Girardi_etal09} and the
LMC's NGC~1751 \citep{Rubele_etal11}.

The presence of MMSTOs is commonly interpreted as the signature of
continued star formation, or multiple events of star formation,
spanning a few 100~Myr in time \citep[e.g.,
][]{Mackey_BrobyNielsen2007, Mackey_etal08, Goudfrooij_etal09,
  Goudfrooij_etal11b, ConroySpergel11, Girardi_etal11,
  Keller_etal11}\footnote{The dispersion in rotational velocities in a
  coeval cluster, advocated by \citet{BastiandeMink09}, was shown by
  not to produce MMSTOs similar to the observed ones, both
  theoretically \citep{Girardi_etal11} and observationally \citep[see
  the recent observations of the open cluster
  Tr\,20 by][]{Platais_etal12}.}. For a well-defined age range between
1.2 and 1.7~Gyr, star clusters with a turn-off age spanning a few
100~Myr will temporarily contain stars that ignited helium under both
non-degenerate and degenerate conditions, and hence naturally develop
a dual red clump \citep{Girardi_etal09}.

The main difficulties with the prolonged-star formation history (SFH)
interpretation of MMSTOs, are related with the theories of star
formation and gas dynamics inside the relatively shallow potential
wells of star clusters.  \citet{ConroySpergel11} and
\citet{Goudfrooij_etal11b} describe scenarios for the continuation of
star formation over long timescales which imply that more massive
clusters may have more extended SFHs.  \citet{ConroySpergel11}
specifically note that all clusters with MMSTOs have masses higher
than $\sim\!10^4$~\Msun, whereas \citet{Goudfrooij_etal11b} find a
correlation between the estimated escape velocities at an age of
10~Myr and the concentration of stars in the brightest half of the
MMSTO region.  Moreover, \citet{Keller_etal11} notice the correlation
between the cluster core radius $r_{\rm c}$ and their SFHs, in the
sense that all known clusters with MMSTOs have $r_{\rm c}>3.7$~pc.
Further testing these trends, and finding additional correlations with
other cluster parameters, are probably necessary steps to clarify the
origin of clusters with MMSTOs.

In this context, the pair of clusters NGC~1846 and NGC~1783 is
extremely interesting. While they have very similar total masses
\citep[$\log(M/\Msun)=5.17\pm0.09$ and $5.25\pm0.09$,
respectively;][]{Goudfrooij_etal11a} and mean ages
\citep[$1.73\pm0.10$ and $1.70\pm0.10$~Gyr,
respectively;][]{Goudfrooij_etal11b}, and a similar location in the
Northwest portion of the LMC (hence similar fore/background, and
likely the same distance), they have very different angular sizes: for
NGC~1846 the core radius is $r_{\rm c}=26.0\arcsec=6.3$~pc, and the
concentration index $c=r_{\rm t}/r_{\rm c}$ (where $r_{\rm t}$ is the
tidal radius) is 6.2 \citep{Goudfrooij_etal09}; for NGC~1783 these
quantities are $r_{\rm c}=37.7\arcsec=9.1$~pc and $c=9.2$
\citep{Goudfrooij_etal11a}, respectively\footnote{The tidal radius and
  concentration index of NGC~1783 are not well constrained from the
  ACS/WFC data alone, due to its large radius.  Its large $r_{\rm c}$,
  instead, is well constrained and evident even from a simple visual
  inspection of the HST ACS/WFC images used by
  \citet{Goudfrooij_etal09}. We note that there are two previous fits
  of King profile to NGC~1783 in the literature, both providing
  smaller values of $r_{\rm c}$: \citet{Elson92} finds $r_{\rm
    c}=4.9~{\rm pc}=20\arcsec$ using ground-based data, while
  \citet{Mucciarelli_etal07} find $r_{\rm c}=24.5\arcsec$ from the
  shallower ACS/WFC images from SNAP~9891 (PI: G.~Gilmore). These
  small values probably explain why, in figure~3 of
  \citet{Keller_etal11}, NGC~1846 and NGC~1783 appear as if they had
  the same $r_{\rm c}$, which is not correct.}.
So, in these clusters we can test whether there is any measurable
difference in their SFHs that can be interpreted as a result of the
different radii, and in the light of the correlation noted by
\citet{Keller_etal11}.

In this paper, we examine the SFHs of NGC~1846 and NGC~1783 using the
same CMD reconstruction method previously applied to NGC~419 and
NGC~1751 \citep{Rubele_etal10, Rubele_etal11}. This method is
significantly different from other analyses of the same clusters that
appeared in the literature \citep{Mackey_BrobyNielsen2007,
  Mucciarelli_etal08, Mackey_etal08, Milone_etal08, Goudfrooij_etal09,
  Goudfrooij_etal11b}, which concentrate on the MSTO region of the
CMD. The CMD reconstruction method finds the best-fitting model for
the entire CMD above a given magnitude cut, without giving particular
weight to any subset of the observed stars.  As such, it may be more
affected by errors in evaluating the contribution from field stars or
the photometric errors of faint stars -- aspects that, in any case,
are taken into account in the method. On the other hand, by using the
entire available data set, the CMD reconstruction method potentially
enhances the statistical significance of the detected SFH features.
This is important considering that the number of stars along the
MMSTOs is anyway limited to a few hundreds, even for the most populous
LMC clusters.  Moreover, this aspect might be crucial in the analysis
of less populous clusters.

In the following, we will use the excellent imaging and photometry of
NGC~1846 and NGC~1783 available from HST/ACS, together with new sets
of evolutionary tracks and isochrones (Sect.~\ref{data}).
Sect~\ref{sec_sfh} will apply the CMD reconstruction method of
SFH-recovery to the surrounding NGC~1846 and NGC~1783 LMC fields
(Sect.~\ref{sec_fieldSFH}) and cluster areas
(Sect.~\ref{sec_sfhclusters}).  Sect.~\ref{conclu} discusses the
results in the framework of present scenarios for the formation of
MMSTOs.

\section{The data}
\label{data}

\subsection{Cluster imaging and photometry}
\label{dataphot}

The data set used in this paper comes from GO-10595 (PI: Goudfrooij),
and consists of one short and two long exposures in F435W, F555W, and
F814W with small dither patterns to avoid the gap between two ACS/WFC
chips. A detailed description of the observations and photometry is
given in \citet{Goudfrooij_etal09} and \citet{Goudfrooij_etal11a}.
Nevertheless, in this paper we use the simultaneous ePSF fitting
technique as it described in \citet{Anderson_etal08}, which fits the
PSF simultaneously on all exposures/observations of the cluster.
Differently from \citet{Goudfrooij_etal09}, the Charge Transfer
Efficiency (CTE) correction was performed using \citet{RiessMack04}
formula (ACS-ISR 2005). The derived photometry was calibrated into the
Vegamag system as described in \citet{Goudfrooij_etal09}.

\begin{figure*}
  \resizebox{0.42\hsize}{!}{\includegraphics{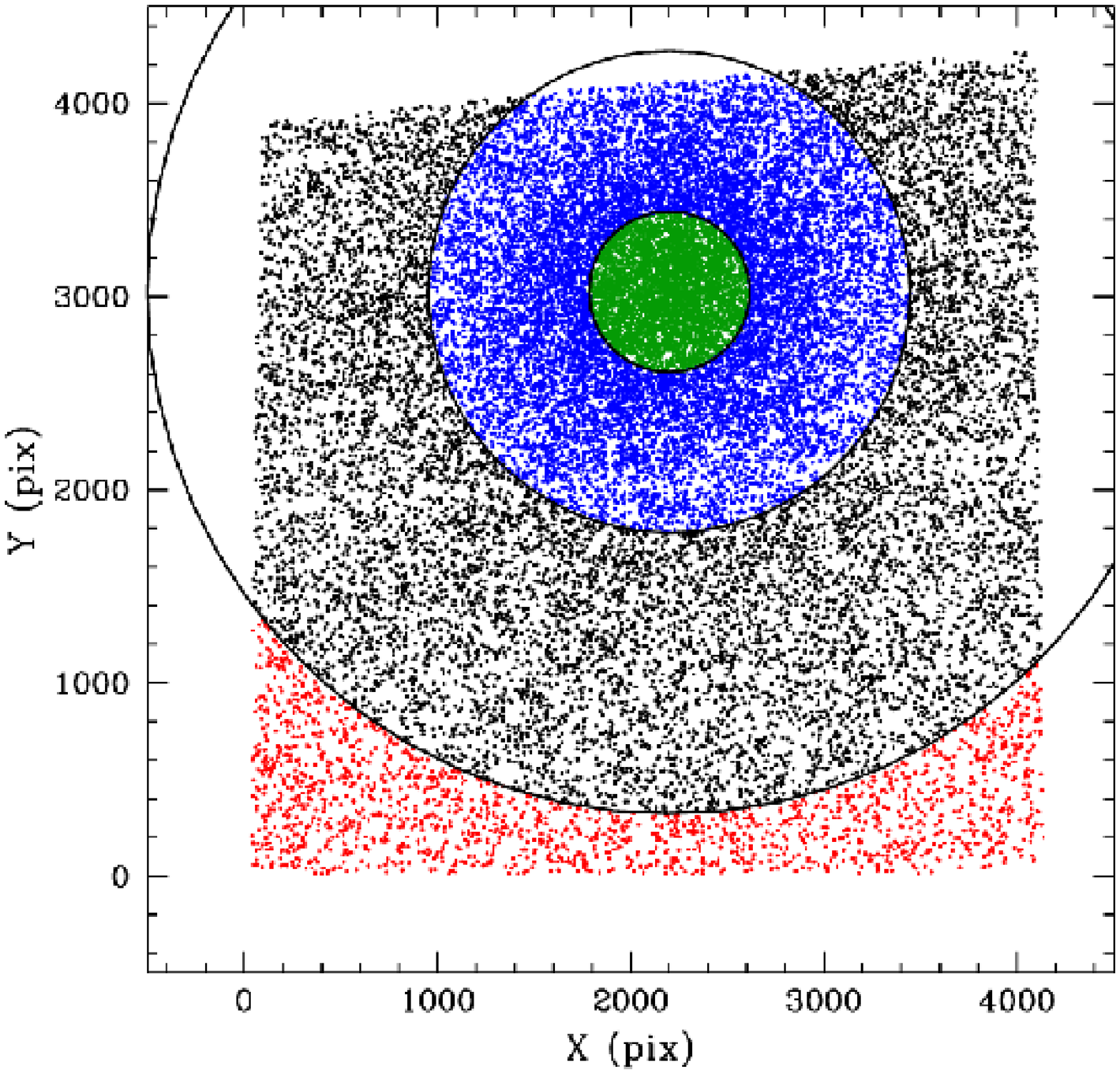}}
  \resizebox{0.42\hsize}{!}{\includegraphics{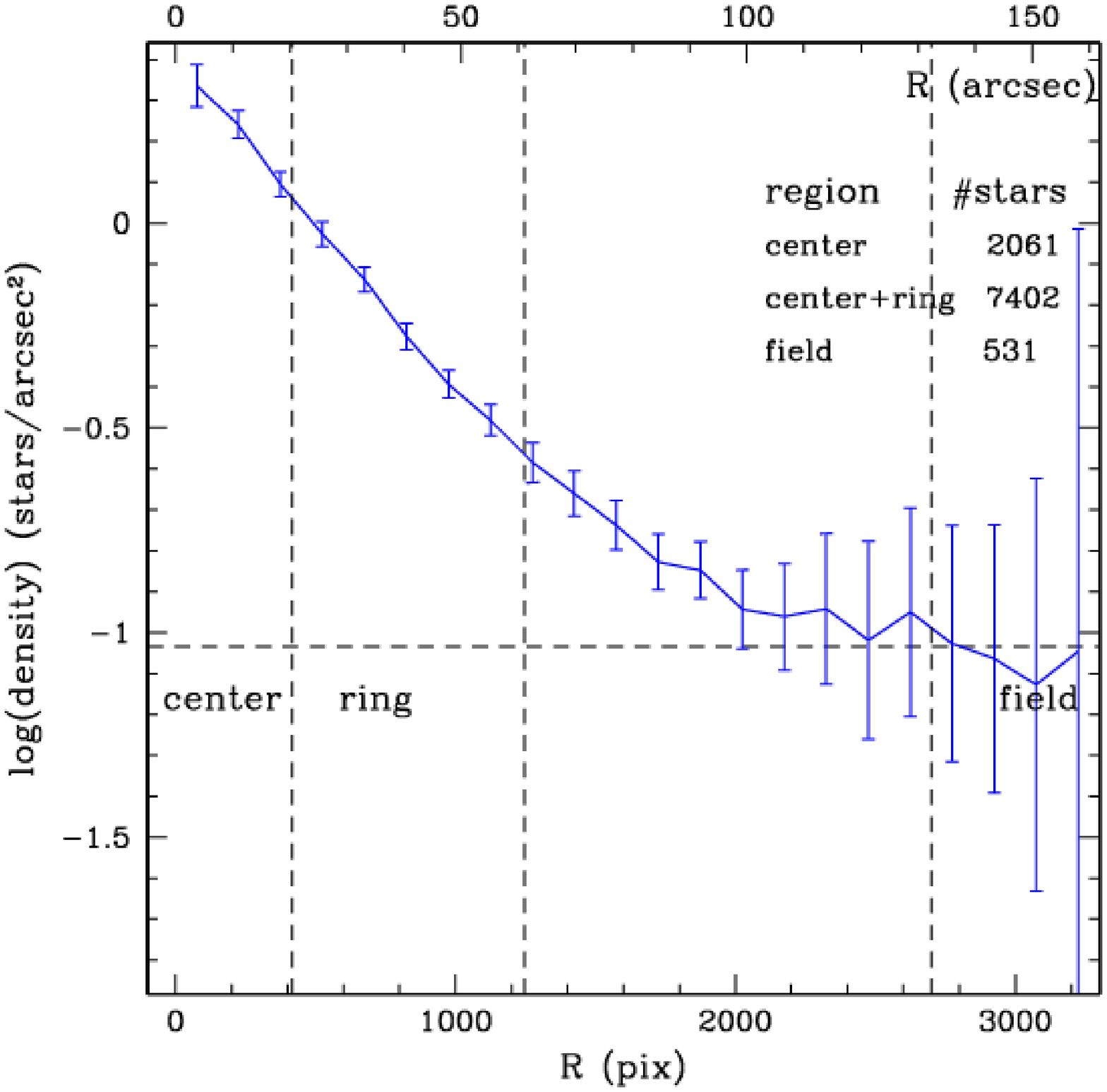}}
  \caption{{\bf Left panel:} Map of the stars used in this work, in
    the $xy$ plane of the ACS/WFC images of NGC~1846. The scale is of
    about $0.05$\arcsec/pix. The observed stars have been grouped in
    areas corresponding to the LMC field (red) and, for NGC~1846, an
    inner ``Centre'' (green) and outer ``Ring'' (blue). {\bf Right
      panel:} The logarithm of stellar density as a function of radius
    from the NGC~1846 centre. Error bars are the random errors. In the
    top right corner we indicate the total numbers of stars used to
    build the profile, selected at ${\rm F814W}<22$~mag.}
  \label{fig_areas1846}
  \label{fig_densityprofile1846}
\end{figure*}

\begin{figure*}
  \resizebox{0.42\hsize}{!}{\includegraphics{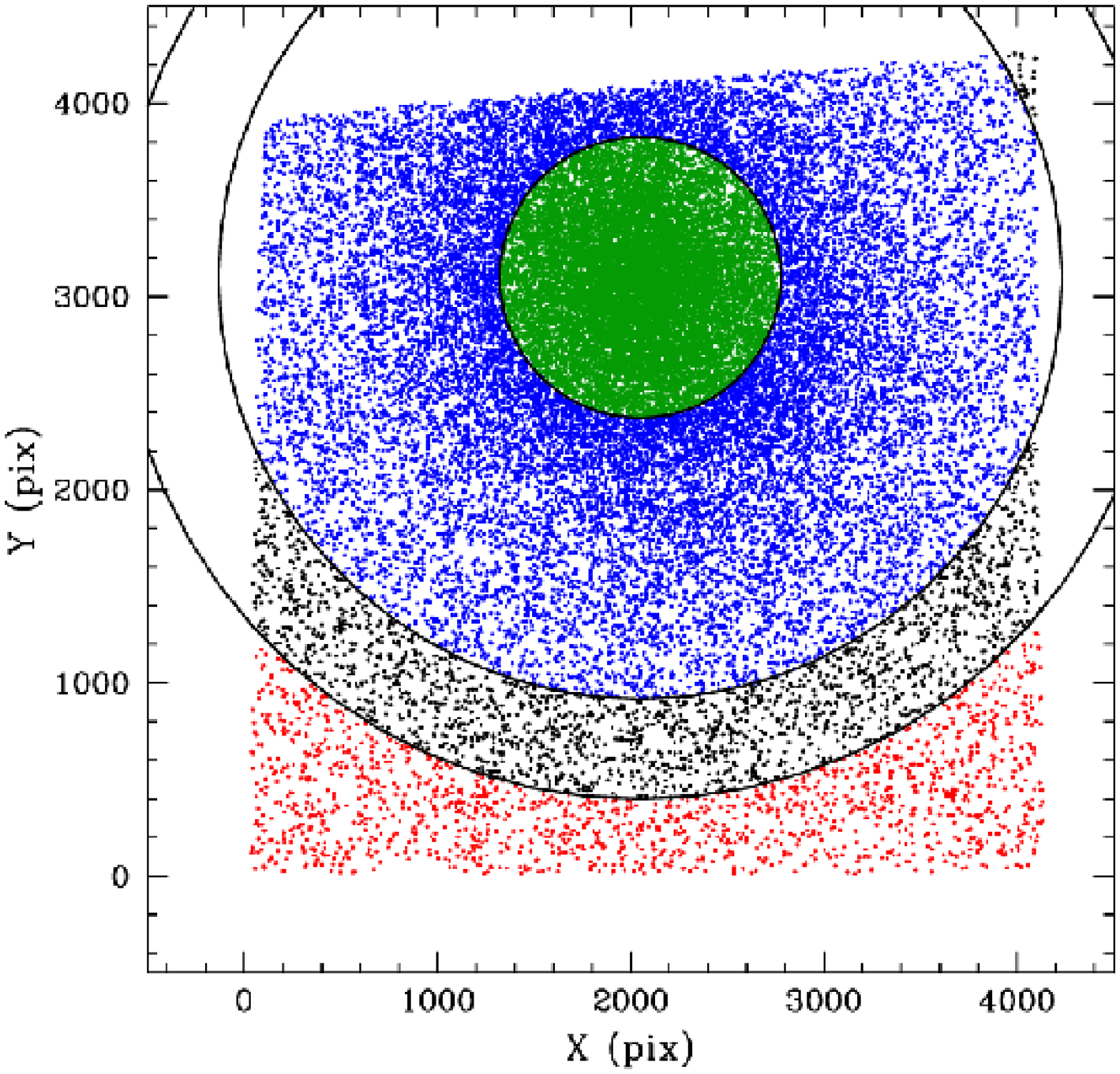}}
  \resizebox{0.42\hsize}{!}{\includegraphics{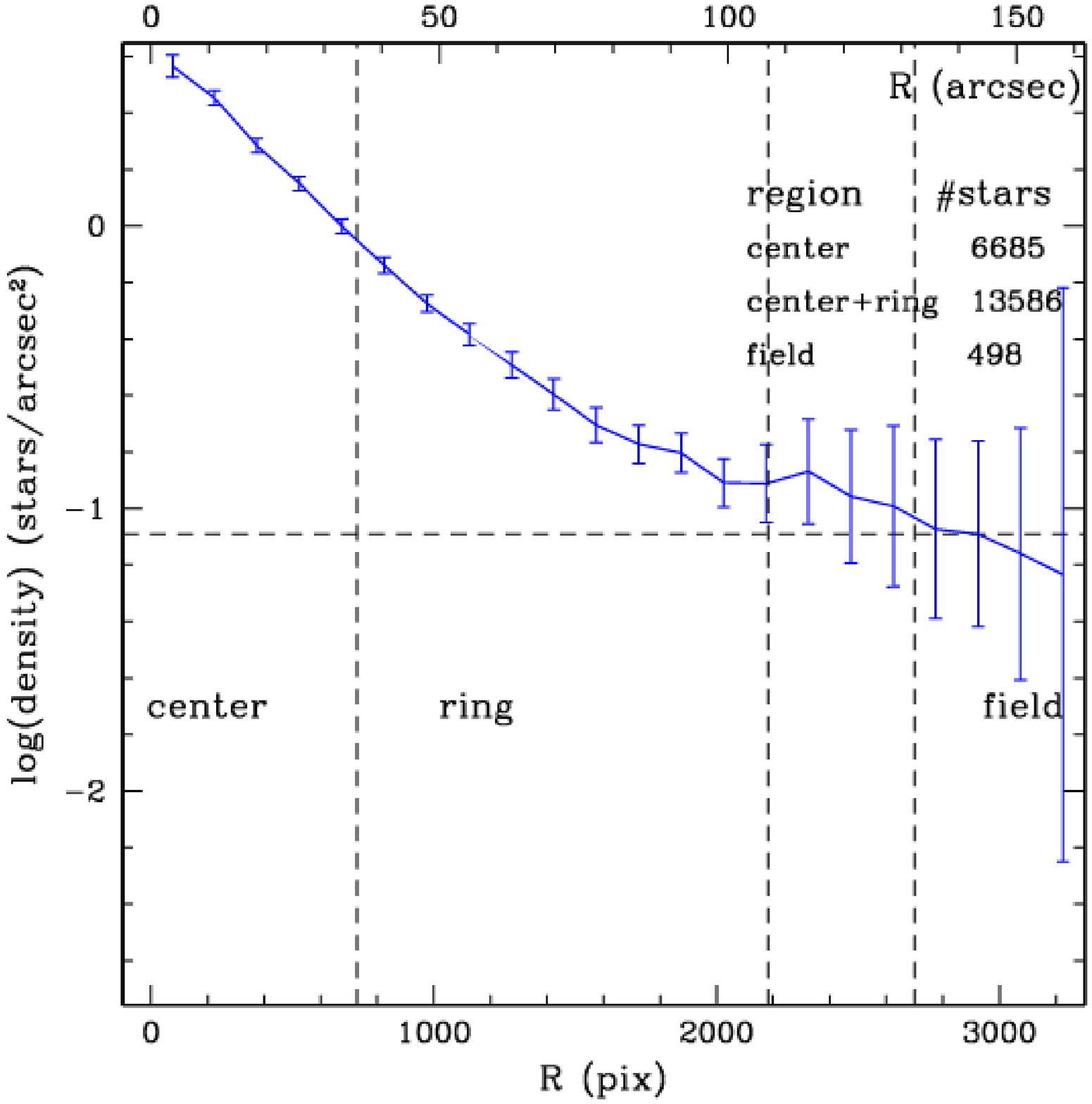}}
  \caption{The same as Fig.~\ref{fig_areas1846}, but for NGC~1783.}
  \label{fig_areas1783}
  \label{fig_densityprofile1783}
\end{figure*}

The left panels in Figs.~\ref{fig_areas1846} and \ref{fig_areas1783}
show the spatial representation of the stars we analyse in this work,
for the two clusters. The right panels show how the stellar density
varies as a function of radius from the NGC~1846 and NGC~1783 centres,
taking into account only the stars of $F814W<22$, for which the
photometry is nearly complete. Based on these figures, we define
Centre and Ring regions having radii and areas as tabulated in
Table~\ref{tab_regions}. For both clusters, the Centre regions have
radii very close to the measured core radii $r_{\rm c}$. The Ring
external radii are selected to be three times the Center radii (or
about $3 \times r_{\rm{c}}$), which include at least twice the number
of stars as in the Centre regions. The figures also indicate the
flattening of the stellar density for radii $r\ga2700$~pix, which
probably represents the regions which start being dominated by LMC
field rather than by cluster stars.  These radii were chosen as the
inner boundary of the LMC Field for each cluster. Although these Field
radii are poorly defined -- especially for NGC~1783 -- their stellar
densities are clearly very small compared to the Centre and Ring
regions. The field stellar densities are also similar, as expected for
clusters located in the same portion of the LMC.

\begin{table*}
  \caption{Selected regions. The last 2 columns refer to stars with
    $F814W\!<\!22$~mag.}
\label{tab_regions}
\begin{tabular}{llllll}
\hline
Region & radii &    & area & \# stars & mean density\\
\cline{2-3} 
       & (\arcsec) & (pc) & (arcmin$^2$) &  & (arcsec$^{-2}$) \\
\hline
NGC~1846 Centre & $\!r\!<20$    & $\!r\!<4.8$     & 0.380 & 2081 & 1.52\\
NGC~1846 Ring   & $20\!<r\!<60$ & $4.8\!<r\!<15.4$& 2.806 & 5321 & 0.527\\
NGC~1846 Field  & $r\!>133$     & $r\!>32.2$      & 2.454 & 531 & 0.060\\
NGC~1783 Centre & $r\!<33$      & $r\!<8.0$       & 1.141 & 6685 & 1.63 \\
NGC~1783 Ring   & $33\!<r\!<107$& $8.0\!<r\!<25.9$& 6.742 & 6901 & 0.284\\
NGC~1783 Field  & $r\!>133$     & $r\!>32.2$      & 2.212 & 498 & 0.063 \\
\hline
\end{tabular}
\end{table*}

\begin{figure*}
  \resizebox{0.47\hsize}{!}{\includegraphics{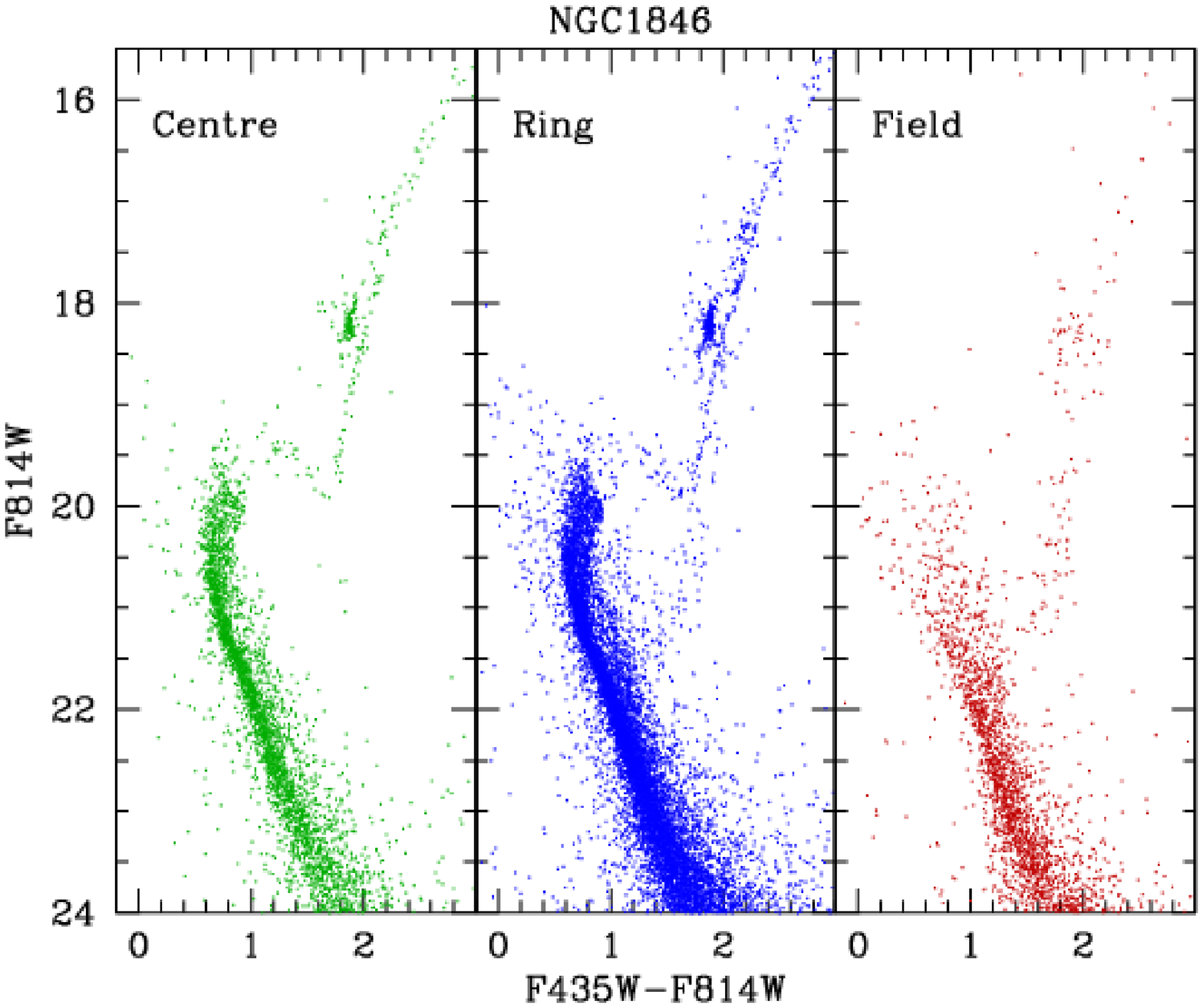}}
  \resizebox{0.47\hsize}{!}{\includegraphics{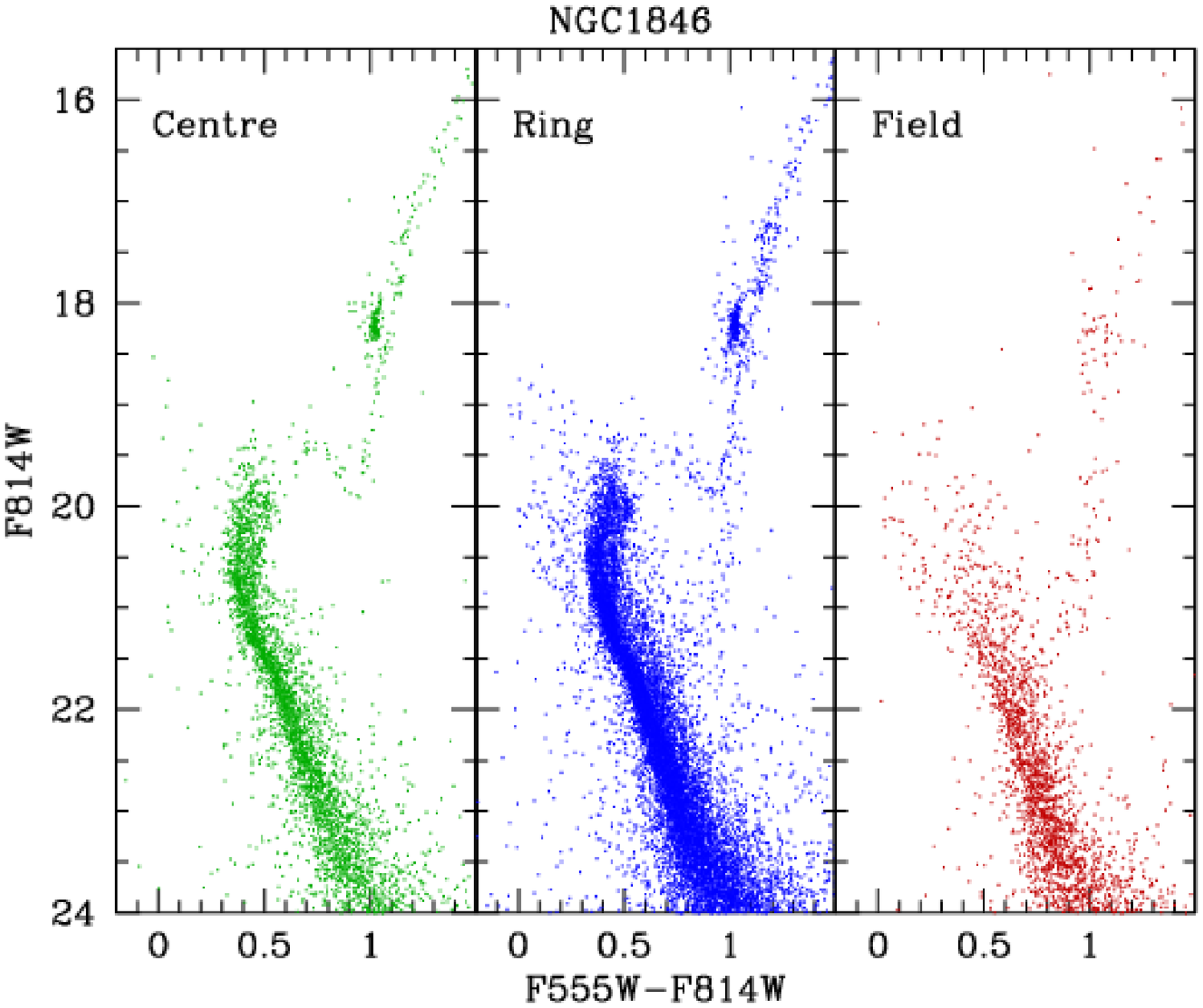}}
\\
  \resizebox{0.47\hsize}{!}{\includegraphics{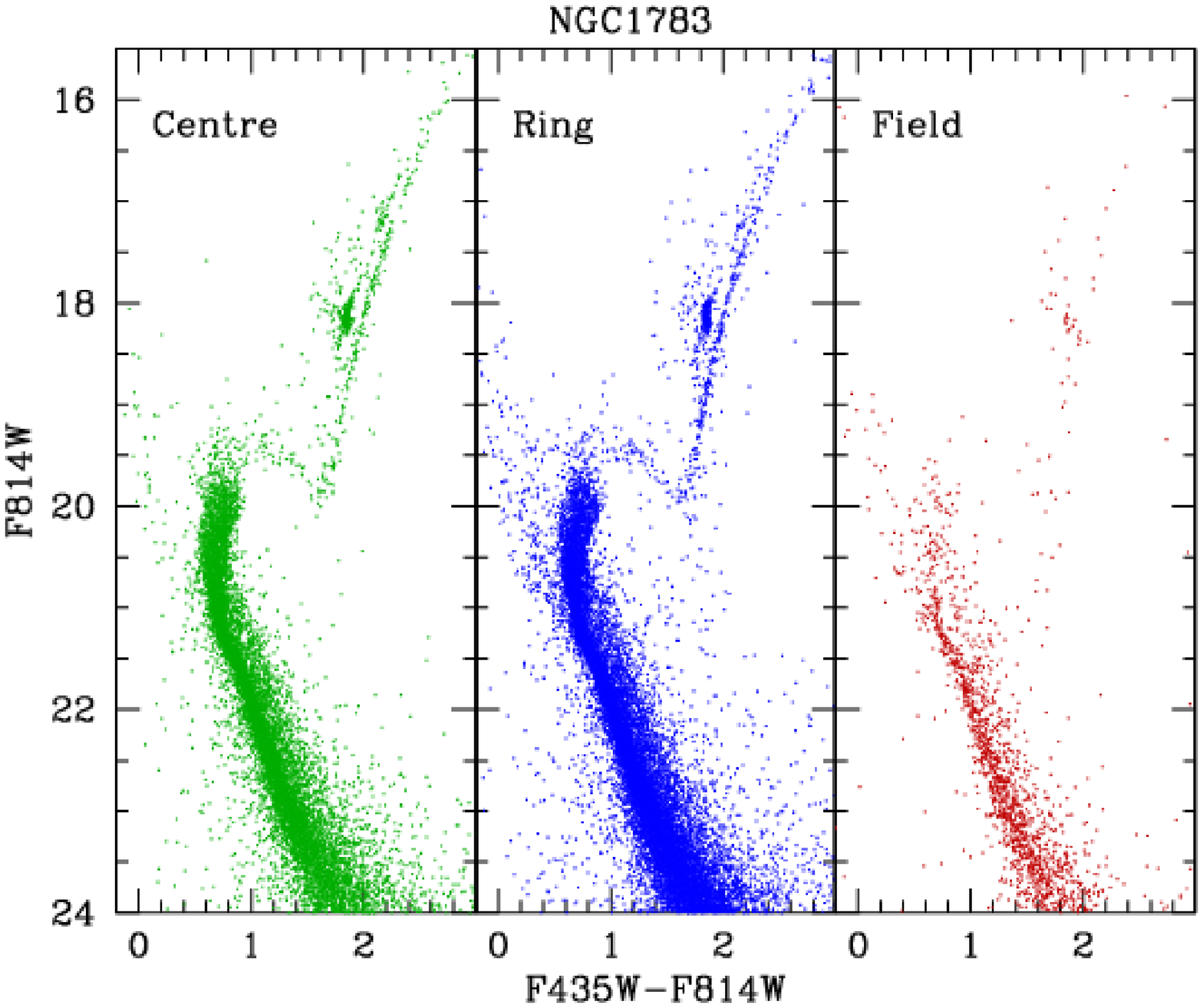}}
  \resizebox{0.47\hsize}{!}{\includegraphics{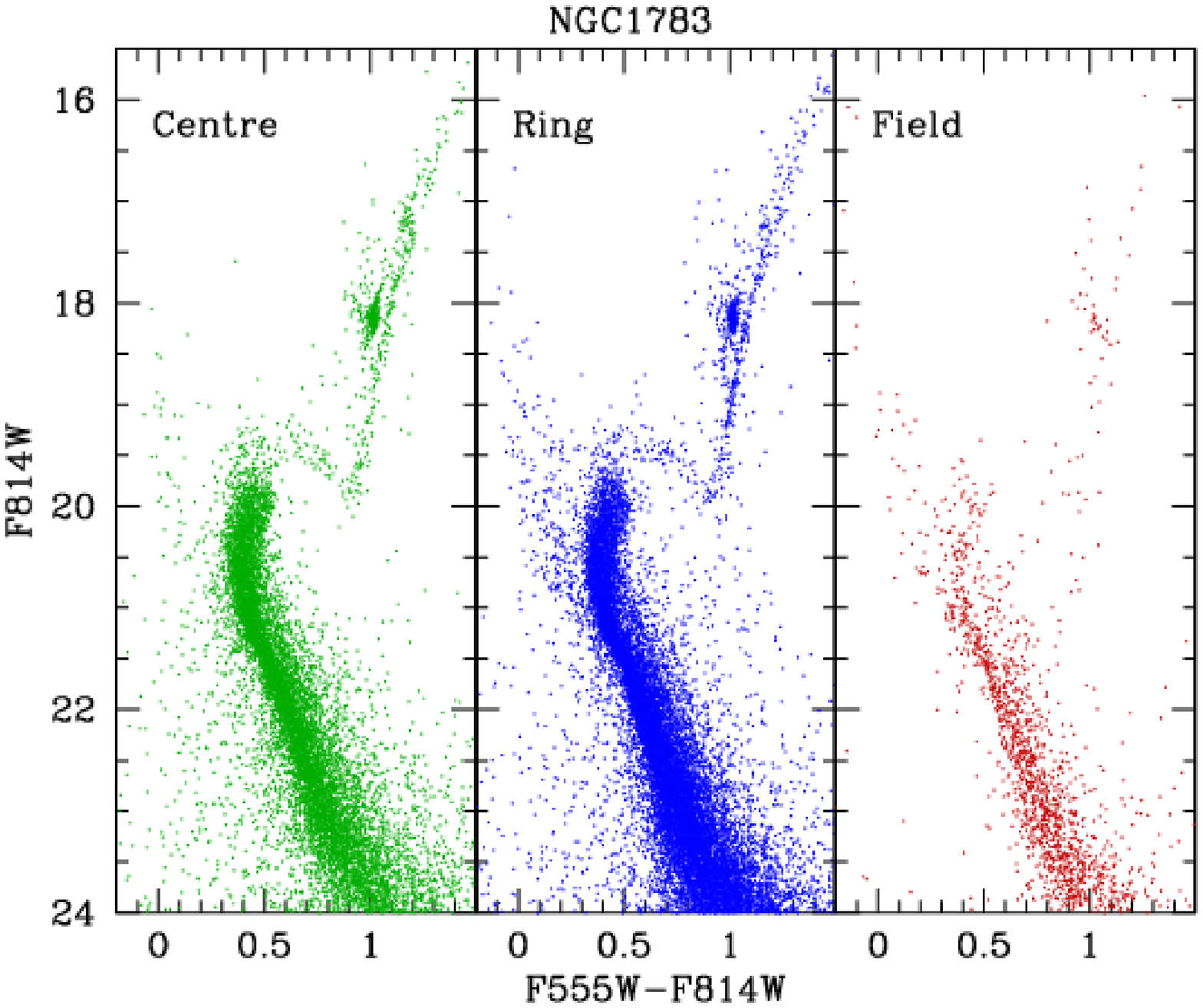}}
  \caption{CMDs for NGC~1846 and NGC~1783. The selected regions are
    the same illustrated in Figs.~\ref{fig_areas1846} and
    \ref{fig_areas1783}.}
  \label{fig_cmd}
\end{figure*}

Fig.~\ref{fig_cmd} shows the ACS data for the different regions of
NGC~1846 and NGC~1783, in the F814W vs.\ F435W\!$-$\!F814W and F814W
vs.\ F555W\!$-$\!F814W CMDs. These plots will be used as a reference
in our analysis.

The CMDs for the clusters show very clearly the broad main sequence
turn-off, the composite structure of the red clump, and other
well-known CMD features such as the sequence of binaries parallel to
the main sequence, and the RGB, subgiants, and early-AGB bump. A
simple comparison between the CMDs for the Centre and Field reveals
that the field contamination in the Centre of both clusters is close
to negligible. Indeed, the stellar densities in the Fields are about
25 times smaller than in the cluster Centres (see last column of
Table~\ref{tab_regions}). Notwithstanding, it appears evident that the
Field region in NGC~1783 is marked by a population with about the same
turn-off as the cluster, which is probably indicating that areas
dominated by the ``pure LMC field'' have not been reached in this
case. We will evaluate the impact of this possibility further down in
our analysis.

\subsection{Assessing photometric errors and completeness}
\label{sec_ast}

In order to characterize the errors in the photometry and the
completeness of the sample, we have performed a series of artificial
star tests (AST) on the reduced images \citep[see
e.g.][]{Gallart_etal99,HZ01}.  The procedure consists of adding stars
of known magnitude and colour at random places in each exposure, and
redoing the photometry exactly in the same way as described in
Sect.~\ref{dataphot}. The artificial stars are considered to be
recovered if the input and output positions are closer than 0.5
pixels, and flux differences are less than 0.5~mag.  In order to avoid
the introduction of additional crowding in the images, artificial
stars are positioned at distances much higher than their PSF width.
So, our AST are distributed on a grid spaced by 20~pix, which is each
time randomly displaced over each set of exposures. Importantly, the
AST tests are repeated many more times in the central cluster regions,
in numbers which are proportional to the density of stars brighter
than ${\rm F814W}\!<\!22.5$~mag.  In this way, we have a better
description of the errors in the most crowded cluster regions, and are
able to accurately describe their decrease with the radial distance
from the centre.

A total of 5.2 million ASTs were performed, with colors and magnitudes
covering in an almost uniform way the CMD area of the observed stars
and of the ``partial models'' to be used in the SFH analysis (see
Sect.~\ref{sec_overview_recovery} below). For the cluster centres, the
90~\% completeness limit turns out to located at ${\rm
  F814W}\!\sim\!24.5$, which is well below the position of the MMSTOs
in both NGC~1846 and NGC~1783, as can be seen in Fig.~\ref{fig_cmd}.

\subsection{Stellar models}
\label{models}

An additional goal of this paper is to employ a new set of
evolutionary tracks derived from the PAdova \& tRieste Stellar
Evolution Code (PARSEC) and extensively described in \citet{parsec}.
They include updated input physics (opacities, equation of state,
neutrino losses, etc.), and revised prescriptions for the convective
processes, including microscopic diffusion in low-mass stars and an
excellent description of helioseismic data. These improvements are
expected to provide a more detailed description of CMD features and a
more robust age scale than previous versions of Padova tracks. In the
mass interval of interest for this work, overshooting is assumed to
operate with an efficiency of $\Lambda_{\rm c}=0.5$ pressure scale
heights \citep[cf.][]{Bressan_etal81}.

The initial chemical composition is derived from the
\citet{Caffau_etal11} new solar composition after enhancing the
abundances of $\alpha$ elements by $+0.2$~dex. This mild $\alpha$
enhancement has just a minor effect in the shape of evolutionary
tracks and isochrones, but it is necessary to reproduce the chemical
composition of AGB stars observed in NGC~1846 by
\citet{Lebzelter_etal08}, as will be discussed in Marigo et al. (in
prep.).

The stellar evolutionary tracks are transformed into isochrones and
converted to the ACS/WFC Vegamag photometric system using the
transformations described in \citet{Girardi_etal08}.


\section{Recovering the SFH}
\label{sec_sfh}
 
\subsection{Overview of the method}
\label{sec_overview_recovery}
 
To recover the SFH from the ACS data, we use the same method applied
to the SMC star cluster NGC~419 \citep{Rubele_etal10} and later
improved with the LMC cluster NGC~1751 \citep{Rubele_etal11}: We use
the StarFISH code \citep{HZ01, HZ04} to look for the linear
combination of ``stellar partial models'' (SPM) that best-fits the
Hess diagram\footnote{The Hess diagram is simply a representation of
  the stellar density -- number of stars per color--magnitude bin --
  across the CMD.} of the observations, via the minimization of a
$\chi^2$-like statistics \citep[cf.][]{Dolphin02}. The coefficients of
this best-fit linear combination directly translated into the SFH, and
can be plotted in different ways -- for instance, as the star
formation rate as a function of age, SFR$(t)$, plus the
age--metallicity relation (AMR), \feh$(t)$.

SPMs are the basic building blocks in the method. They are
theoretically-derived Hess diagrams of simple stellar populations
spanning very small ranges of age and metallicity. They are initially
produced in a purely theoretical way, with the aid of the TRILEGAL
population synthesis code \citep{Girardi_etal05} and the PARSEC
stellar evolutionary tracks.  Then, these ``perfect'' SPMs are
displaced by the distance modulus and reddening to be tested, and
degraded using the distributions of incompleteness and photometric
errors as derived from the ASTs. Specifically, all ASTs falling in the
spatial region under consideration, and for each small box in the CMD,
are grouped together and used to derive the two-dimensional error and
completeness distributions, that then are used to blur the same boxes
in the theoretical SPMs. Examples of this procedure are presented in
figure 4 of \citet{Rubele_etal10} .

In the case of star clusters, our customized version of StarFISH
performs the following steps:
\begin{enumerate}
\item For each set of ACS frames, we first recover the best-fitting
  SFH of the Field region, exploring different values of $\av$ and
  $\dmo$, as described in Sect.~\ref{sec_fieldSFH}.
\item From this best fitting solution and our set of stellar models,
  we generate a Field Stellar Partial Model (FSPM). This special
  partial model is (a) scaled to the cluster area to the analysed, and
  (b) degraded using the ASTs performed {\em in} the cluster area.
  Finally, this FSPM is included as a fixed component during the
  SFH-recovery of the Center and Ring regions. In this way, the
  SFH-recovery of the cluster area includes the best possible estimate
  for the LMC field contamination.
\item We perform the SFH-recovery of the cluster area exploring a wide
  range of extinction, distant modulus, and metallicity values
  (Sect.~\ref{sec_sfhclusters}).
\item This process is initially performed assuming a single value of
  binary fraction $f$ (for binaries with mass ratios in the range
  between 0.7 and 1.0), and a fixed present-day mass function (PDMF)
  from \citet{chabrier01}. Variations in these parameters are later
  explored. The default value of $f$ is 0.3 for the LMC field, and 0.2
  for the star clusters \citep[cf.\ sect.~4.1 in][]{Rubele_etal11}.
\end{enumerate}

For the specific case of our NGC~1846 and NGC~1783 data, we will use
the entire CMD regions above ${\rm F814W}\!<\!22.5$~mag. This ensures
that we will be dealing with near-complete CMDs including all main
sequence turn-offs up to the oldest possible ages.

\subsection{The SFH in the LMC Fields }
\label{sec_fieldSFH}

\begin{figure*}
\resizebox{0.4\hsize}{!}{\includegraphics{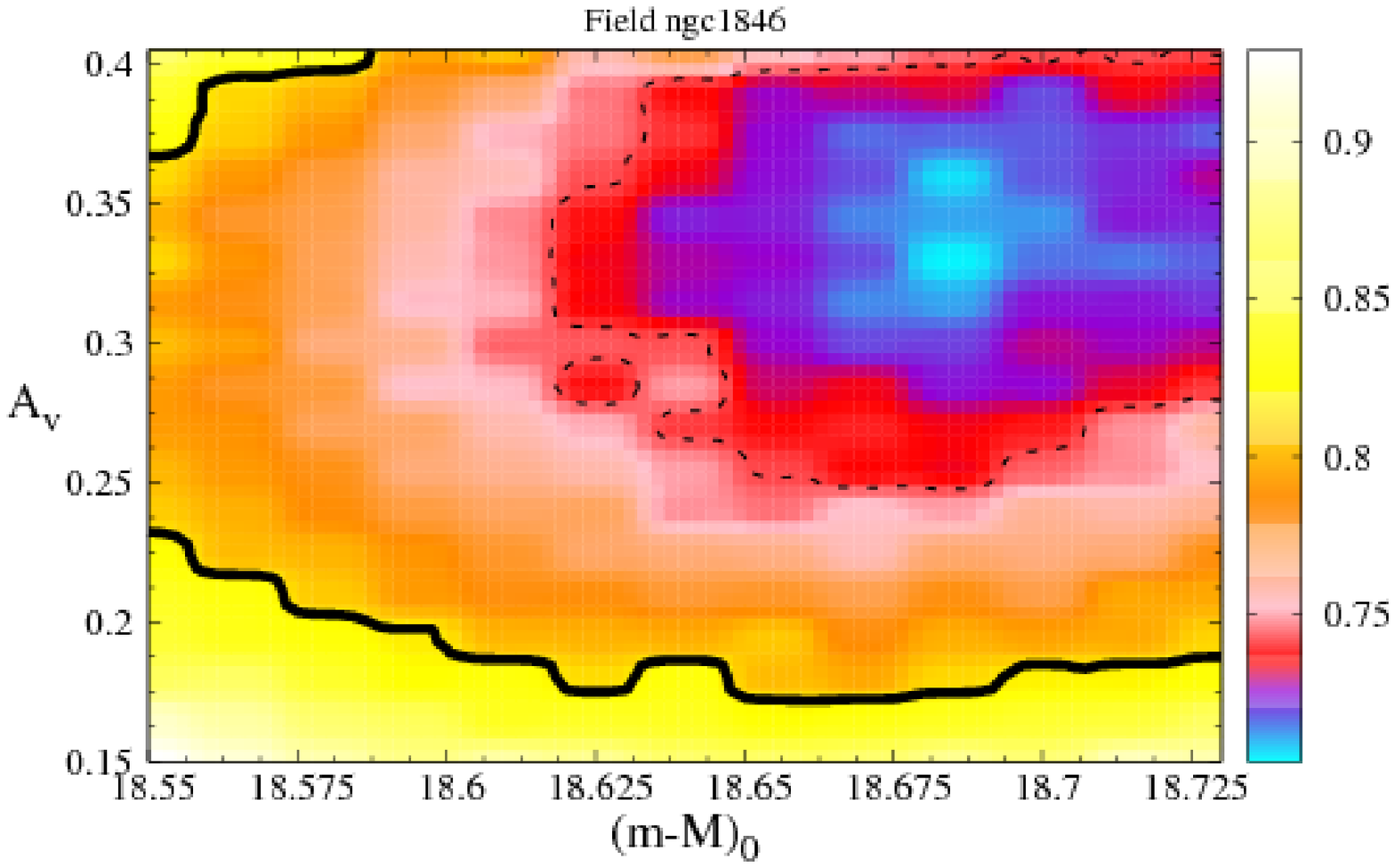}}
\resizebox{0.4\hsize}{!}{\includegraphics{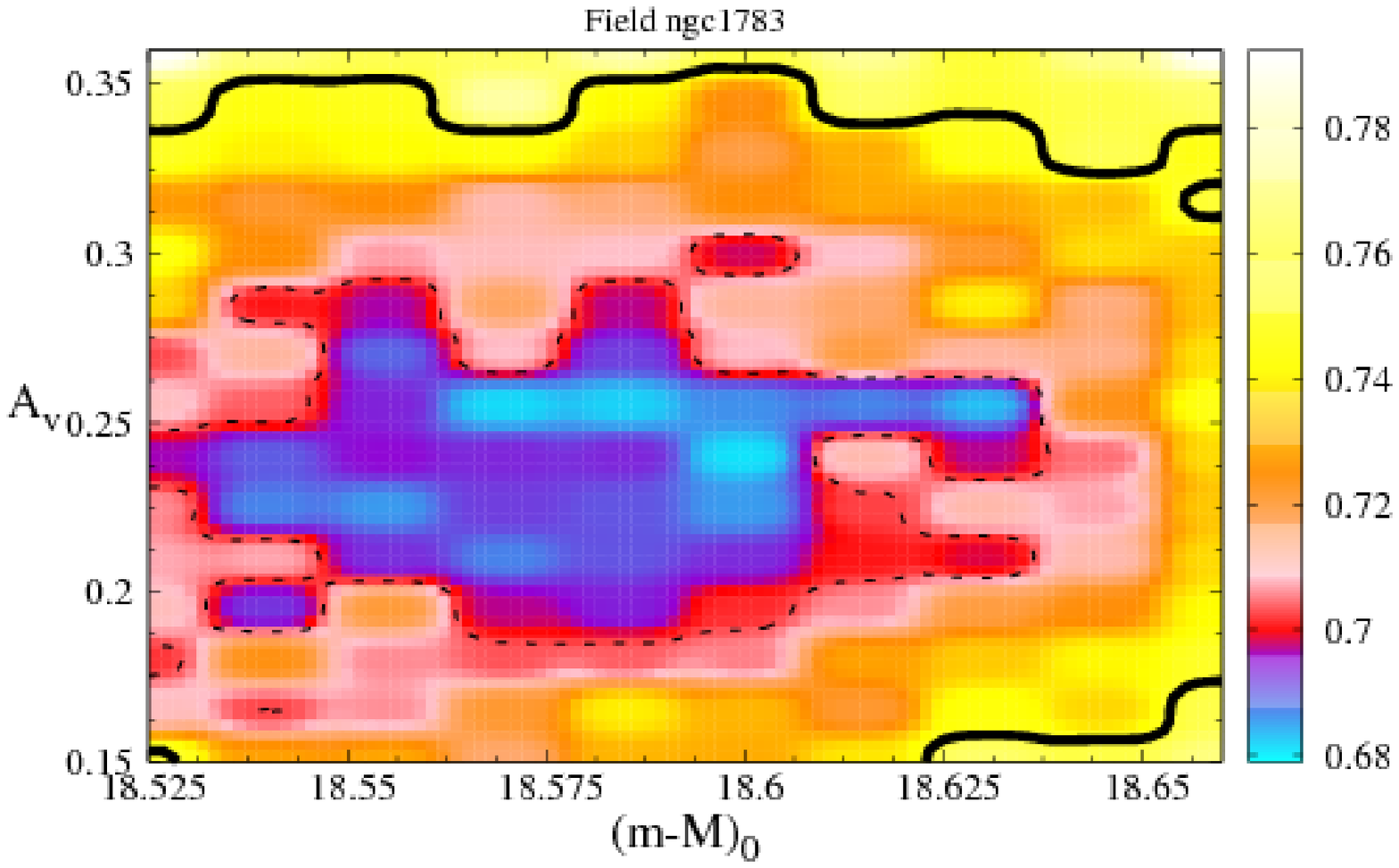}}
\caption{$\chi^{2}$ map for the Field best-fitting solutions, as a
  function of distance modulus and $V$-band extinction. The dashed and
  continuous black line show the $68~\%$ and 95~\% confidence levels
  for the overall best-fitting solution. The best-fitting model is
  located at $\dmo=18.685, \av=0.33$ for the NGC~1846 Field ({\bf left
    panel}), and at $\dmo=18.60, \av=0.225$ for the NGC~1783 Field
  ({\bf right panel}).}
\label{fig_fieldchimap}
\end{figure*}

\begin{figure*}
\resizebox{0.33\hsize}{!}{\includegraphics{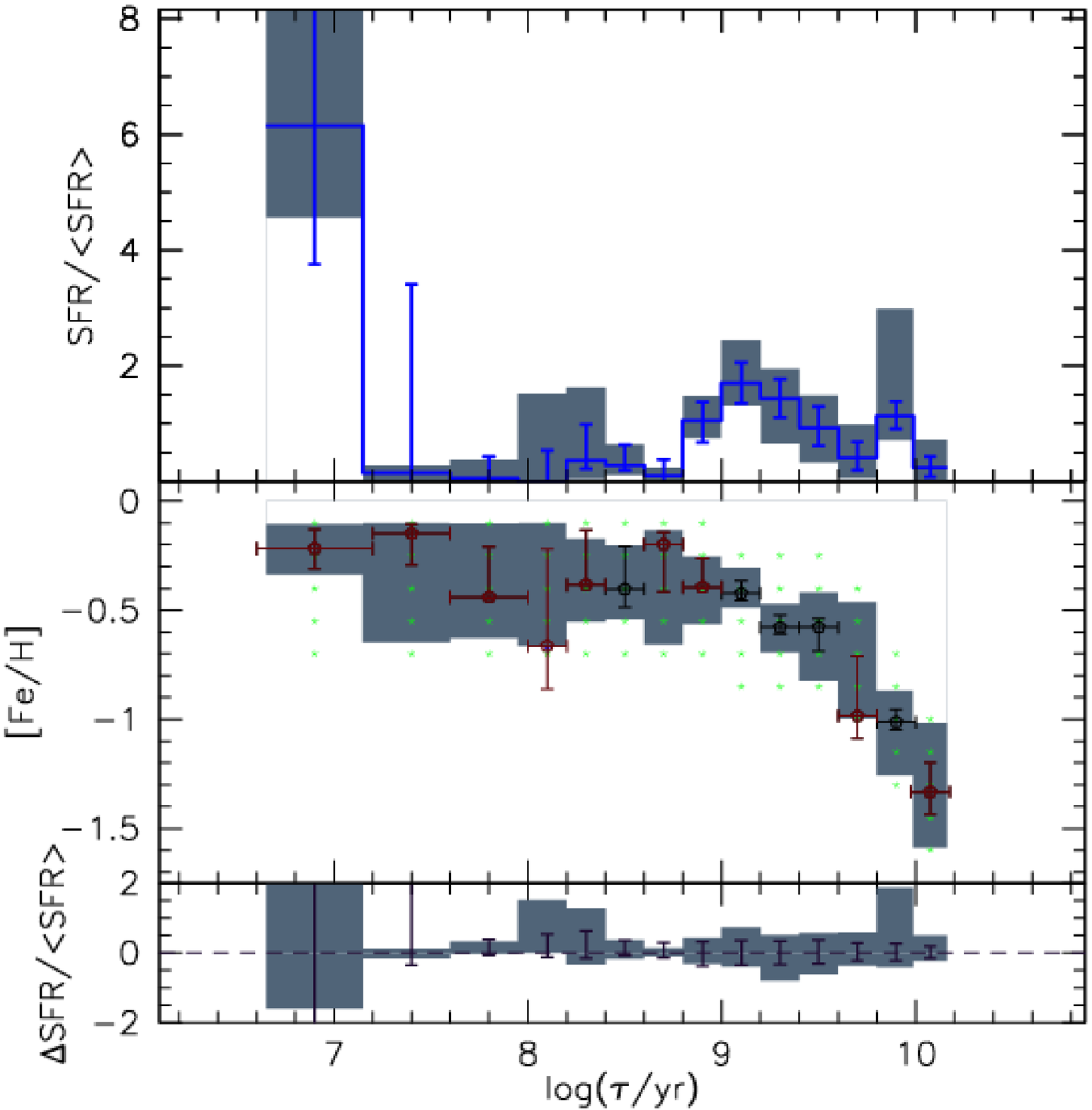}}
\resizebox{0.33\hsize}{!}{\includegraphics{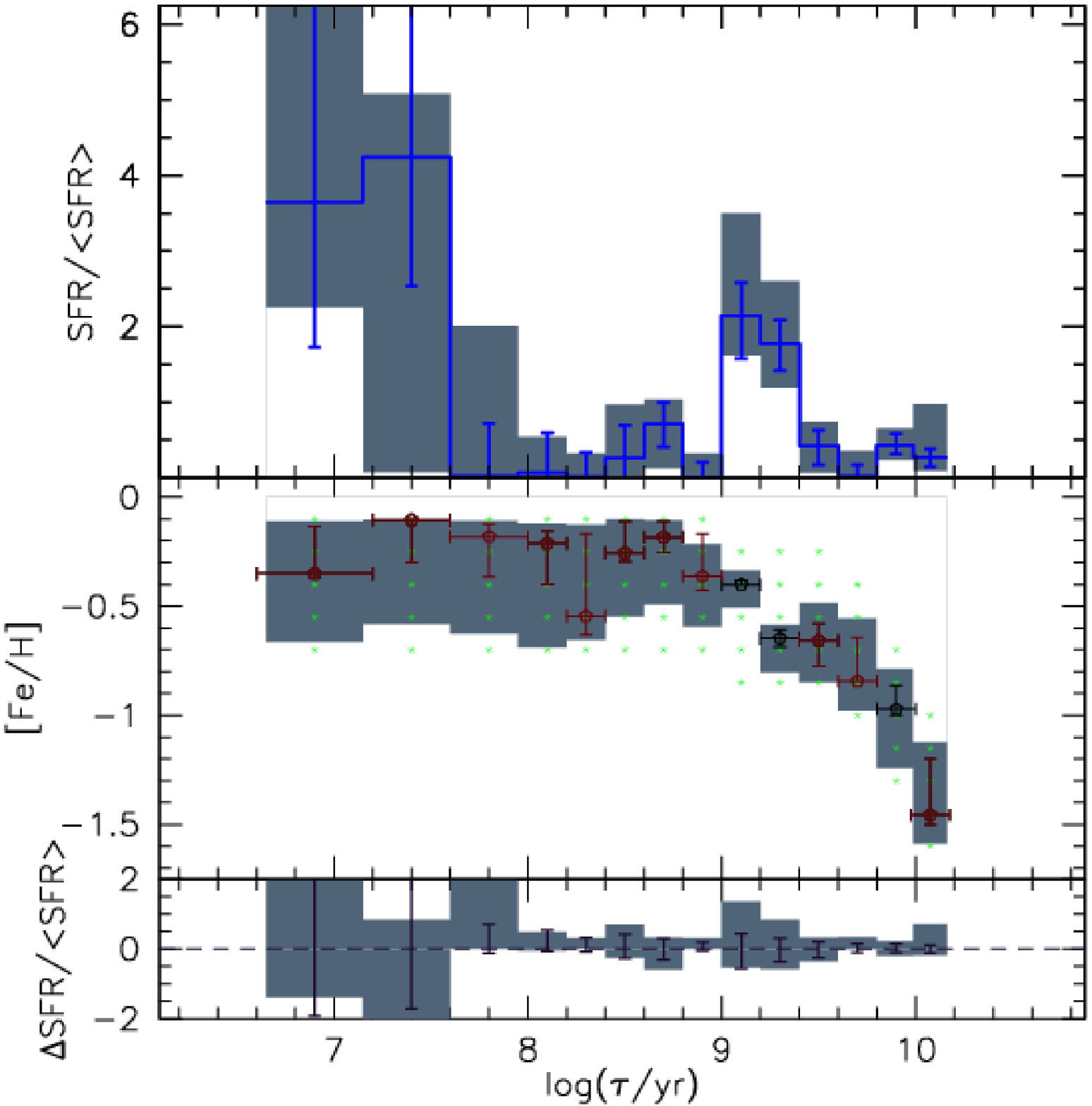}}
\caption{The SFH derived for the NGC~1846 and NGC~1783 Fields ({\bf
    left and right panels}, respectively). In both cases, the {\bf top
    panel} presents the best-fitting SFR$(t)$ for the field (blue
  histogram), together with the random errors ($1\sigma$ blue bar) and
  systematic errors (gray shadow) as estimated from the entire $68~\%$
  confidence level interval in the $\dmo, \av$ plane.  The {\bf bottom
    panels} present the mean age--metallicity relation (red and black
  point) with stochastic errors (red and black) and systematic errors
  (shaded regions). The green points show the center of distributions
  of the SPMs used to derive the SFH.}
\label{fig_fieldSFH}
\end{figure*}

Deriving the SFH in the LMC fields is not a main goal in this paper,
but it is both an interesting side-product of the present analysis,
and a necessary step to reduce systematic errors in the cluster SFH.

For both the NGC~1846 and NGC~1783 Fields, we derive the SFH using the
same set of SPMs as defined in \citet{Rubele_etal11}. We then run
StarFISH to find the best-fitting solution to the observed CMDs, for a
given value of distance modulus \dmo\ and extinction \av. Both F814W
vs.\ F435W$\!-\!$F814W and F814W vs.\ F555W$\!-\!$F814W Hess diagrams
are used simultaneously in the process of $\chi^2$ minimization.

Figure~\ref{fig_fieldchimap} shows the map of $\chi^2_{\rm min}$ --
that is, the $\chi^2$ value to which StarFISH converges -- for the
solutions in the $\dmo\times\av$ plane. We explored a range in both
parameters just extended enough to allow a clear identification of the
absolute minimum and most of its 68~\% confidence level interval.
There are a couple of remarkable aspects in our results.

First, the best-fitting models for the NGC~1846 and NGC~1783 Fields
(left and right panels of Fig.~\ref{fig_fieldchimap}, respectively)
present quite a different mean extinction, with $\av=0.33$~mag and
$\av=0.225$~mag, respectively.  In comparison, for a radius of
12~arcmin around the clusters NGC~1846 and NGC~1783,
\citet{Zaritsky_etal04} derive extinction values of
$\langle\av\rangle=0.59\pm0.40$ and $\langle\av\rangle=0.41\pm0.42$,
respectively ($\langle\av\rangle=0.48\pm0.32$ and
$\langle\av\rangle=0.31\pm0.26$ for cool stars). The reddening maps by
\citet{Haschke_etal11} cover only the region of NGC~1846, providing
$\langle E_{V\!-\!I}\rangle\ge0.06\pm0.075$, which translates into
$\langle\av\rangle\ga 0.14$~mag. Considering the errors and the large
dispersion in extinction values, all these values are consistent with
each other.

Second, the best-fitting distances for both fields just marginally
agree one with each other, considering their $68$~\% confidence level:
indeed the distance modulus of $\dmo\sim18.625$~mag represents, at the
same time, a lower limit to the distance of the NGC~1846 field, and an
upper limit to the distance of the NGC~1783 one. Although these
distance measurements are perfectly compatible when we consider their
$95$~\% confidence levels, we cannot refrain from noticing this
unexpected result. Both fields have similar location in the Northwest
portion of the LMC, relatively close to the line of nodes of the LMC
disk. So, according to recent results for the LMC disk geometry
\citep[e.g.][]{vdMC01, vanderMarel_etal02, Nikolaev_etal04,
  Rubele_etal12}, they would be expected to have the same distance of
the LMC center (of $\dmo\sim18.46$~mag, see \citealt{Ripepi_etal12}
and references therein). Although the relative distances between the
cluster fields and the LMC centre could be affected by systematic
errors, the relative distances between both cluster fields should be
quite solid.

We note however that both the distances and extinction values for the
field populations are consistent, within their 1$\sigma$
uncertainties, with the values found for the cluster centres, as
detailed in the next subsection.

Finally, Fig.~\ref{fig_fieldSFH} presents the SFR$(t)$ and AMR
corresponding to the best-fitting solutions for both Fields. It is
remarkable that the SFR$(t)$ recovered for the NGC~1846 Field presents
features that are consistent with those commonly found in previous
works, based on both HST \citep{Olsen99, Holtzman_etal99,
  Smecker-Hane_etal02, Javiel_etal05} and ground-based data
\citep{HZ01, HZ09, Rubele_etal12}. There is an initial period of star
formation at $\sim\!10$~Gyr, followed by a minimum at $\log(t/{\rm
  yr})\simeq9.7$, and then a more extended period of star formation
for ages between about 1 and 4~Gyr. For younger ages, our data
includes too small an area to set stringent constraints on the SFH;
however, there are hints for significant SFR at ages of about 300~Myr
($\log(t/{\rm yr})=8.5$), and a well detected burst of formation of
stars with $\sim\!10$~Myr ($\log(t/{\rm yr})=7.0$). Error bars are too
large to allow a meaningful quantitative comparison with the results
from \citet{HZ09} and \citet{Rubele_etal12}, for nearby regions of the
LMC.

\begin{figure*}
\resizebox{0.33\hsize}{!}{\includegraphics{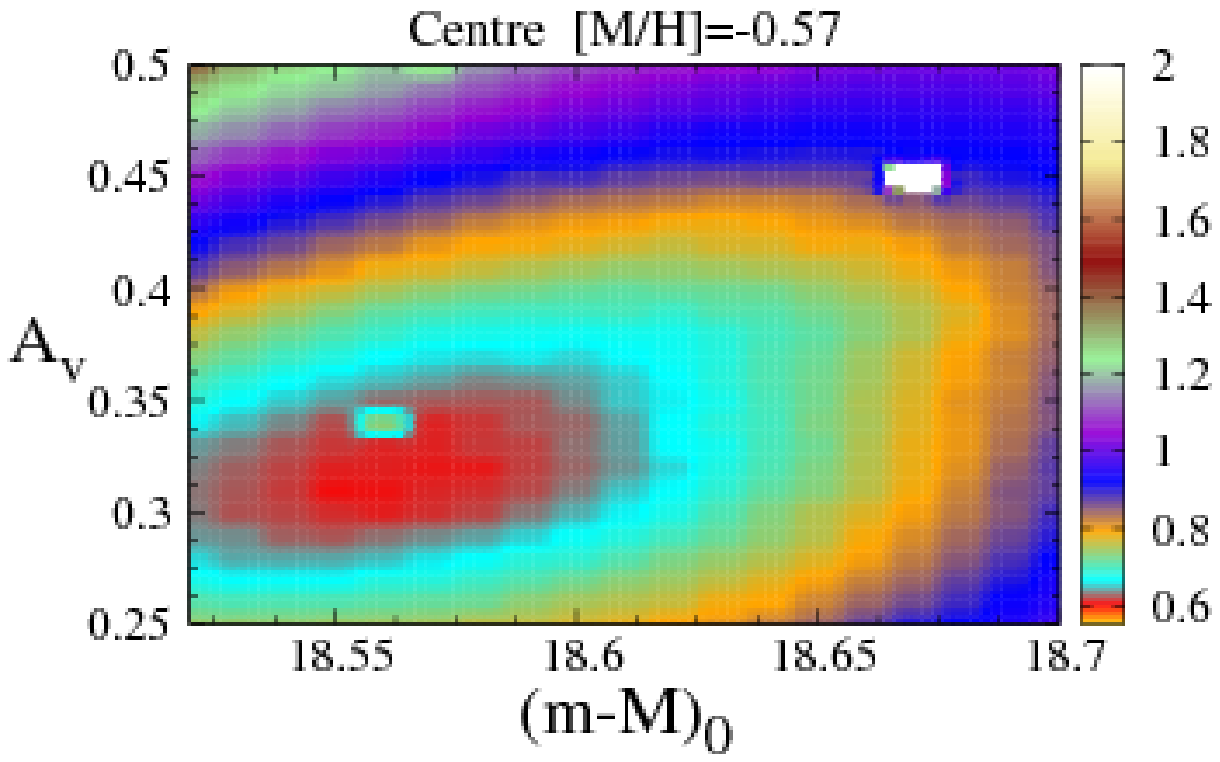}}
\resizebox{0.33\hsize}{!}{\includegraphics{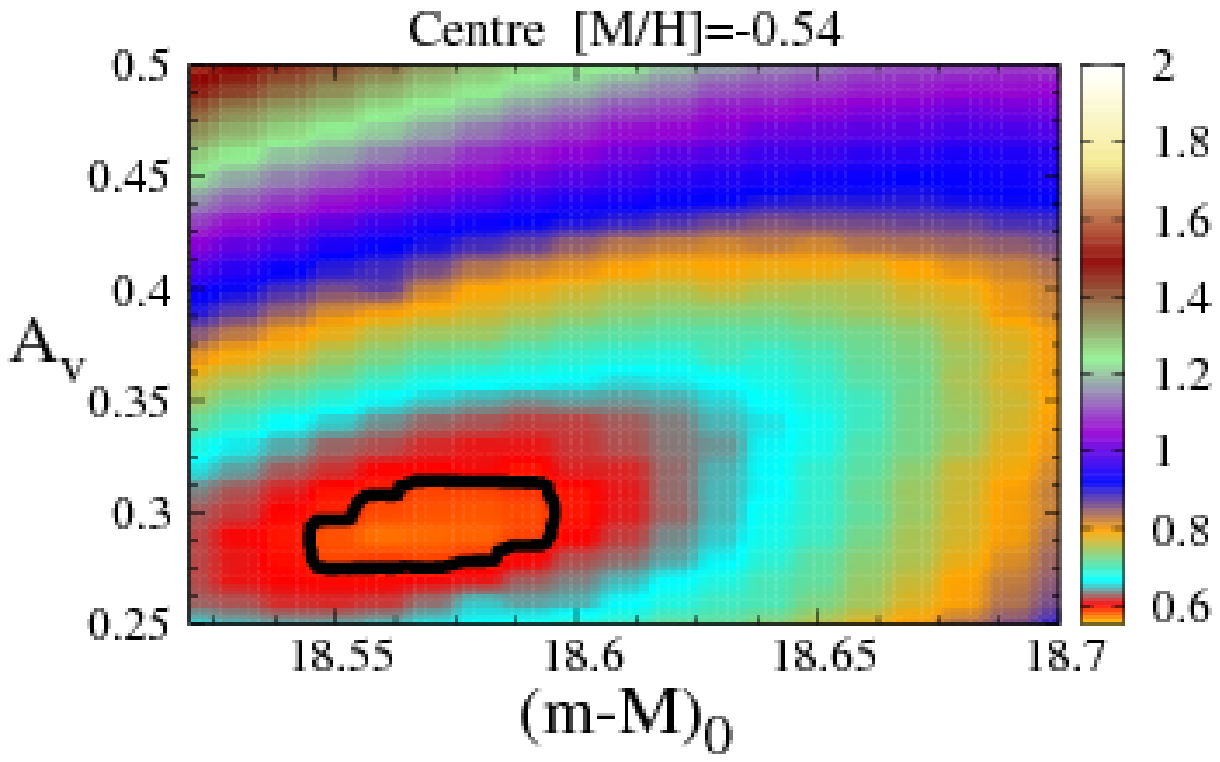}}
\resizebox{0.33\hsize}{!}{\includegraphics{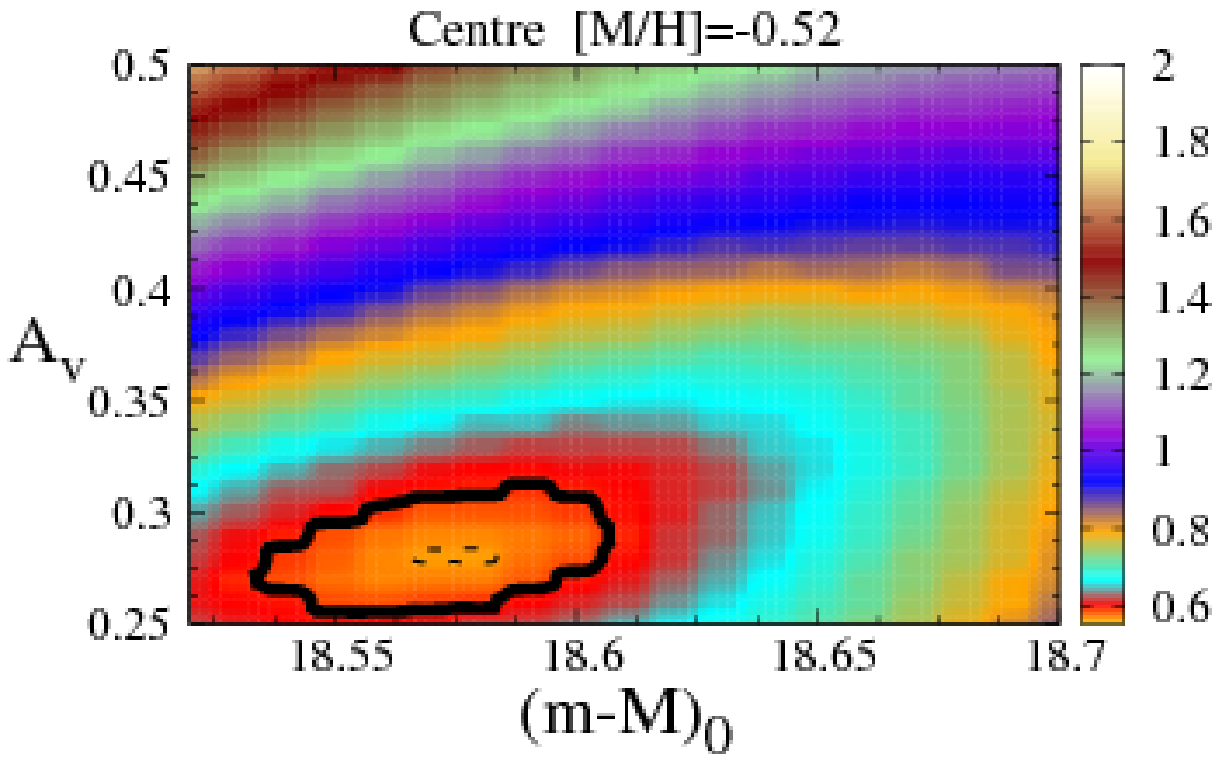}}
\hfill\\
\resizebox{0.33\hsize}{!}{\includegraphics{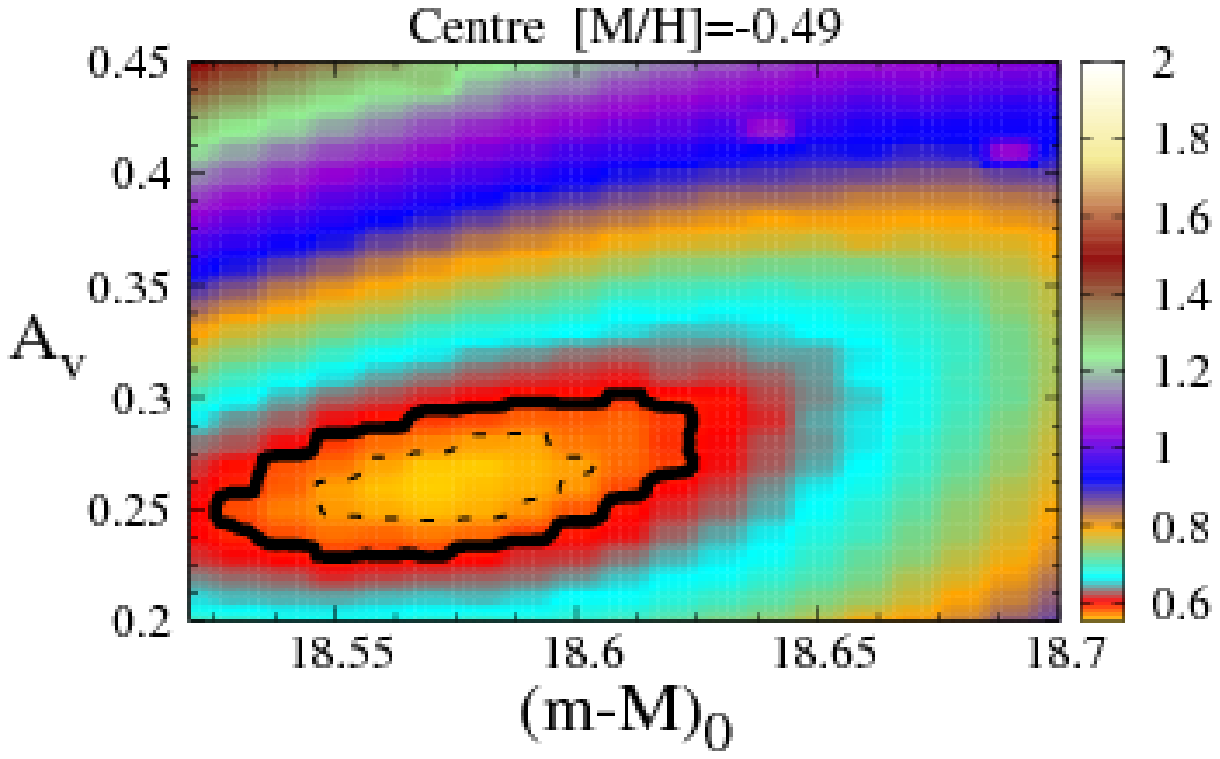}}
\resizebox{0.33\hsize}{!}{\includegraphics{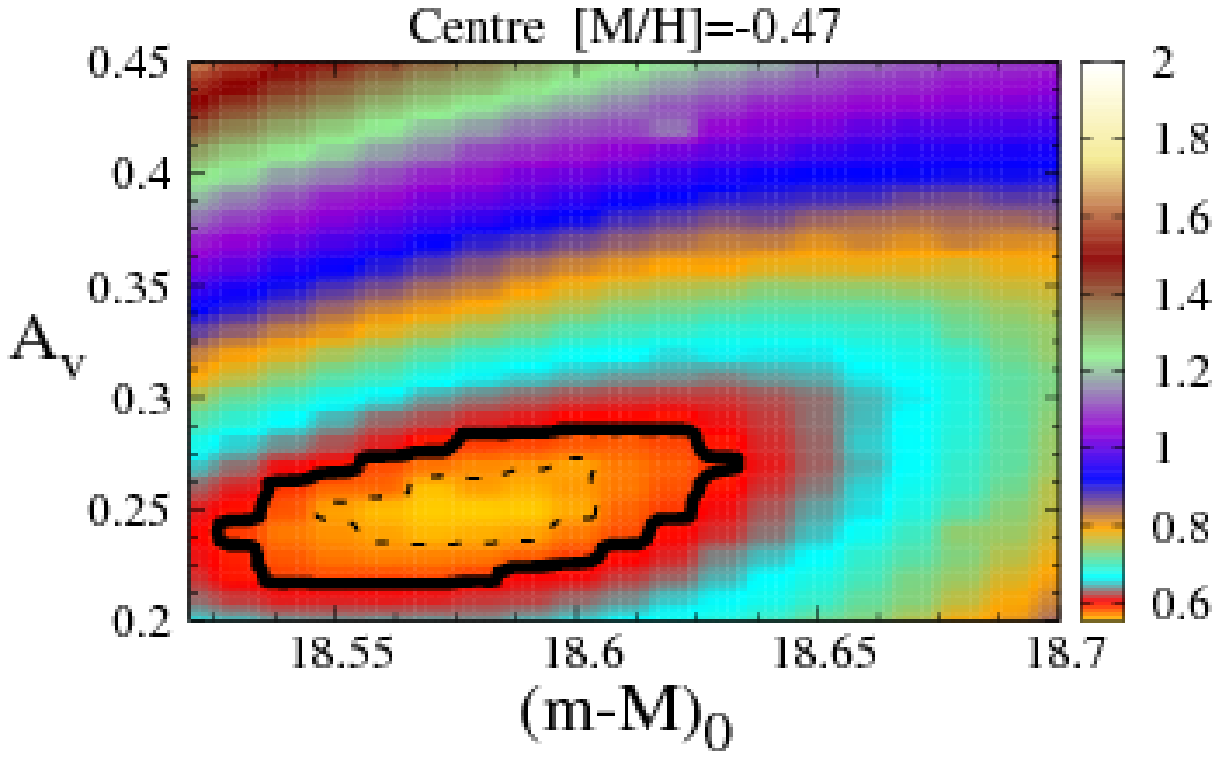}}
\resizebox{0.33\hsize}{!}{\includegraphics{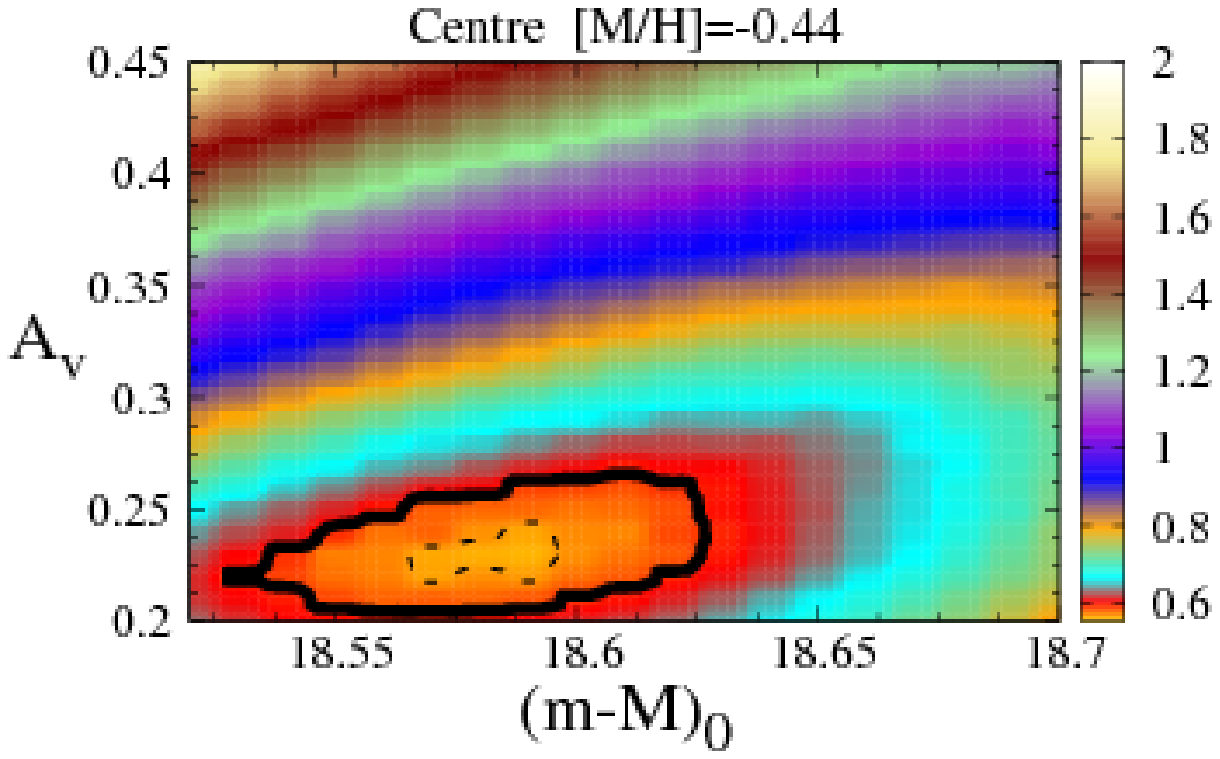}}
\hfill\\
\resizebox{0.33\hsize}{!}{\includegraphics{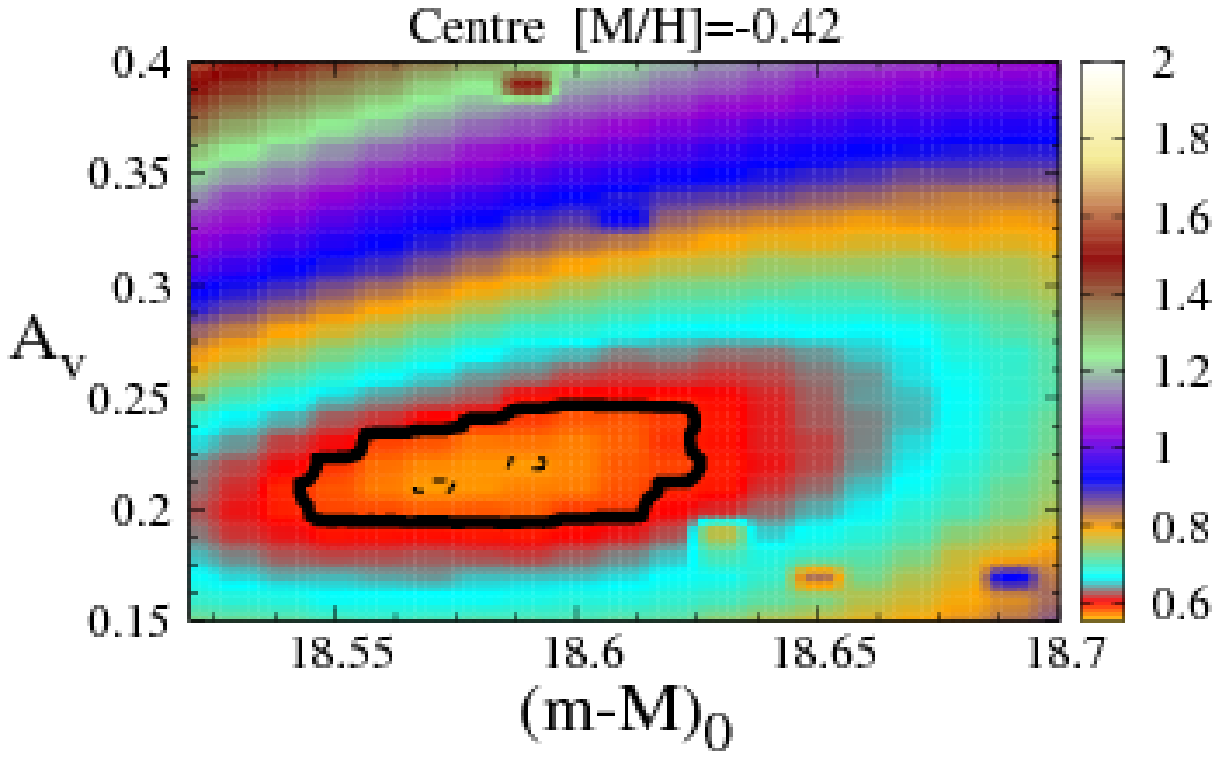}}
\resizebox{0.33\hsize}{!}{\includegraphics{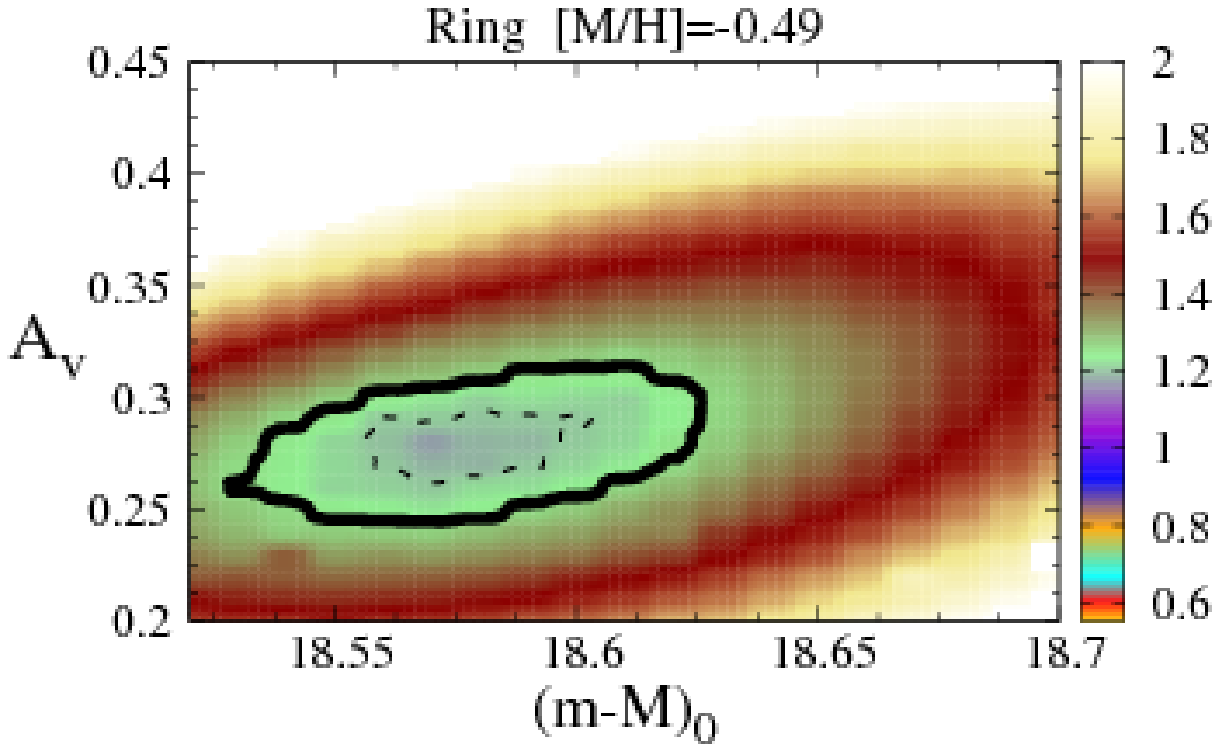}}
\hfill
\caption{The {\bf first 7 panels} show the maps of the $\chisqmin$
  obtained from the SFH-recovery of the NGC~1846 Centre, as a function
  of $\dmo$ and $\av$, for several $\mh$ values (from $-0.57$ to
  $-0.42$).  The black lines delimit the regions within the 68~\%
  (dashed lines) and 95~\% confidence levels (continuous lines) of the
  absolute best solution, which is found at $\mh=-0.49$ and is shown
  in the fourth panel. The {\bf last panel} shows the same for the
  NGC~1846 Ring, but only for the metallicity of $\mh=-0.49$. The
  minimum $\chisqmin$ are of 0.55 and 1.1, respectively, for the
  Centre and Ring. In both cases, the best-fitting solutions are found
  at $\dmo=18.57, \av=0.26$.}
\label{chi2_map}
\end{figure*}


The SFR$(t)$ for the NGC~1783 Field, instead, presents a marked
``burst'' at ages between 1 and 2 Gyr, which, as we will see later,
coincides with the ages of cluster formation. This is a further
evidence that in NGC~1783 Field we are still sampling the cluster
population.

Concerning the AMR, the results for the NGC~1846 and NGC~1783 fields
are quite similar and consistent with those derived from LMC star
clusters \citep{Kerber_etal07, HZ09} and for the LMC field using
different sets of data \citep[e.g.][]{Carrera_etal08, Rubele_etal12}.


\subsection{The SFHs in NGC~1846 and NGC~1783}
\label{sec_sfhclusters}

\begin{figure*}
\begin{minipage}{\hsize}
\subfigure[Centre of NGC~1846]{
\resizebox{0.15\hsize}{!}{\includegraphics[trim=1.0cm 1.0cm 1.0cm 1.0cm]{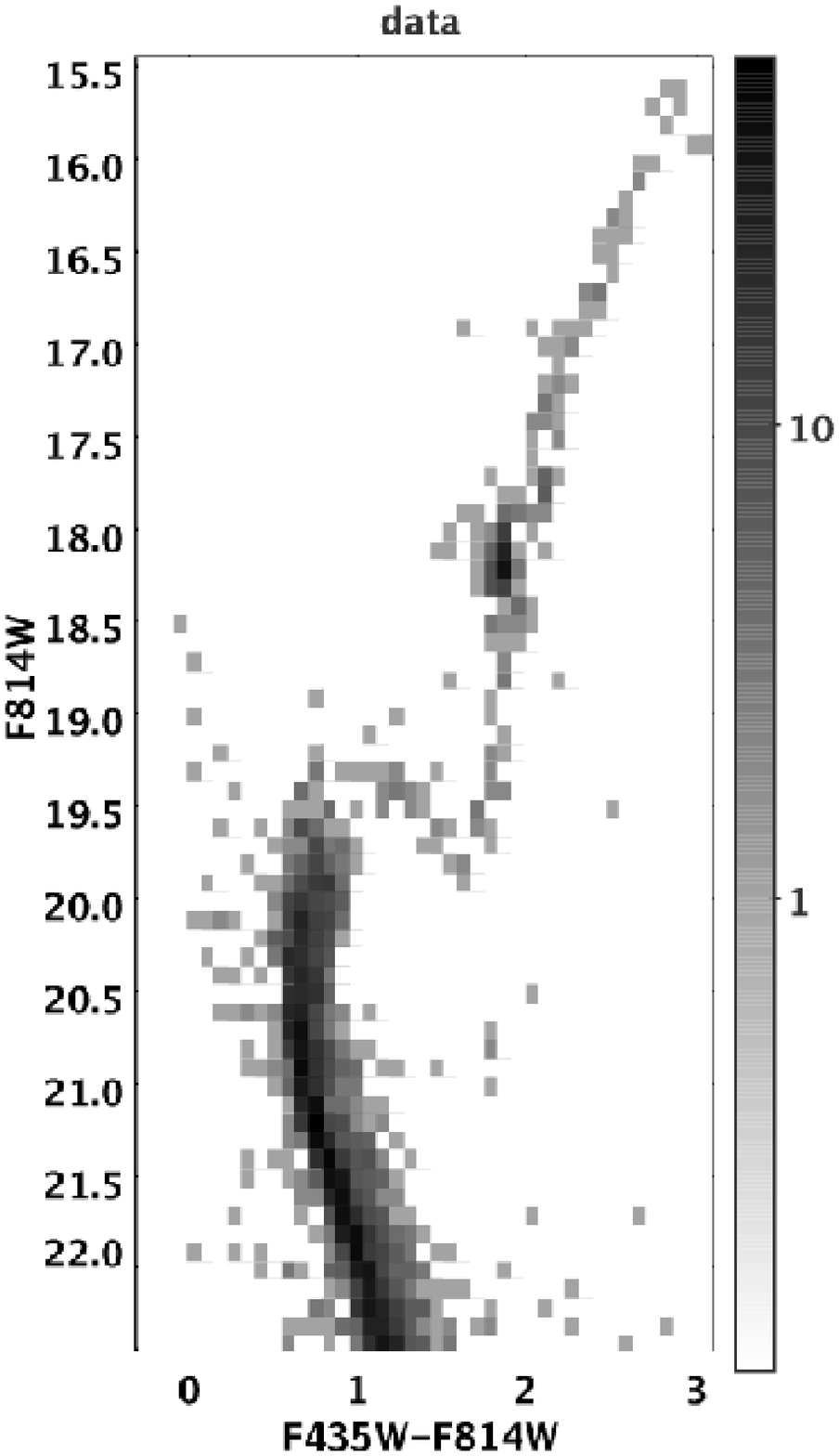}}
\resizebox{0.15\hsize}{!}{\includegraphics[trim=1.0cm 1.0cm 1.0cm 1.0cm]{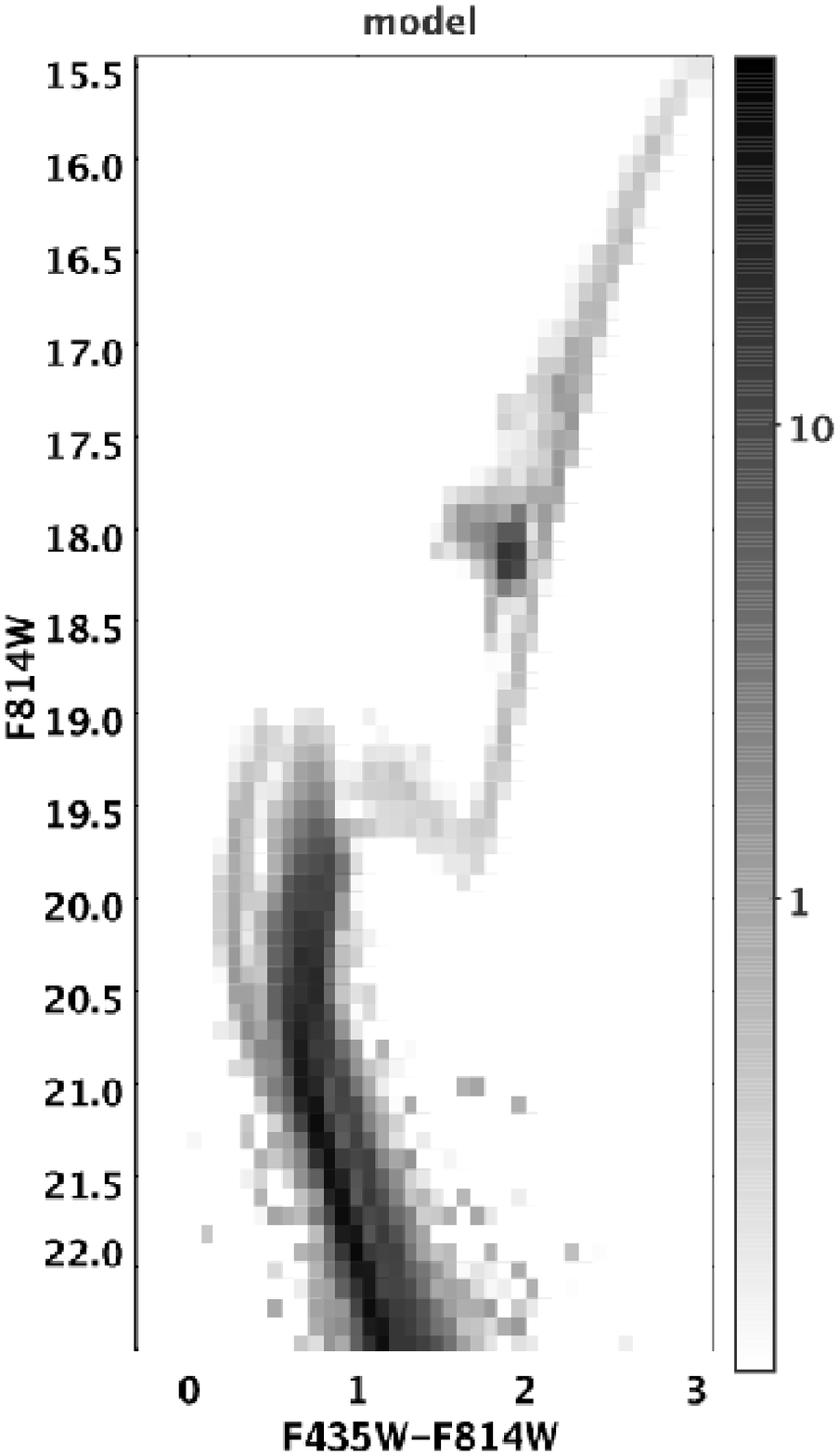}}
\resizebox{0.15\hsize}{!}{\includegraphics[trim=1.0cm 1.0cm 1.0cm 1.0cm]{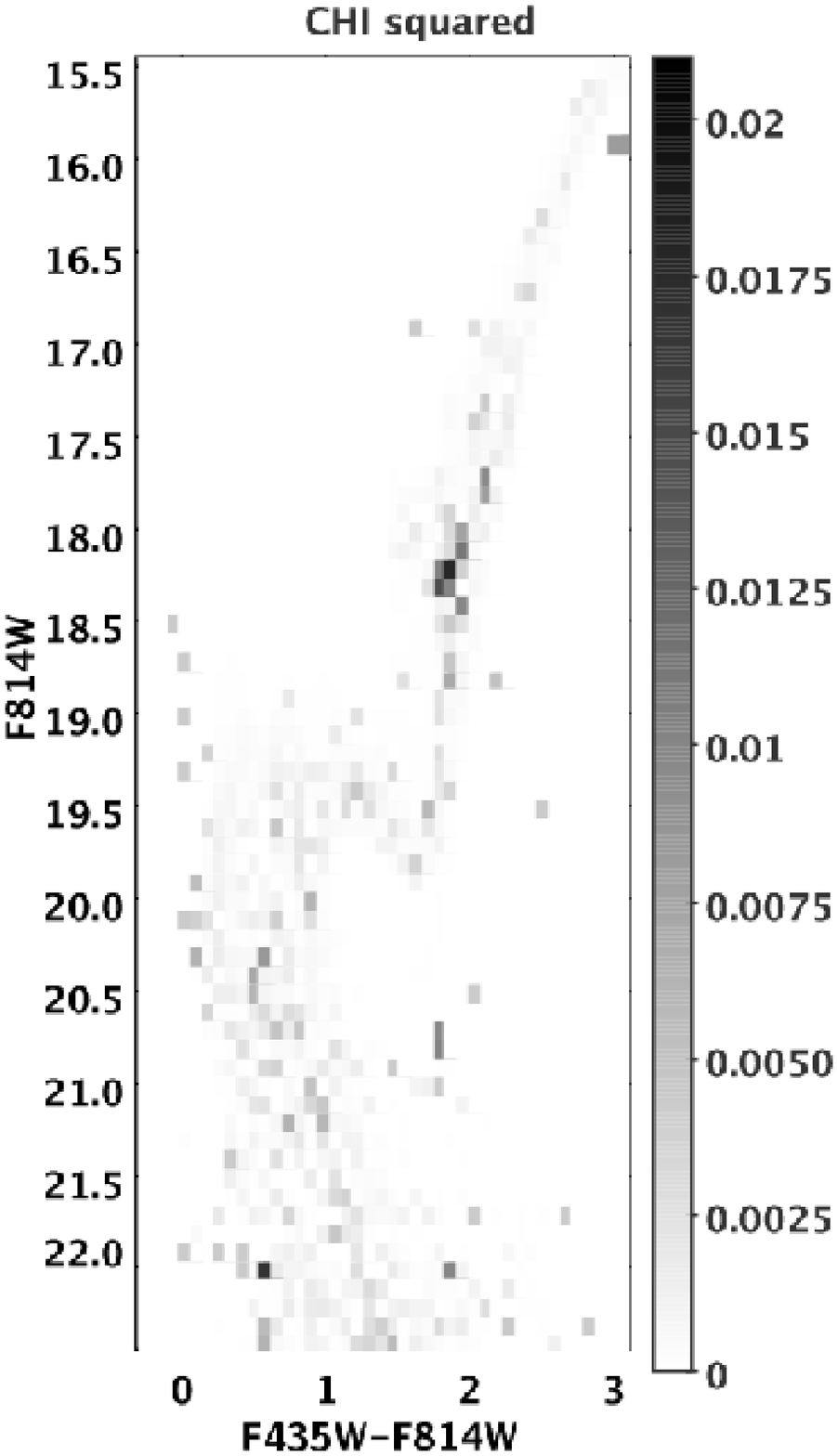}}
\hfill\hspace{0.5cm}
\resizebox{0.15\hsize}{!}{\includegraphics[trim=1.0cm 1.0cm 1.0cm 1.0cm]{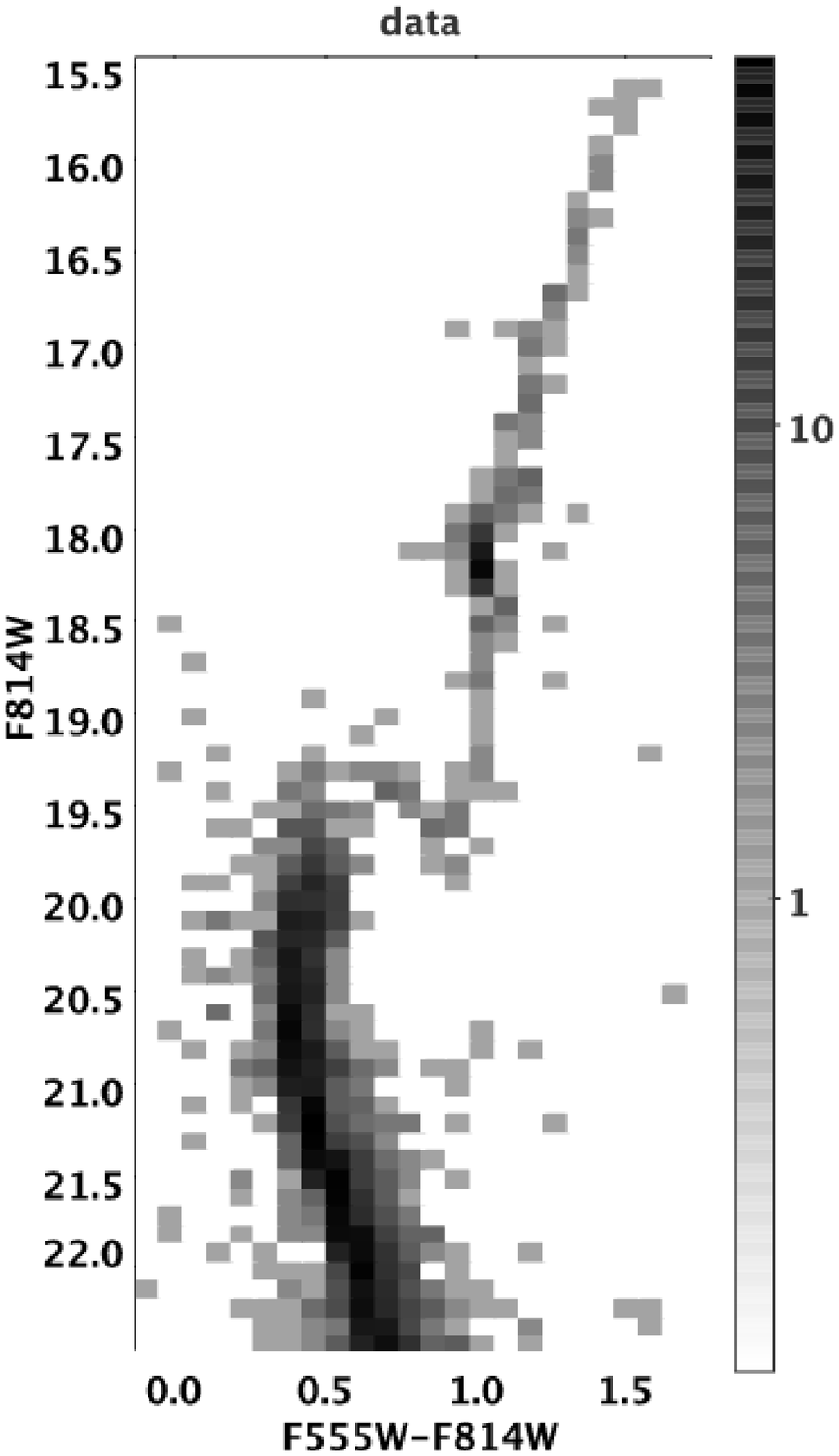}}
\resizebox{0.15\hsize}{!}{\includegraphics[trim=1.0cm 1.0cm 1.0cm 1.0cm]{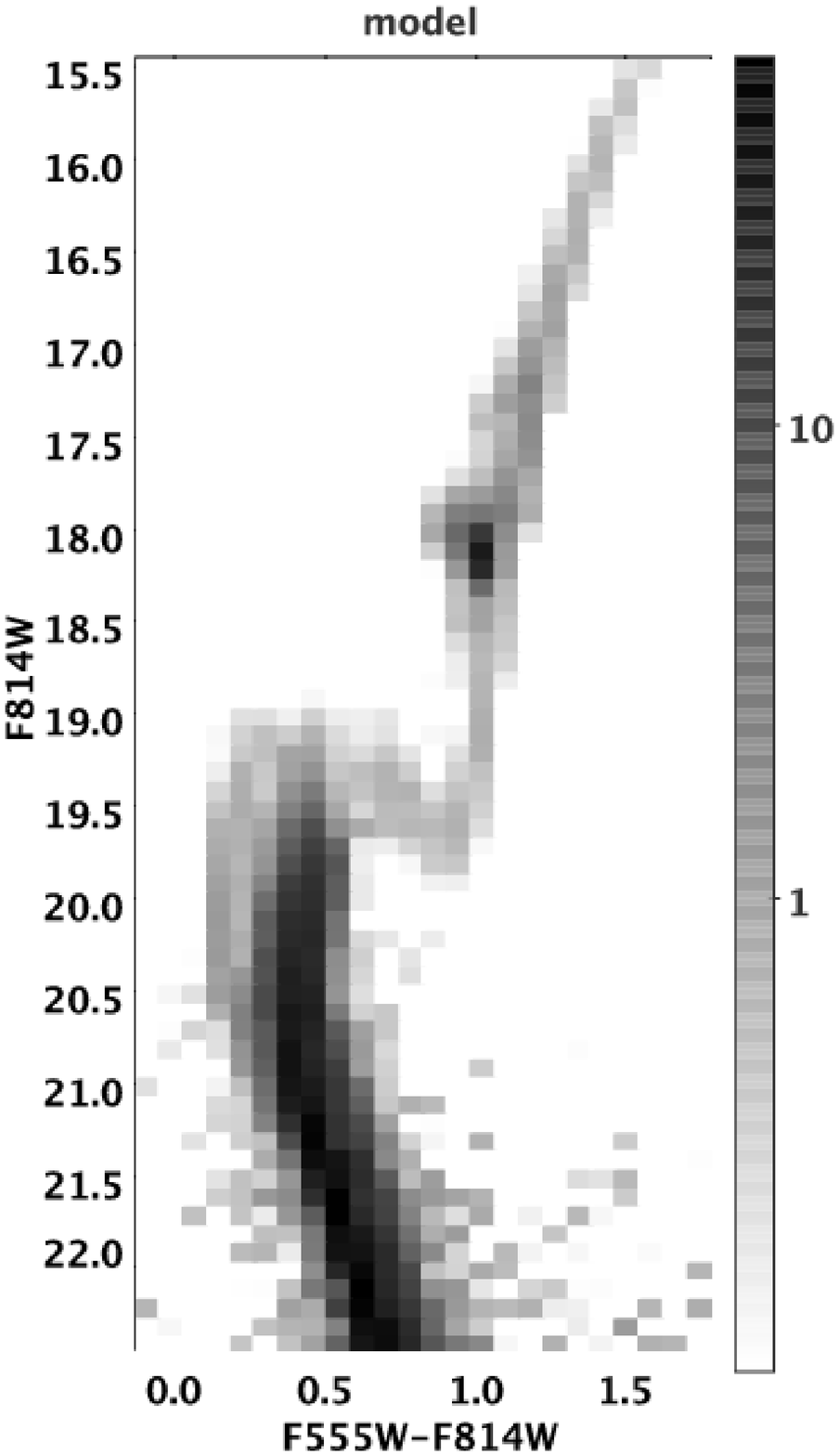}}
\resizebox{0.15\hsize}{!}{\includegraphics[trim=1.0cm 1.0cm 1.0cm 1.0cm]{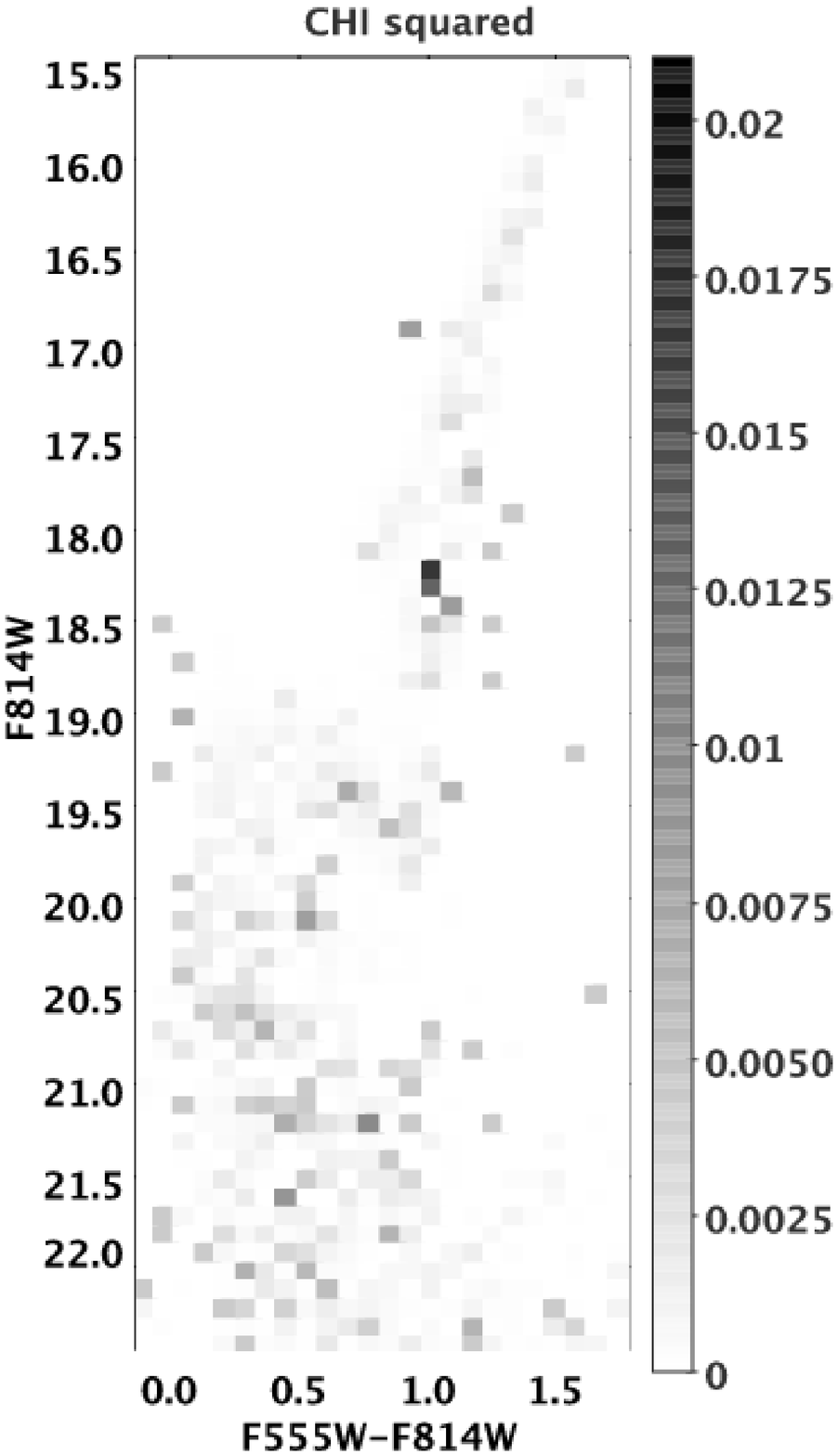}}
}
\end{minipage}
\begin{minipage}{\hsize}
\subfigure[Ring of NGC~1846]{
\resizebox{0.15\hsize}{!}{\includegraphics[trim=1.0cm 1.0cm 1.0cm 1.0cm]{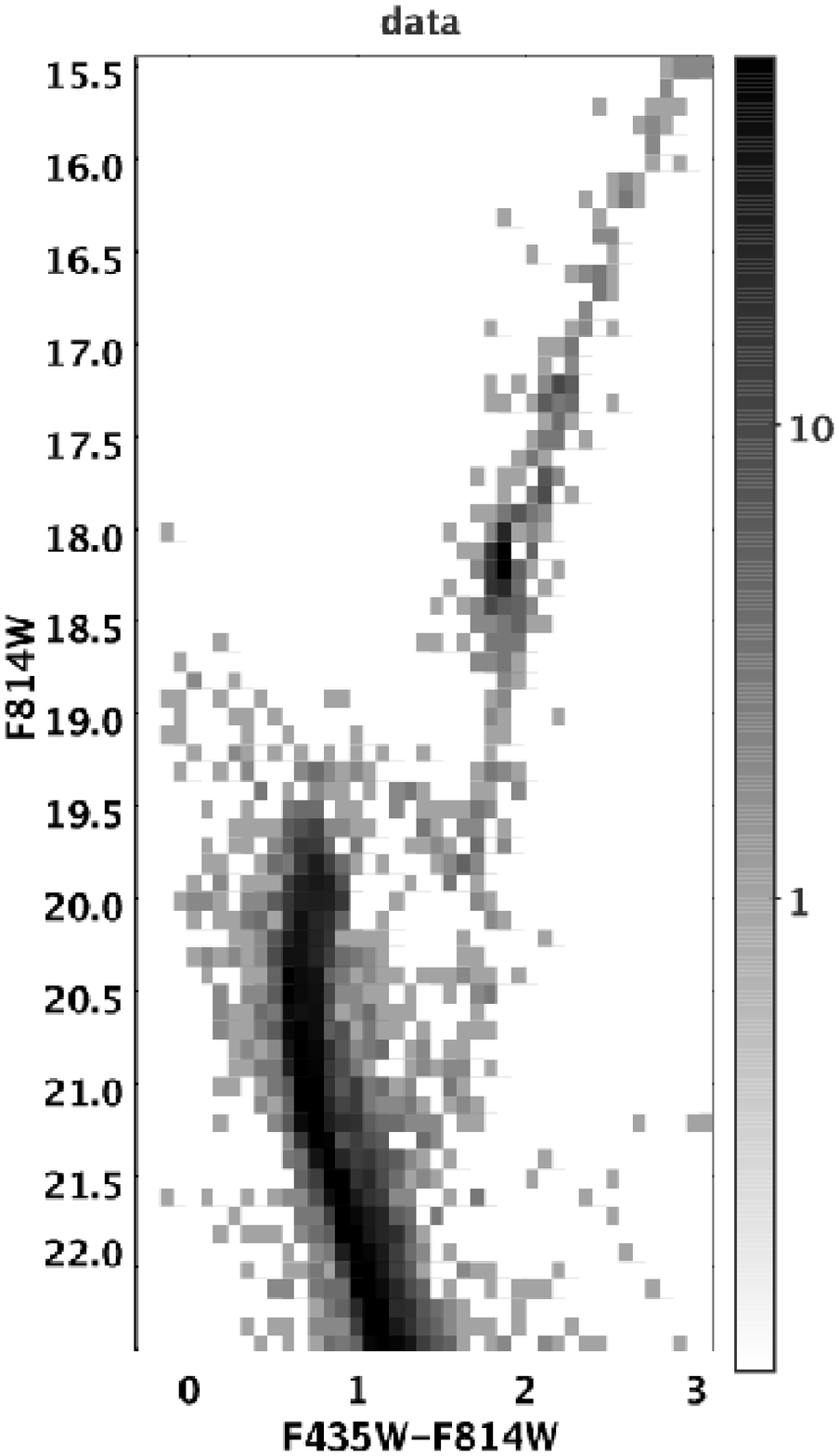}}
\resizebox{0.15\hsize}{!}{\includegraphics[trim=1.0cm 1.0cm 1.0cm 1.0cm]{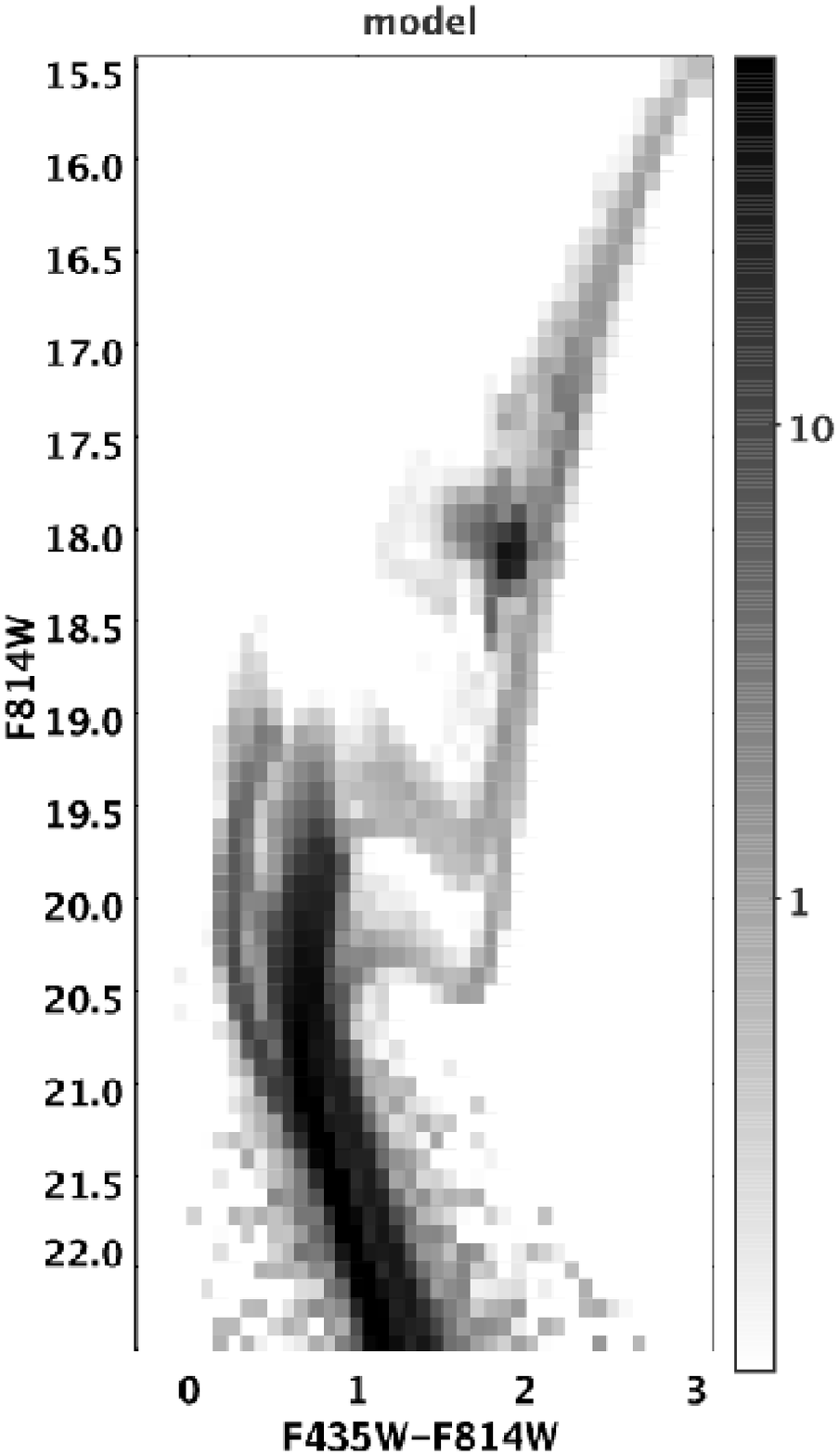}}
\resizebox{0.15\hsize}{!}{\includegraphics[trim=1.0cm 1.0cm 1.0cm 1.0cm]{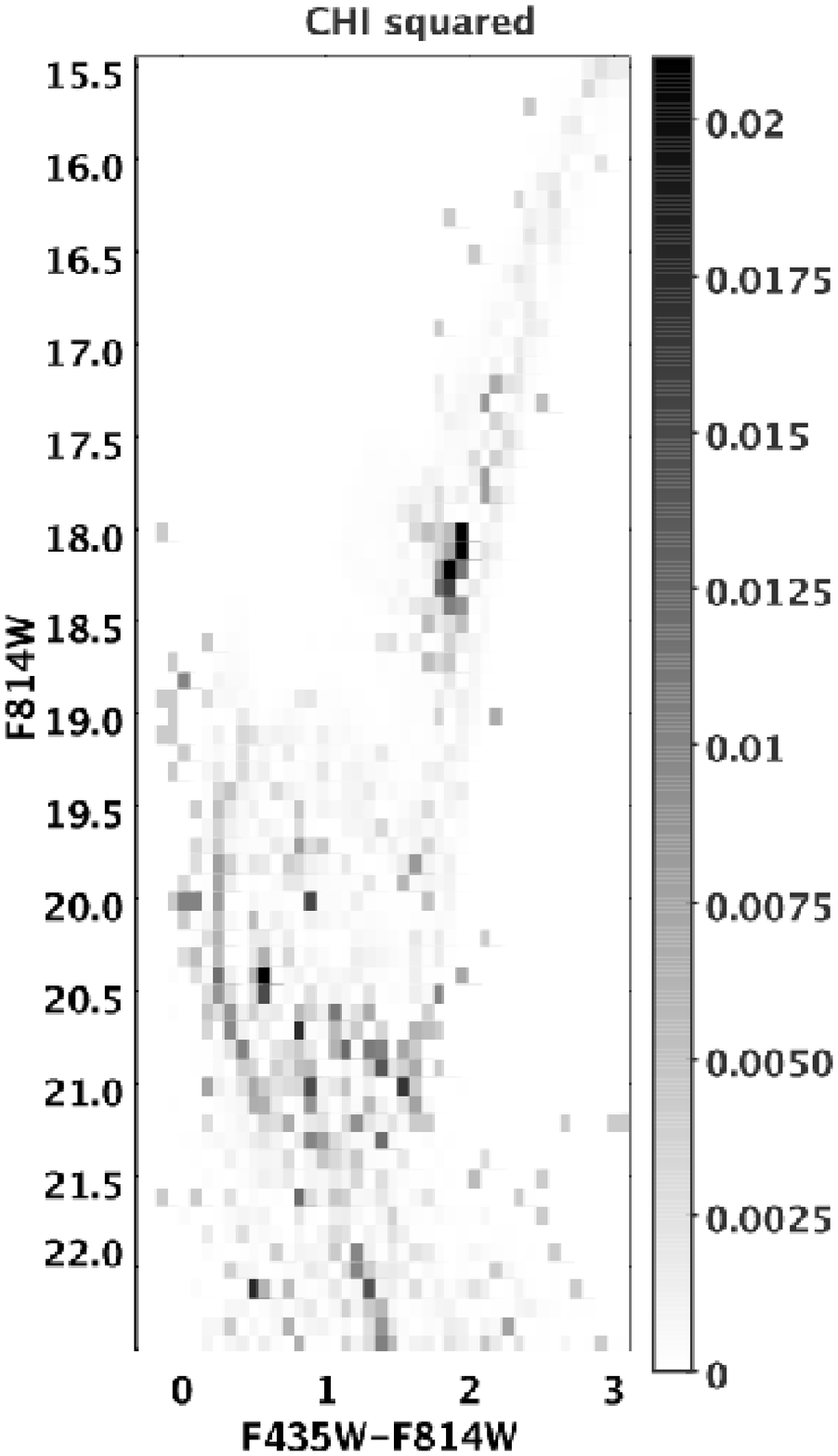}}
\hfill\hspace{0.5cm}
\resizebox{0.15\hsize}{!}{\includegraphics[trim=1.0cm 1.0cm 1.0cm 1.0cm]{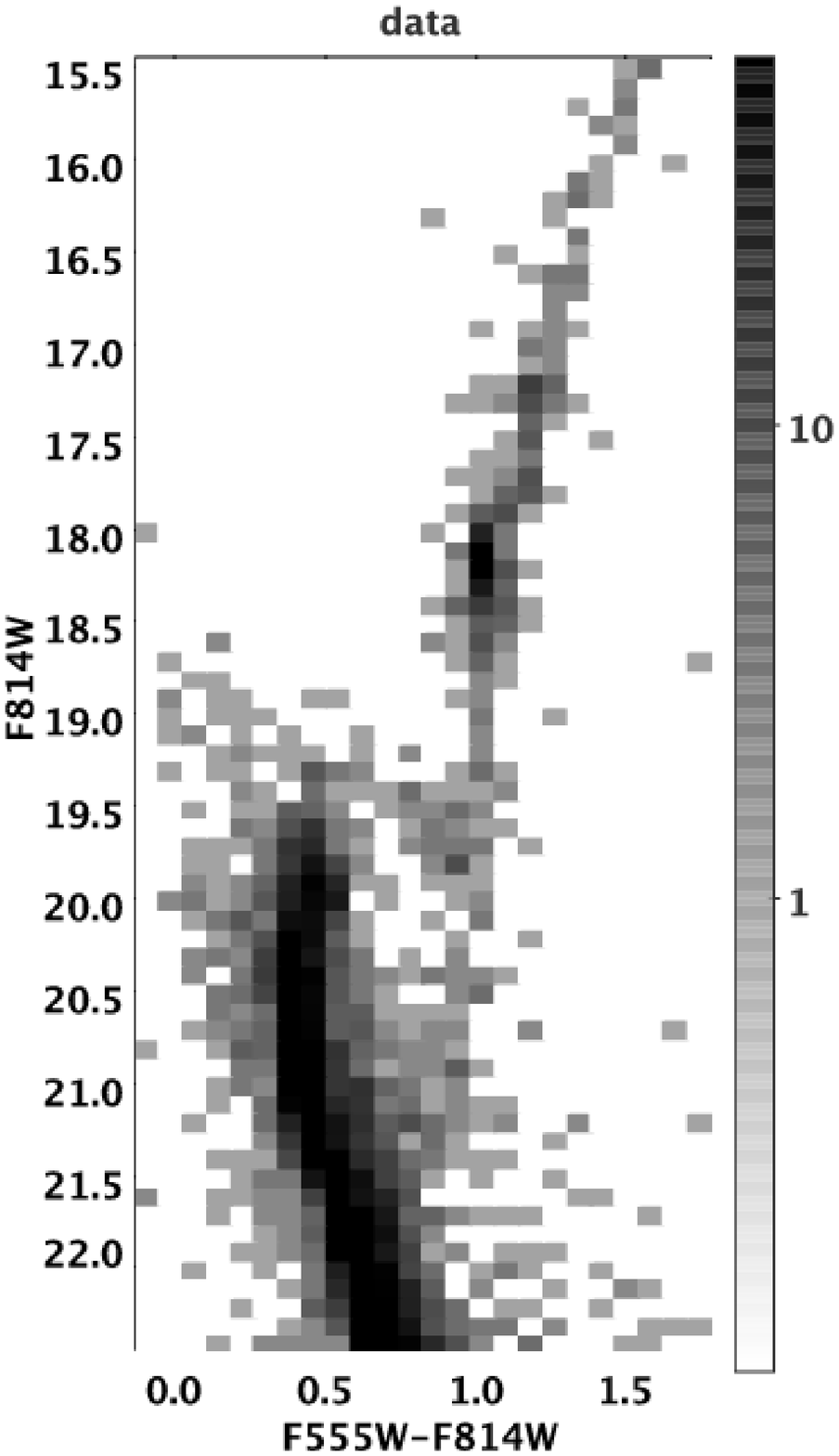}}
\resizebox{0.15\hsize}{!}{\includegraphics[trim=1.0cm 1.0cm 1.0cm 1.0cm]{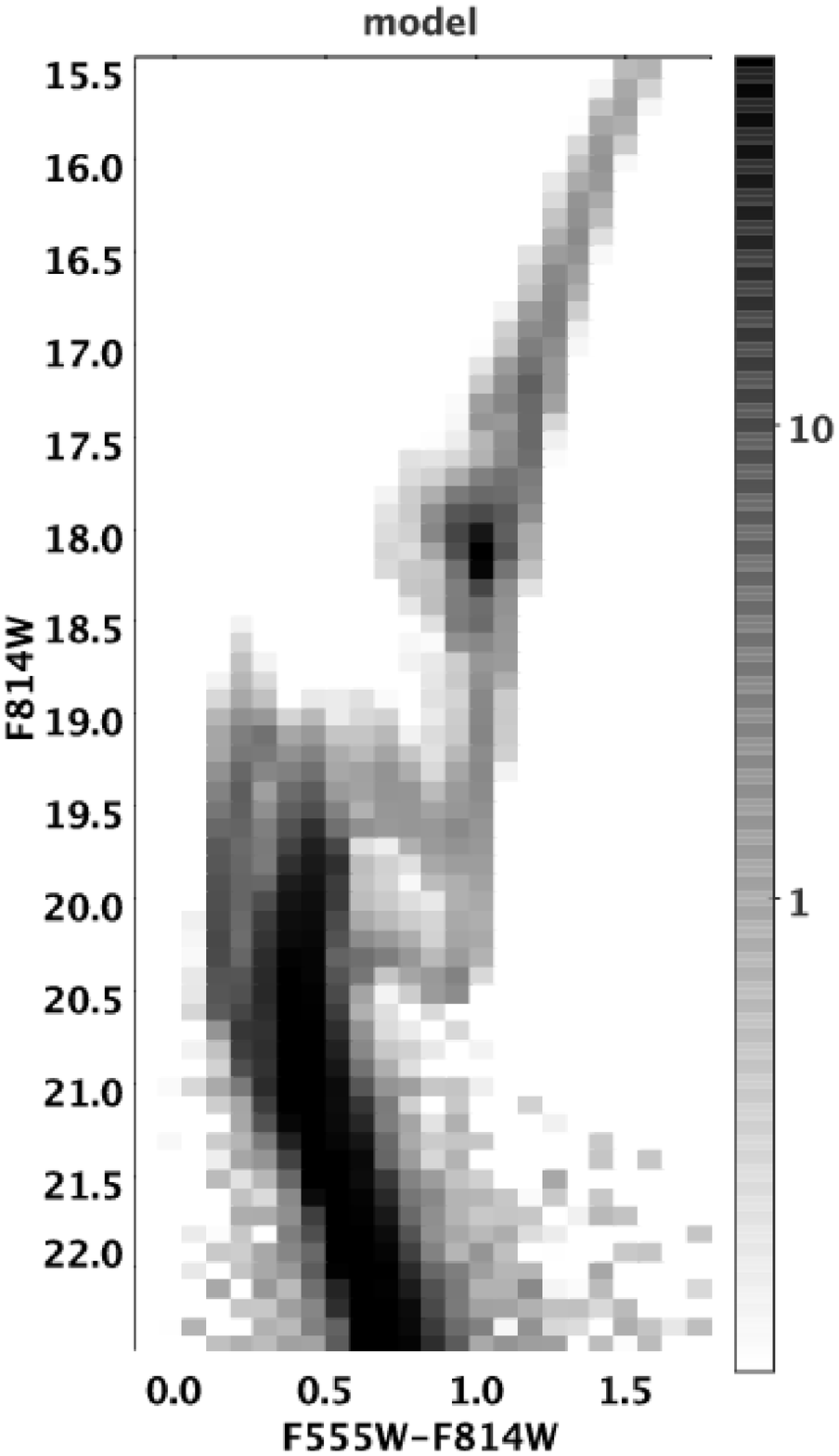}}
\resizebox{0.15\hsize}{!}{\includegraphics[trim=1.0cm 1.0cm 1.0cm 1.0cm]{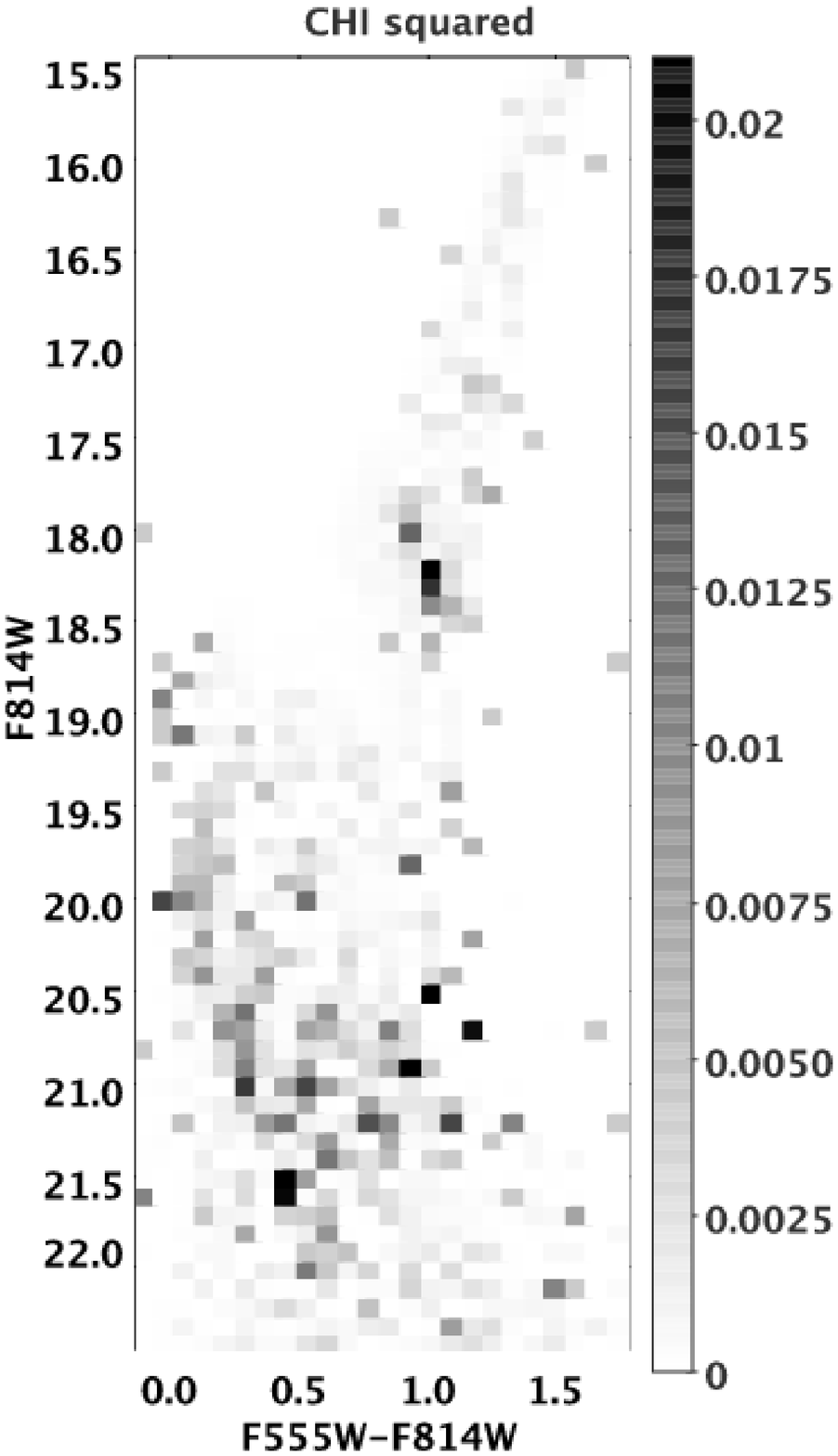}}
}
\end{minipage}
\caption{The Hess diagrams for the NGC 1846 Centre region (top panels)
  and Ring region (bottom panels). From left to right, the panels show
  the cluster F435W$\!-\!$F814W vs.\ F814W diagram, its best-fitting
  solution model, and the chi2 map. The same is done for the
  F555W$\!-\!$F814W vs.\ F814W diagram.}
\label{data_model}
\end{figure*}

For NGC~1846 and NGC~1783 we proceed deriving the best-fitting SFH in
the same way as performed for NGC~1751 \citep{Rubele_etal11}. In these
cases, we assume that cluster stars present the same mean \mh\ value
for all ages, since so far there is no evidence for significant
spreads in metallicity in such star clusters
\citep[e.g.][]{Mucciarelli_etal08, Rubele_etal10, Rubele_etal11}. We
have explored seven $\mh$ values: $-0.57$, $-0.54$, $-0.52$, $-0.49$,
$-0.47$, $-0.44$, $-0.42$.  For each one of these mean $\mh$ values, a
box-shaped metallicity distribution is assumed, with a total width of
$\Delta\mh=0.025$~dex. This spread is similar to the separation
between the mean $\mh$ values, and contributes to produce results that
vary smoothly as a function of metallicity.

The age interval covered by our SPMs goes from $\log(t/{\rm yr})=8.9$
to $9.4$, which is much wider than the interval suggested by the
position of NGC~1846 and NGC~1783 MMSTOs. We initially adopt an age
resolution (bin width of SPMs) of $\Delta\log t=0.05$~dex. So, for
each set of parameters, we have a total of 11 partial models -- 10 for
the cluster, plus the FSPM described in
Sect.~\ref{sec_overview_recovery} -- completely encompassing the age
interval of interest.


\begin{figure*}
\subfigure[SFR$(t)$ for NGC~1846 Centre]{
\resizebox{0.3\hsize}{!}{\includegraphics{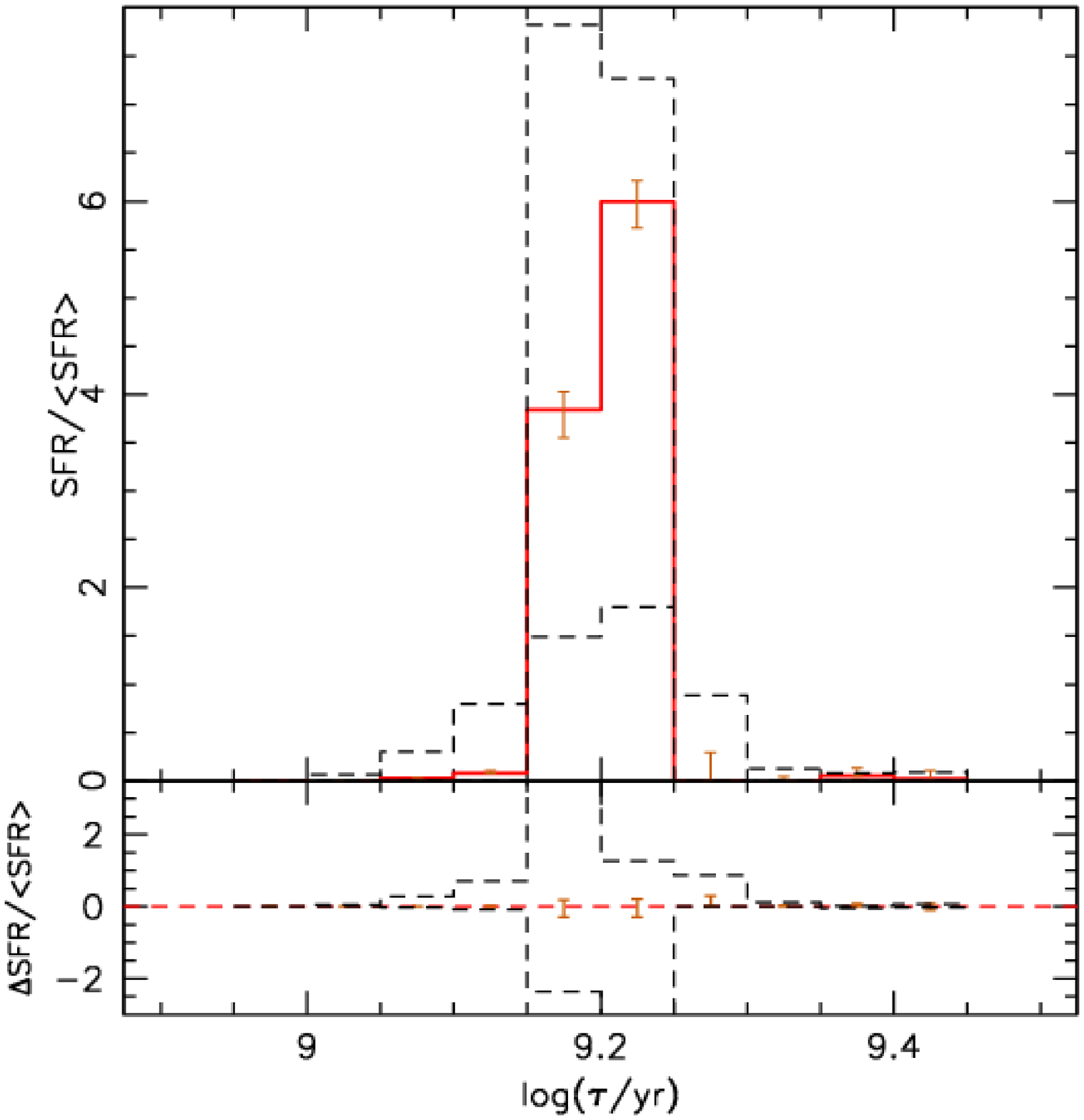}}
\resizebox{0.3\hsize}{!}{\includegraphics{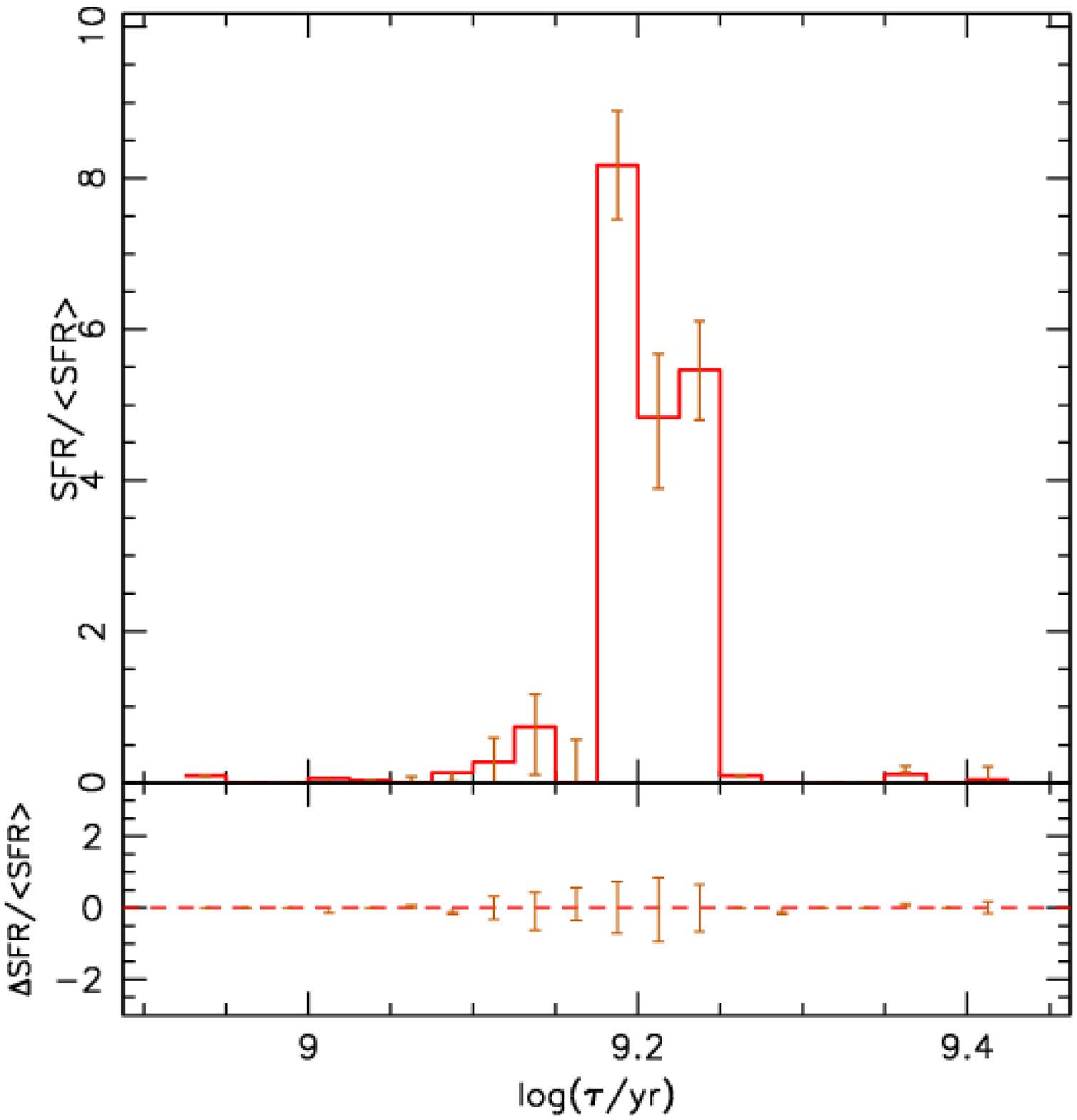}}
\resizebox{0.3\hsize}{!}{\includegraphics{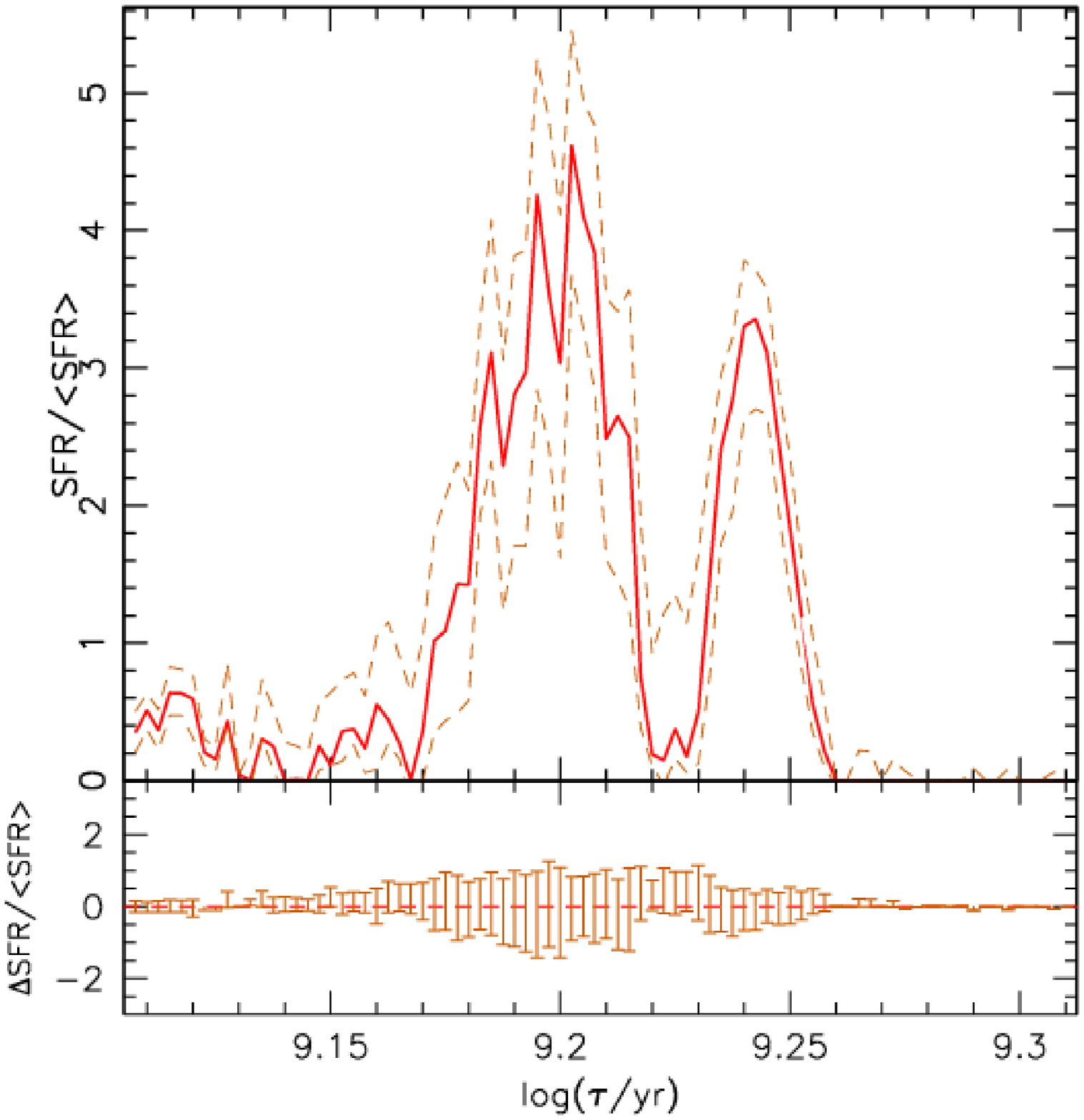}}
}
\subfigure[SFR$(t)$ for NGC~1846 Ring]{
\resizebox{0.3\hsize}{!}{\includegraphics{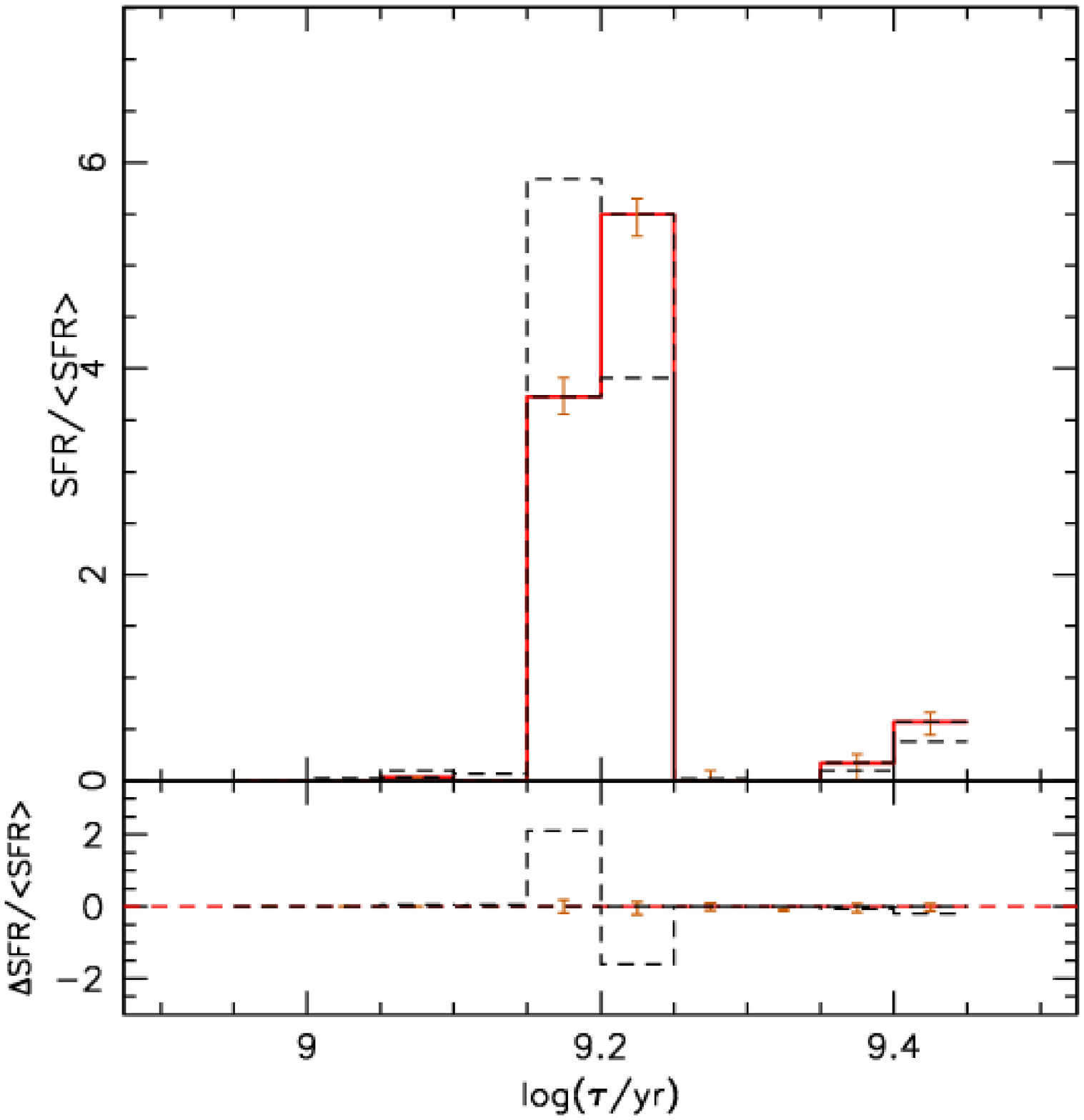}}
\resizebox{0.3\hsize}{!}{\includegraphics{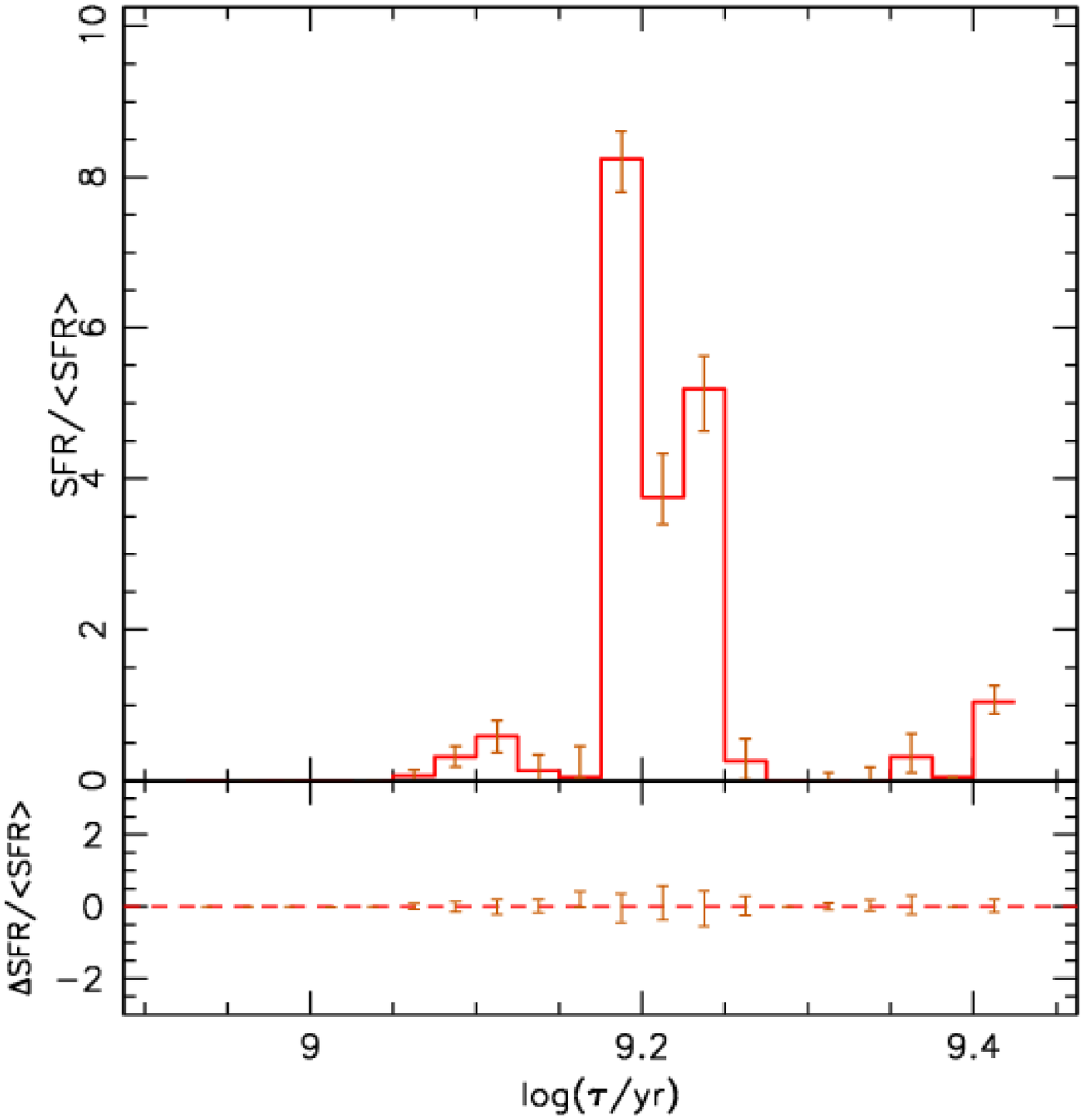}}
\resizebox{0.3\hsize}{!}{\includegraphics{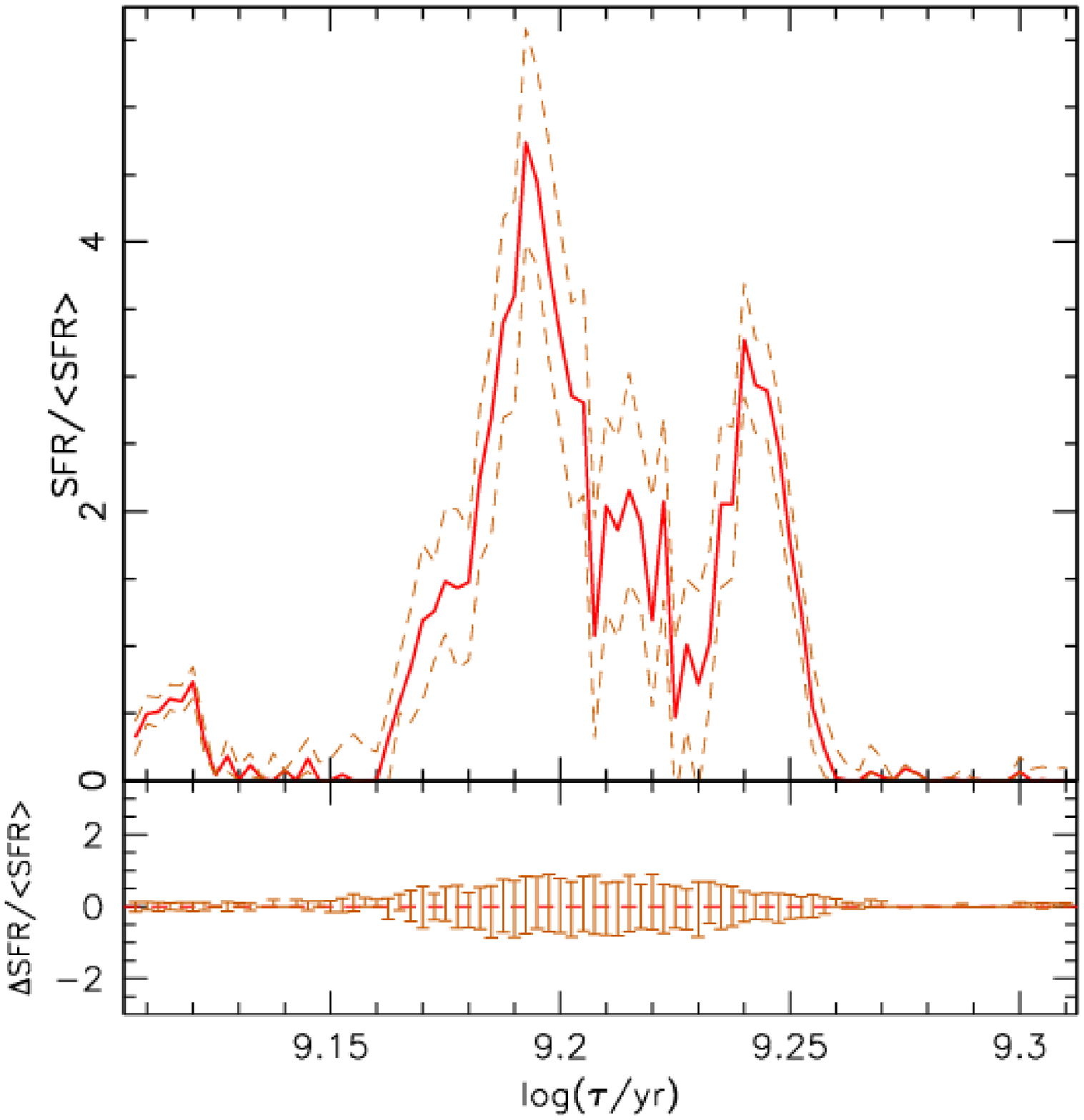}}
}
\caption{{\bf Top row (a):} The SFR$(t)$ normalized to the average SFR
  over the complete age interval, for the NGC~1846 Centre, and derived
  for three different widths of the age bins, namely $\Delta\log
  t=0.05, 0.025$ and $0.015$ (from left to right, respectively). In
  the {\bf leftmost panel}, the red histogram shows the SFR$(t)$ for
  the best-fitting solution, while the orange bars show the random
  errors at the 68~\% confidence level.  The black dashed lines show
  the systematic errors inside the area of 68~\% confidence level
  shown in Fig.~\ref{chi2_map}. The {\bf central panel} shows the same
  for a resolution twice as better, resulting on a more detailed
  SFR$(t)$, but with random errors about twice as large.  Finally, the
  {\bf rightmost panel} shows the SFR$(t)$ computed with a fixed
  resolution of $\Delta\log t=0.015$, but moving the age bin centers
  at small steps of 0.003~dex in $\log t$, so that the SFR$(t)$
  appears more continuous -- and indeed, we change the SFR$(t)$ plot
  to use continuous lines instead of histograms, in order to better
  illustrate these particular results.  The continuous red line is the
  SFR$(t)$ while the dashed orange lines illustrate the interval of
  random errors, at the 68~\% confidence level.  {\bf Bottom row (b):}
  The same for the NGC~1846 Ring region.  To provide a better
  comparison between the error bars among the different cases, the
  {\bf smaller sub-panels} at the bottom of each panel show the errors
  (random, and also systematic in the case of $\Delta\log t=0.05$~dex)
  as referred to the mean SFR$(t)$ line.  }
 \label{sfr1846}
\end{figure*}

\subsubsection{The SFH for  the NGC~1846 Centre}
\label{sec_clusterSFH1846}

Complete maps of $\chisqmin$ for the NGC~1846 Centre, as a function of
\dmo, \av\ and metallicity, are presented in the first seven panels of
Fig.~\ref{chi2_map}. These results are obtained for the age resolution
of $\Delta\log t=0.05$~dex. We can notice that the best solutions are
found in the metallicity interval between $\mh=-0.52$ and $-0.42$. The
last panel shows the same kind of map for the Ring, but limited to the
metallicity that provides the best fit for the Centre, namely
$\mh=-0.49$. This value is in excellent agreement with the one derived
from the Ca~{\sc ii} triplet of cluster members by
\citet{Grocholski_etal06}.

The best solution for the Centre is for $\dmo=18.57, \av=0.26$, with a
$\chisqmin=0.55$. Such a small \chisqmin\ is already an indication of
an excellent fit to the observational data.  This best-fitting
solution and map of residuals are also presented in the Hess diagrams
of Fig.~\ref{data_model}.  Finally, the best-fitting SFR$(t)$ is shown
in the upper left panel of Fig.~\ref{sfr1846}.

To evaluate the errors for all involved parameters, we first find the
correspondence between the $\chisqmin$ value for each model and its
confidence level.  This correspondence was estimated simulating 100
synthetic CMDs generated with a number of stars equal to the observed
CMD, using the best-fitting SFR$(t)$ and its parameters as the input
for the simulations. So, after recovering the SFH for this sample of
synthetic CMDs, it was possible to build the $\chisqmin$ distribution
and to establish the relation between their values and the confidence
level.

In the $\chisqmin$ maps of Fig.~\ref{chi2_map}, we superimposed the
68~\% and 95~\% significance levels for all the solutions for the
Centre. Only for \mh\ values between $-0.52$ and $-0.42$ we have
solutions within the 68~\% significance level of the best solution.
Based on this figure, we determine $\dmo=18.57\pm0.07$ and
$\av=0.26\pm0.05$ for the cluster Centre (with random errors at the
68~\% significance level).
  
The left panel of Fig.~\ref{sfr1846} shows the SFR$(t)$ for the
cluster Centre together with error bars, as derived for the initial
age resolution of $\Delta\log t=0.05$~dex. The most basic feature in
this plot is that the SFR$(t)$ is clearly non-null for two age bins,
spanning the $\log(t/{\rm yr})$ interval from 9.15 to 9.25 (ages from
1.41 to 1.78 Gyr). This result is not only valid for the best fitting
model, but also across the entire 68~\% significance level volume of
the $\av$ vs. $\dmo$ and \mh\ diagrams. Solutions within this volume
are used to define the range of systematic errors, which is also
depicted in the figure. Finally, we note that the two bins of non-null
star formation are found even if we adopt less restrictive limits for
the random errors, i.e.\ if we plot all solutions inside the 95~\%
significance level.

However, the very small random errors for the $\Delta\log t=0.05$~dex
solution clearly suggest that the data presents the potential for a
more detailed determination of the age distribution. The middle panel
of Fig.~\ref{sfr1846} shows the solution when we adopt 20 SPMs
separated by $\Delta\log({\rm age})=0.025$~dex, as obtained with the
same data and methods, but for a fixed value of $\dmo=18.57, \av=0.26,
\mh=-0.49$. Now the solution is clearly non-null for 3 age bins,
spanning the $\log(t/{\rm yr})$ interval from 9.175 to 9.25 (ages from
1.50 to 1.78 Gyr). As a result of the smaller number of stars per bin,
random errors are larger than in the previous $\Delta\log t=0.05$~dex
case\footnote{Note that systematic errors are computed only in the
  case of the default age resolution of $\Delta\log t=0.05$~dex, for
  which we fully explored the possible interval of $\dmo, \av, \mh$.
  The basic reasons for not recomputing the systematic errors when
  adopting a better age resolution, are essentially: (1) The large CPU
  times needed to explore the entire $\dmo, \av, \mh$ interval. (2)
  Test runs of our software indicate that the best-fitting values of
  these parameters depend little on the age resolution $\Delta\log t$
  being adopted.  Indeed, they are strongly constrained by the CMD
  portions corresponding to the lower main sequence, RGB, and main
  body of the red clump, which are equally well fitted at all age
  resolutions.  Therefore, the systematic errors obtained at
  $\Delta\log t=0.05$~dex shall be considered as indicative of those
  expected at all age resolutions.}.

We proceed with this experiment, determining the SFH for an even
better age resolution, namely $\Delta\log t=0.015$~dex. The results
are in the right panel of Fig.~\ref{sfr1846}. This time, the random
errors are really large, as a consequence of the quite small number of
stars per age bin. Under these conditions, the result of the SFH
analysis could depend very much on the particular choice of limits
for the age bins.  In order to appreciate the SFR$(t)$ for several bin
positions (and always with the same resolution, of $\Delta\log
t=0.015$~dex), the age bins have been shifted progressively by steps
equal to 1/5 of the total bin width, that is, by $0.003$~dex.  The
right panel of Fig.~\ref{sfr1846} is the result of plotting all these
results together as a function of the bin central age.  The SFR$(t)$
appears much more continuous than in previous cases, and presents
clear indications about two points:
\begin{enumerate}
\item The SFR$(t)$ is non-null in the complete age interval from
  $\log(t/{\rm yr})=9.18$ to 9.25 (ages from 1.51 to 1.78 Gyr), which
  is in perfect agreement with the interval revealed by the SFR$(t)$
  of intermediate resolution, $\Delta\log t=0.025$~dex.
\item There is a marked minimum in the SFR$(t)$ for the age bins with
  $\log(t/{\rm yr})$ going from 9.22 to 9.23 (ages 1.66 to 1.70 Gyr).
  Despite the large error bars, this minimum is statistically
  significant. It indicates, surprisingly, that there might have been
  a hiatus in the SFR$(t)$ of NGC~1846, starting $\sim150$~Myr after
  the first episode of star formation, and lasting for about 50~Myr.
\end{enumerate}

\subsubsection{The SFH for the NGC~1846 Ring}

In the case of the NGC~1846 Ring we {\em assume} it has the same \mh\
as the cluster Centre, and explore the solutions in the \dmo\ vs.\
\av\ plane, as shown in the last panel of Fig.~\ref{chi2_map}. The
assumption of the same $\mh$ is natural for stars belonging to the
same (presumably chemically-homogeneous) cluster. In this case, the
cluster Centre is taken as the reference because it suffers less from
field contamination and differential reddening, which are two
potential sources of systematic errors in the determination of \mh.
Moreover, the cluster Centre presents low \chisqmin\ values, which
means very good overall solutions.

The best solution for the Ring turns out to be located at the same
\dmo\ as the cluster Centre, and at virtually the same $\av$, as shown
in Fig.~\ref{chi2_map}.  Experiments of SFH recovery at varying age
resolution, as performed for the Centre, were repeated in the Ring.
Also these results turned out to be remarkably similar to the Centre
ones, as revealed by the bottom row of Fig.~\ref{sfr1846}. In
particular, the hiatus in star formation at $\log(t/{\rm yr})=9.22$ is
clearly present also in the Ring. We will further comment on this in
Sect.~\ref{conclu}.

\begin{figure}
\resizebox{0.75\hsize}{!}{\includegraphics{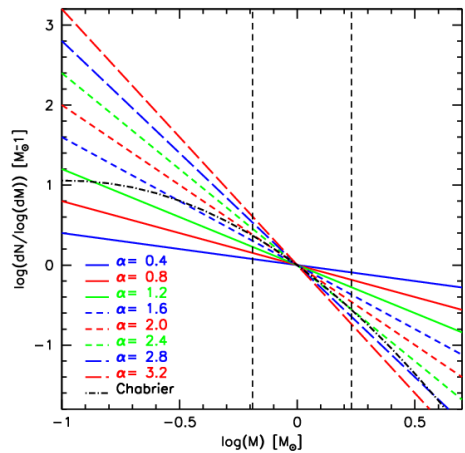}}
\caption{PDMFs used to derive the best-fit solutions for the clusters.
  The red, blue and green lines show the PDMF with 8 different slopes,
  $\alpha$, from 0.4 to 3.6. The dot-dashed line shows the
  \citet{chabrier01} PDMF previously used to recover the detailed SFHs
  for the clusters and their fields. The dashed vertical lines indicate
  the mass interval relevant to this work.}
\label{fig_alpha}
\end{figure}

\subsubsection{The binary fraction and mass function in NGC~1846}

Apart from the already-mentioned assumptions (constant \mh, no
differential reddening, etc.) our analysis also adopted SPMs built for
a fixed value for the PDMF slope and fraction of unresolved binaries.
The question arises whether our results can help to better constrain
these parameters, and their possible radial variations within the
clusters.  We test this by varying these parameters while keeping the
same \dmo, \av, and \mh\ as in the best-fitting solution. In order to
be more sensible to differences in the PDMF slope, we now cut the CMDs
at a limiting magnitude of ${\rm F814W}\!<\!24.25$~mag, which
corresponds to main sequence stars with $\sim\!0.7$~\Msun. Needless to
say, by going to deeper magnitudes we are also including CMD regions
of smaller completeness and with higher photometric errors. Although
these processes are properly modelled by our ASTs, it is also true
that the PDMF determinations we are doing here shall be considered of
a more exploratory nature, than our previous determinations of the
cluster SFHs.

\begin{figure*}
\resizebox{0.33\hsize}{!}{\includegraphics{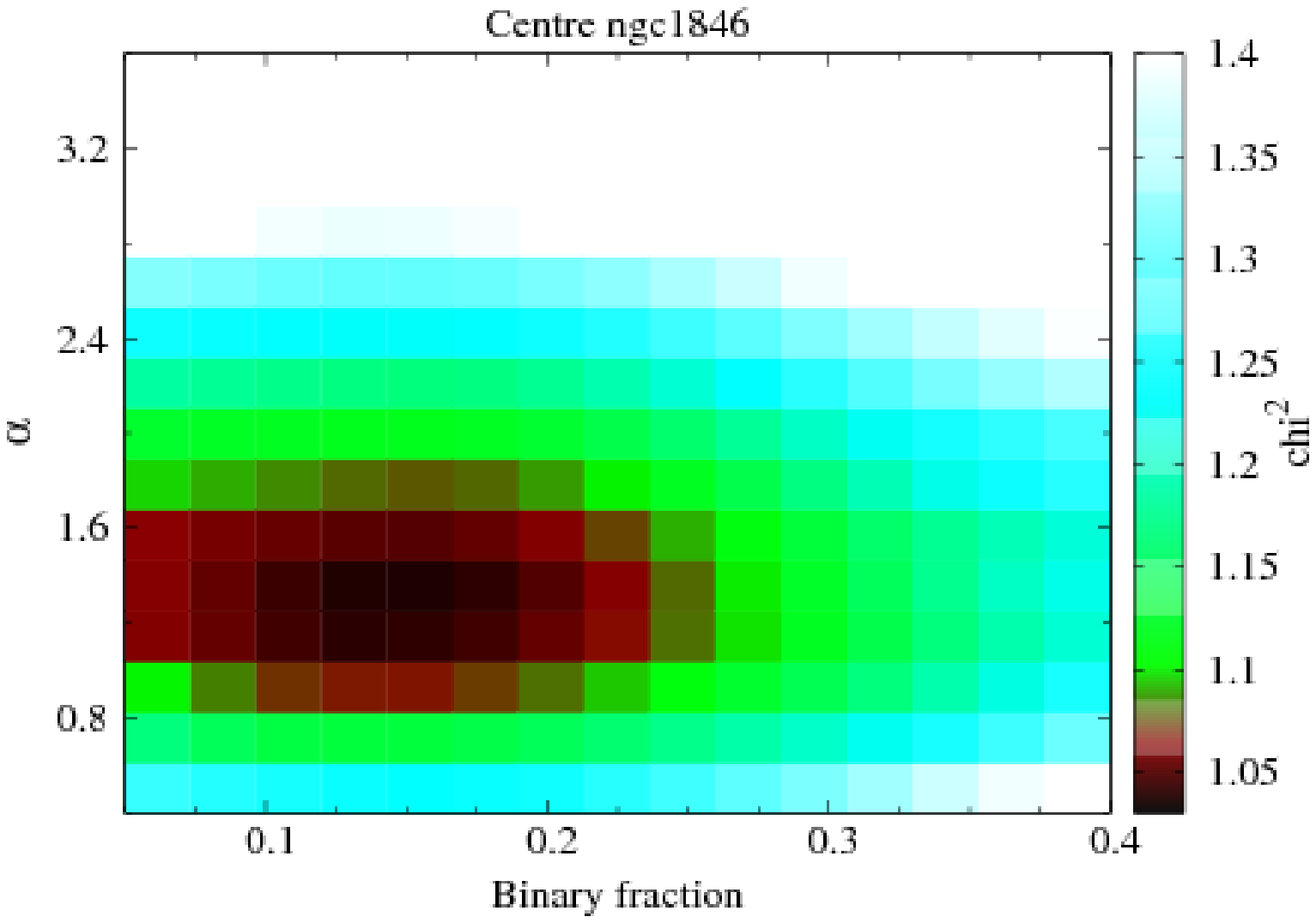}}
\resizebox{0.33\hsize}{!}{\includegraphics{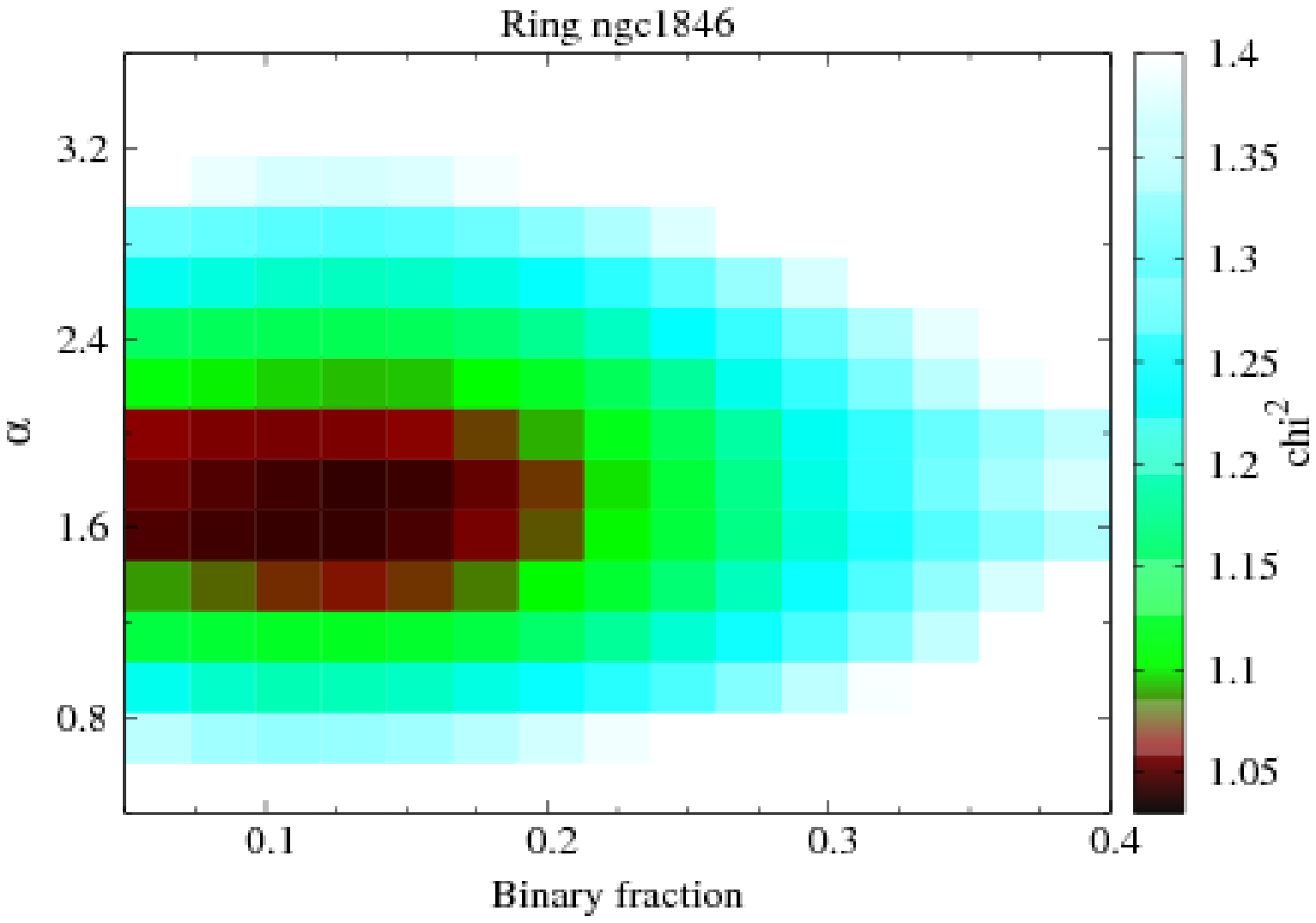}}
\caption{$\chi^{2}_{\rm min}$ map normalized to the minimum
  $\chi^{2}_{\rm min}$ for the NGC~1846 Centre (left panel) and the
  Ring (right panel) solutions, as a function of binary fraction $f$
  and PDMF slope $\alpha$.}
\label{fig_IMF1846}
\end{figure*}

The binary fraction is parametrized by the fraction $f$ of detached
binaries with a mass ratio in the interval from 0.7 to 1. These
binaries cause the well-known sequence parallel to the main sequence
in the CMD, while binaries with a smaller mass ratio leave hardly any
signature in optical CMDs.  The PDMF slope is parametrized by the
\citet{salpeter} slope, $\alpha$ (with $dN/dM\propto M^{-\alpha}$),
which changes the number ratio between the stars at and above the
MSTO, and the fainter main sequence. PDMF slopes tested in this work
are illustrated in Fig.~\ref{fig_alpha}. As a reference, the main
sequence stars at ${\rm F814W}\!\sim\!24.25$~mag, have
$\sim\!0.7$~\Msun. For the mass range of interest here, the
\citet{chabrier01} PDMF has a slope of $\alpha\simeq2.2$.

In this exercise, we start from the best fitting solutions for an age
resolution of 0.05~dex, and run StarFISH for all $f$ and $\alpha$
combinations, for a fixed distance and reddening.  The results are
illustrated in the $\chi^2_{\rm min}$ map of Fig.~\ref{fig_IMF1846},
separately for the NGC~1846 Centre and Ring. In this figure, the
$\chi^2_{\rm min}$ maps were normalized to the minimum value of
$\chi^2_{\rm min}$, to allow a better comparison between our results.
The best-fitting solutions are found for about the same values of
$\alpha$ and $f$ for both Centre and Ring: ($\alpha=1.2,f=0.15$) and
($\alpha=1.6,f=0.13$), respectively. Since we do not perform the error
analysis, it is impossible to tell whether these small differences
between Centre and Ring are statistically significant or not. Anyway,
these values seem to indicate flatter PDMFs than the ones commonly
found in galactic or extragalactic stellar clusters, which typically
present PDMF slopes close to the \citet[][$\alpha\!=\!2.35$]{salpeter}
or \citet[][$\alpha\!\sim\!2.2$]{chabrier01} one \citep[see also
e.g.][]{Kroupa01, Kroupa02, Bastian_etal10}.



\subsubsection{The SFH of NGC~1783 Centre and Ring}
\label{sec_clusterSFH1783}

\begin{figure*}
\resizebox{0.33\hsize}{!}{\includegraphics{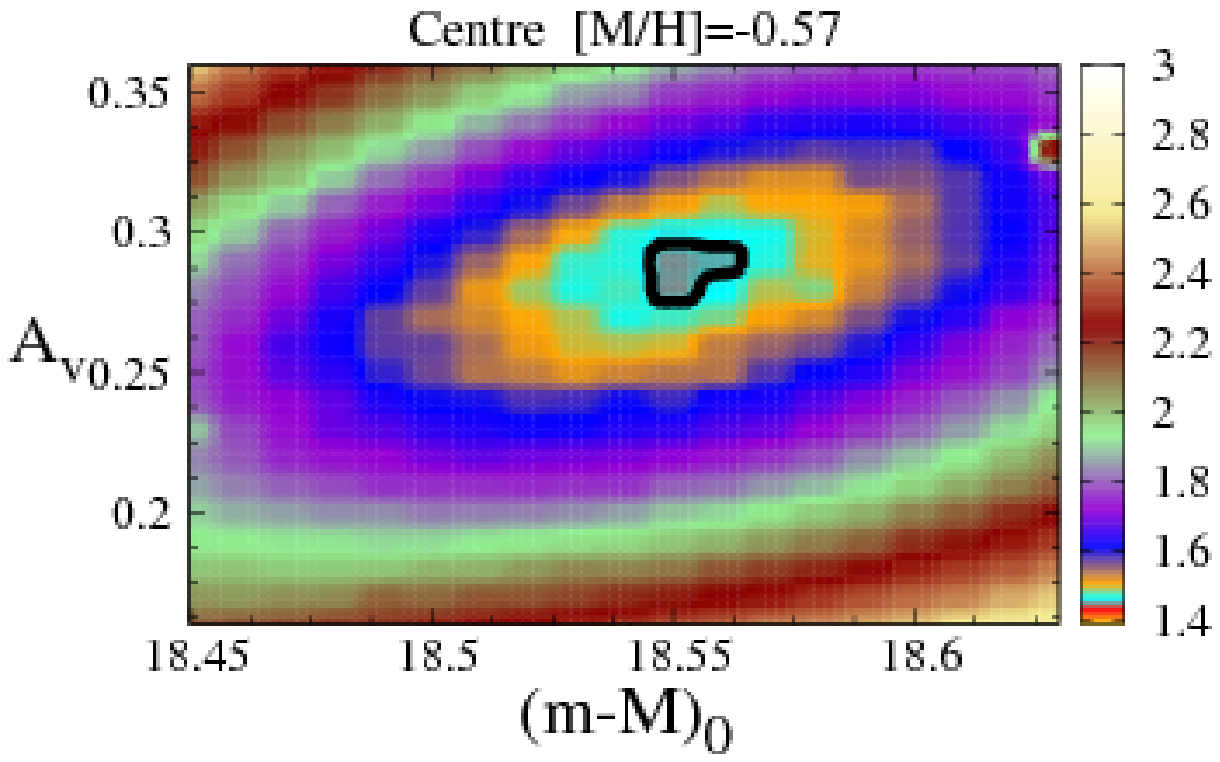}}
\resizebox{0.33\hsize}{!}{\includegraphics{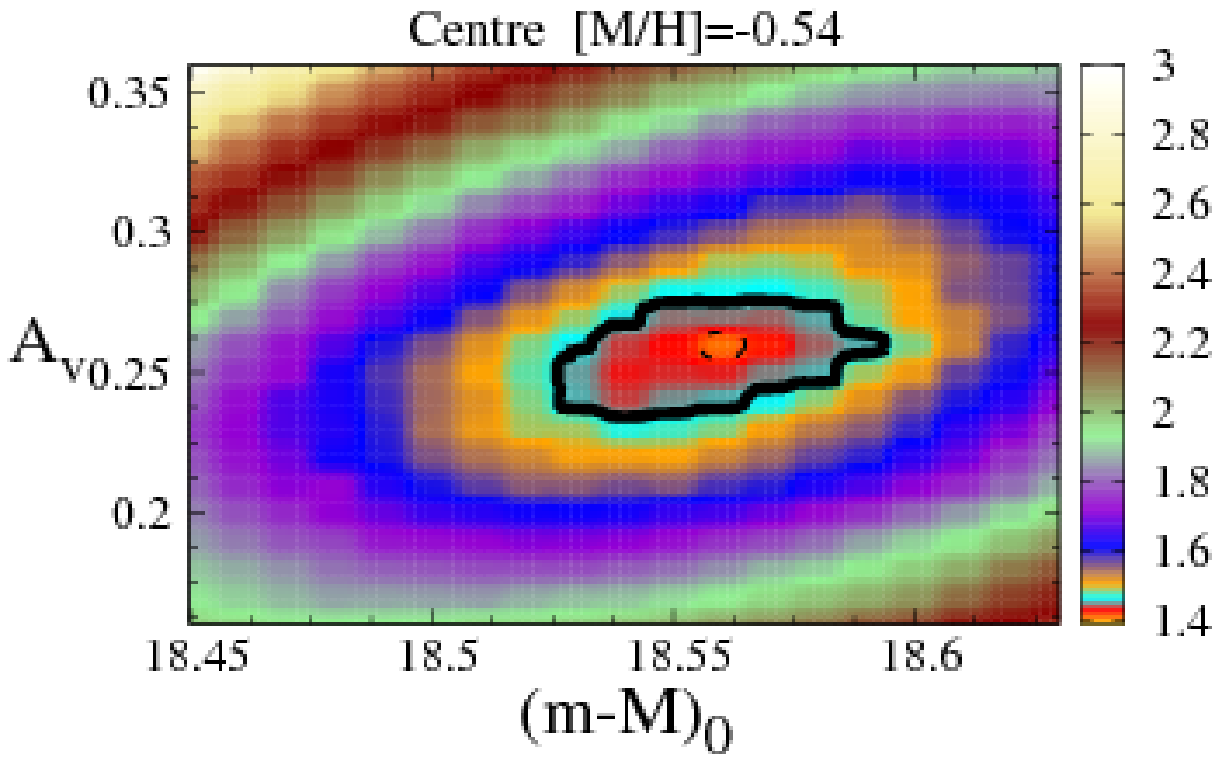}}
\resizebox{0.33\hsize}{!}{\includegraphics{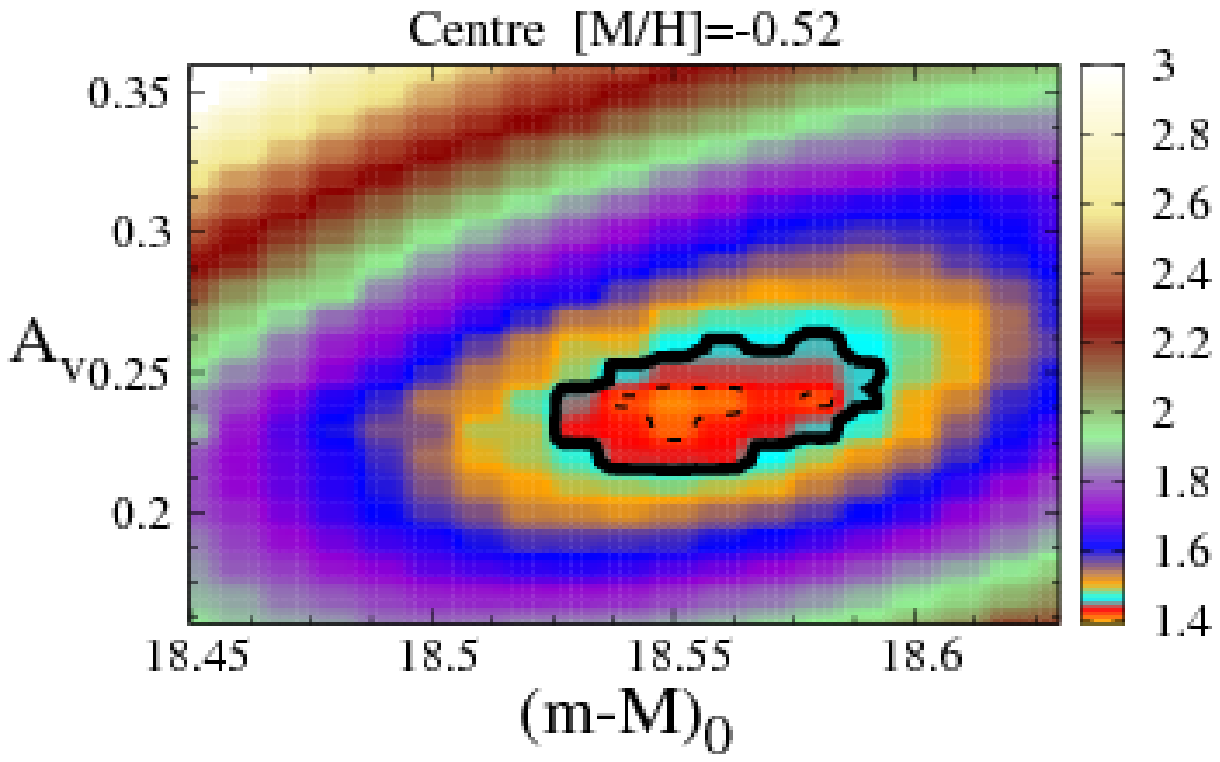}}
\hfill\\
\resizebox{0.33\hsize}{!}{\includegraphics{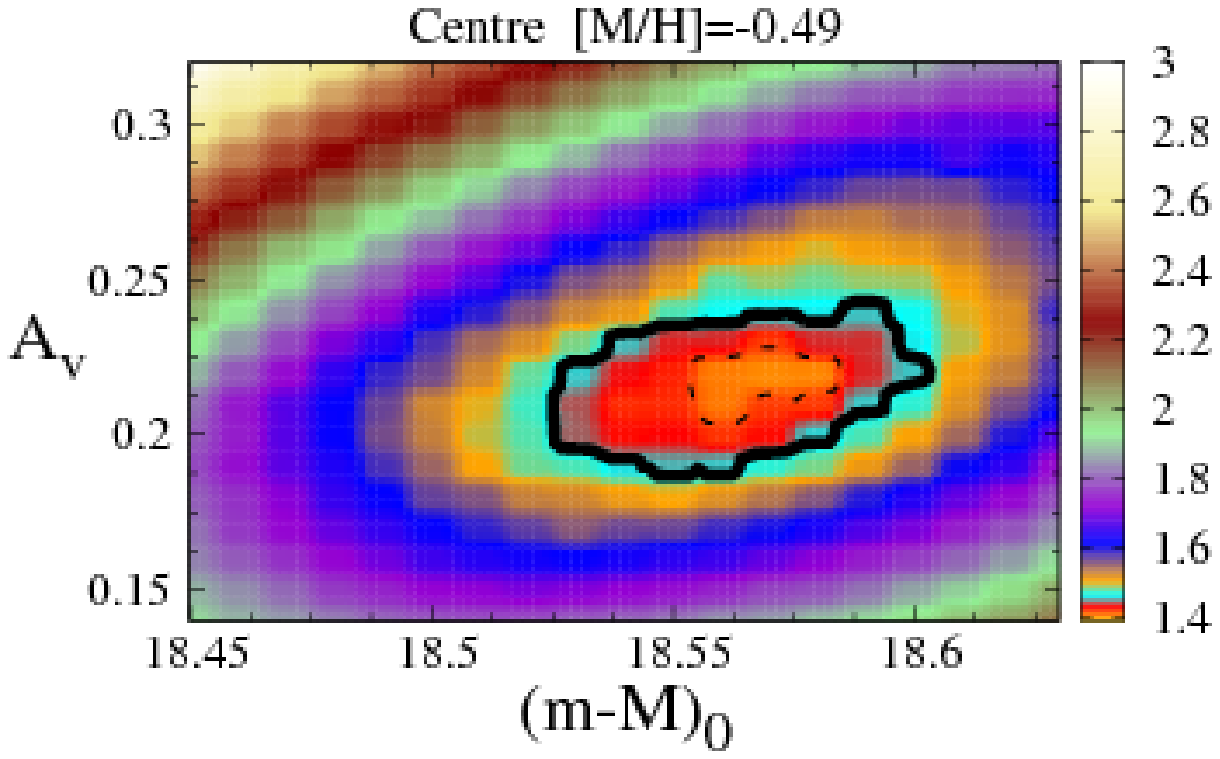}}
\resizebox{0.33\hsize}{!}{\includegraphics{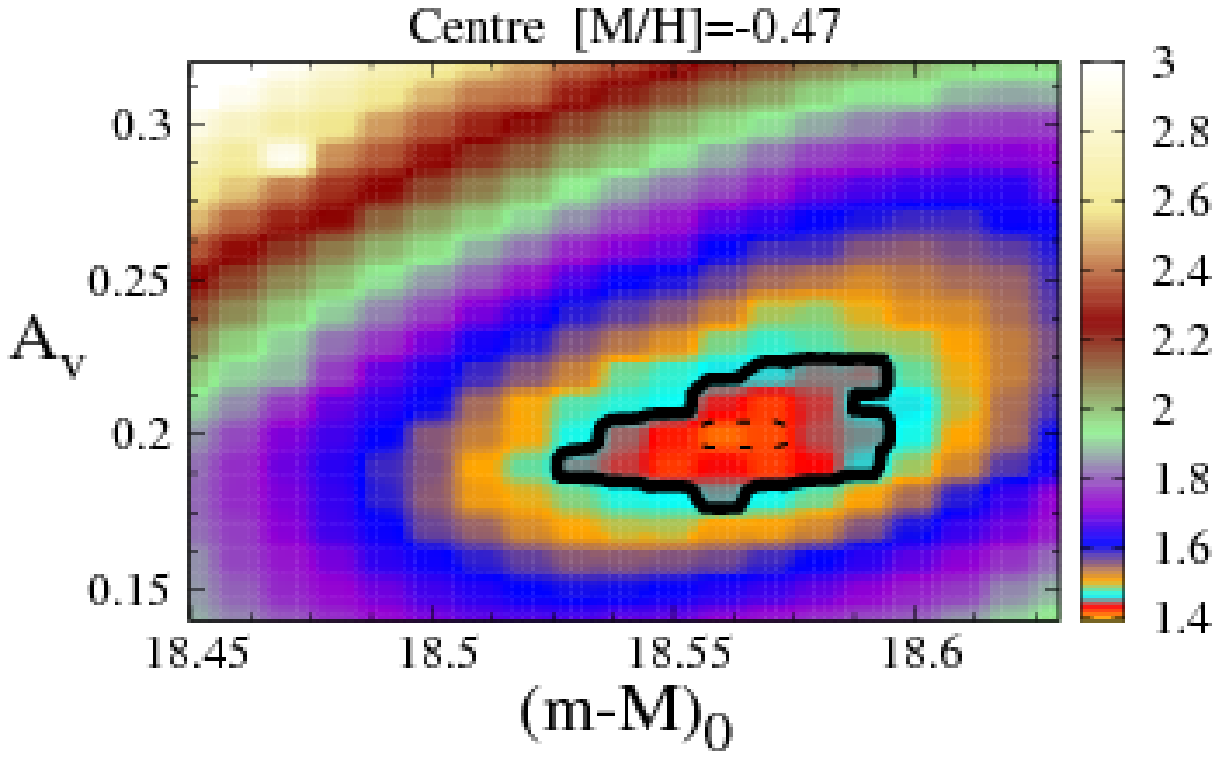}}
\resizebox{0.33\hsize}{!}{\includegraphics{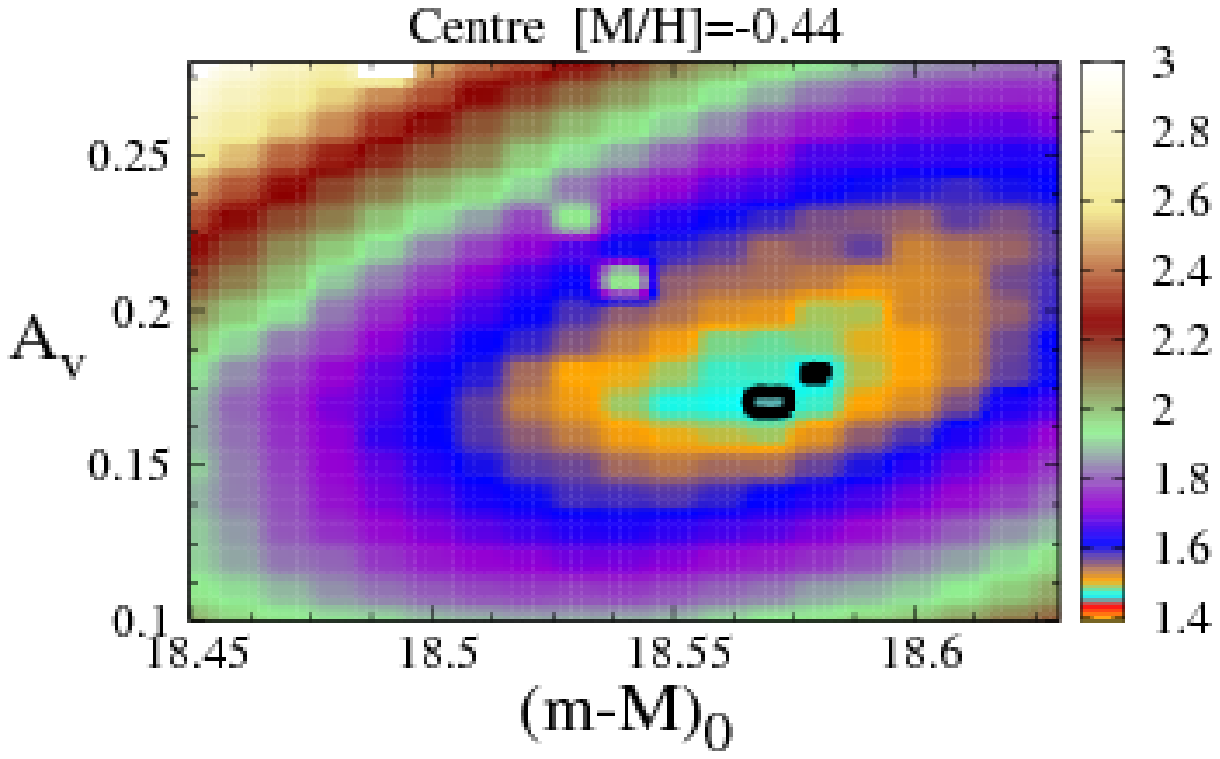}}
\hfill\\
\resizebox{0.33\hsize}{!}{\includegraphics{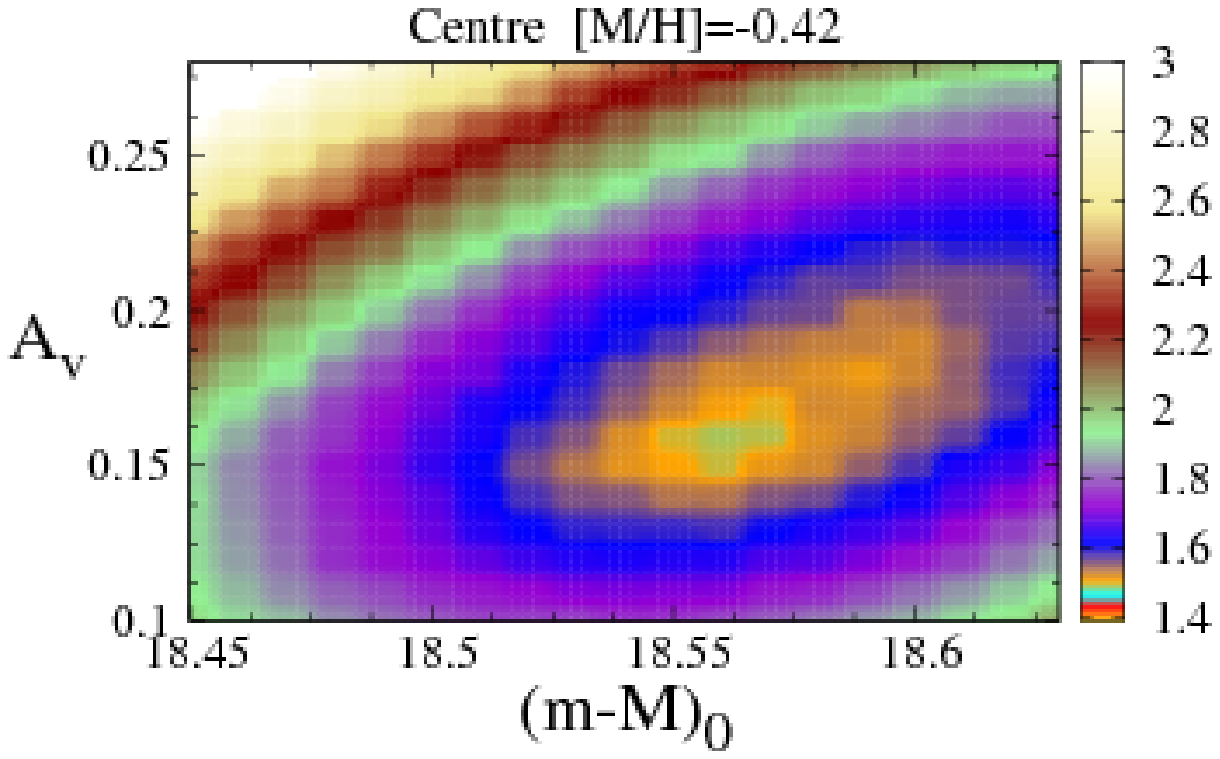}}
\resizebox{0.33\hsize}{!}{\includegraphics{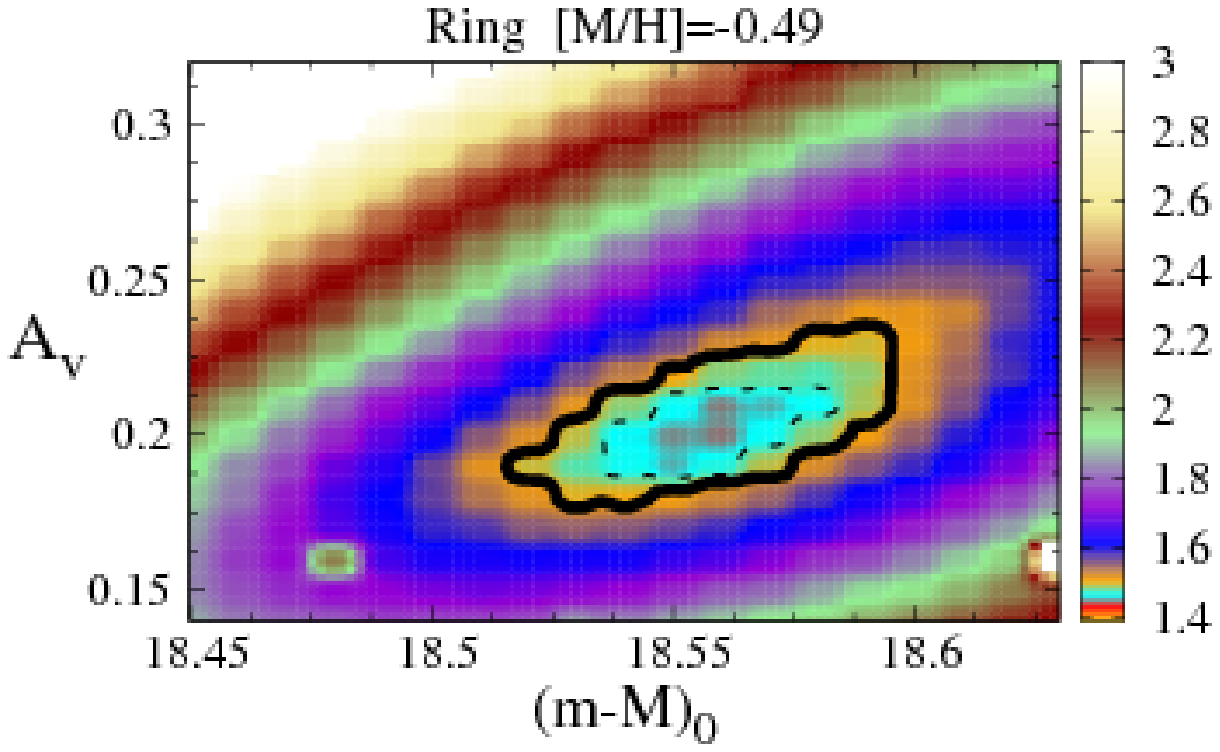}}
\hfill\\
\caption{The same as Fig.~\ref{chi2_map} but for NGC~1783. The
  minimum $\chisqmin$ is of 1.39 and 1.45 for Centre and Ring,
  respectively.}
\label{chi2_map1783}
\end{figure*}

\begin{figure*}
\begin{minipage}{0.96\hsize}
\subfigure[Centre of NGC~1783]{
\resizebox{0.16\hsize}{!}{\includegraphics[trim=1.0cm 1.0cm 1.0cm 1.0cm]{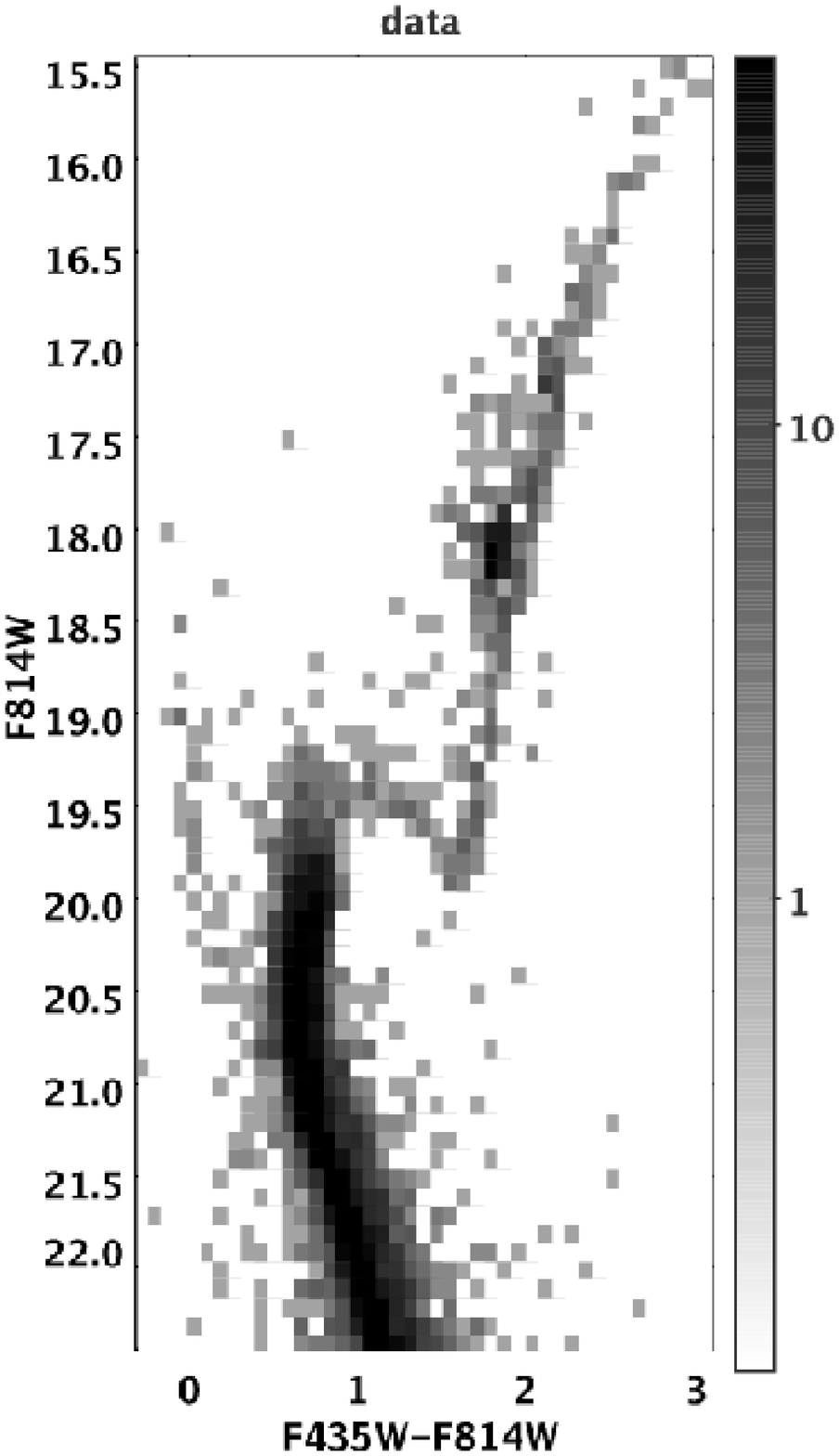}}
\resizebox{0.16\hsize}{!}{\includegraphics[trim=1.0cm 1.0cm 1.0cm 1.0cm]{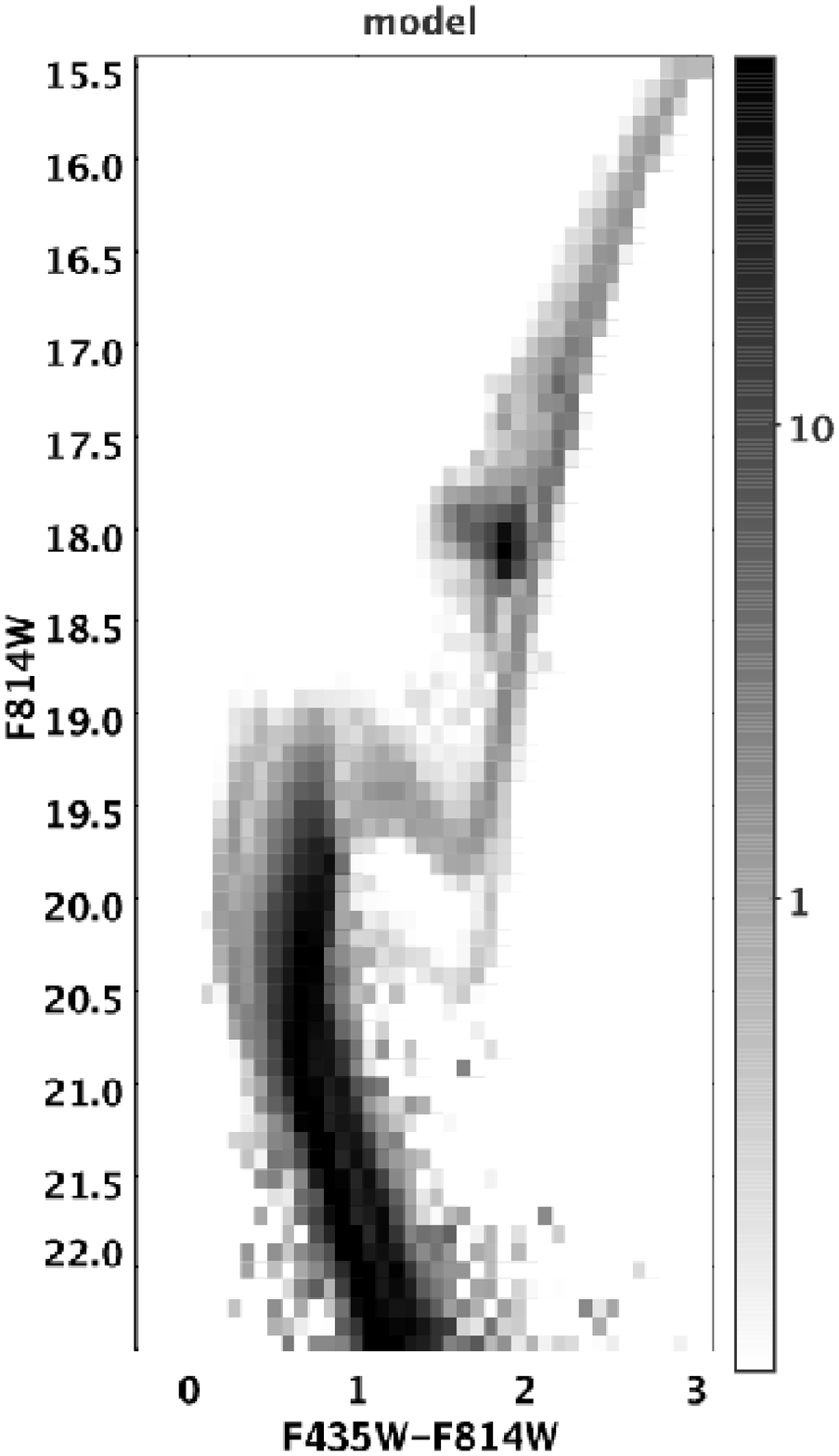}}
\resizebox{0.16\hsize}{!}{\includegraphics[trim=1.0cm 1.0cm 1.0cm 1.0cm]{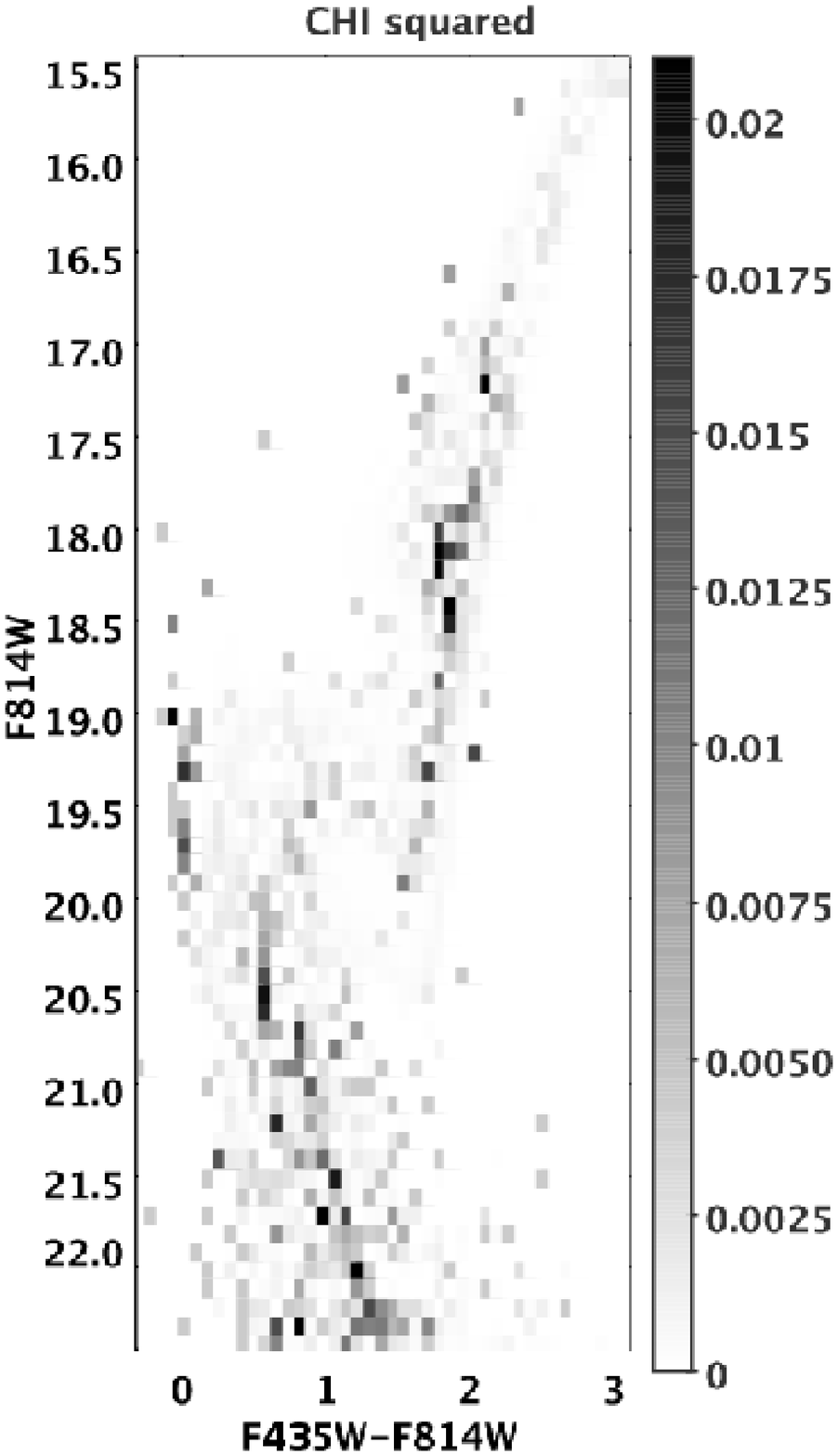}}
\hfill\hspace{0.5cm}
\resizebox{0.16\hsize}{!}{\includegraphics[trim=1.0cm 1.0cm 1.0cm 1.0cm]{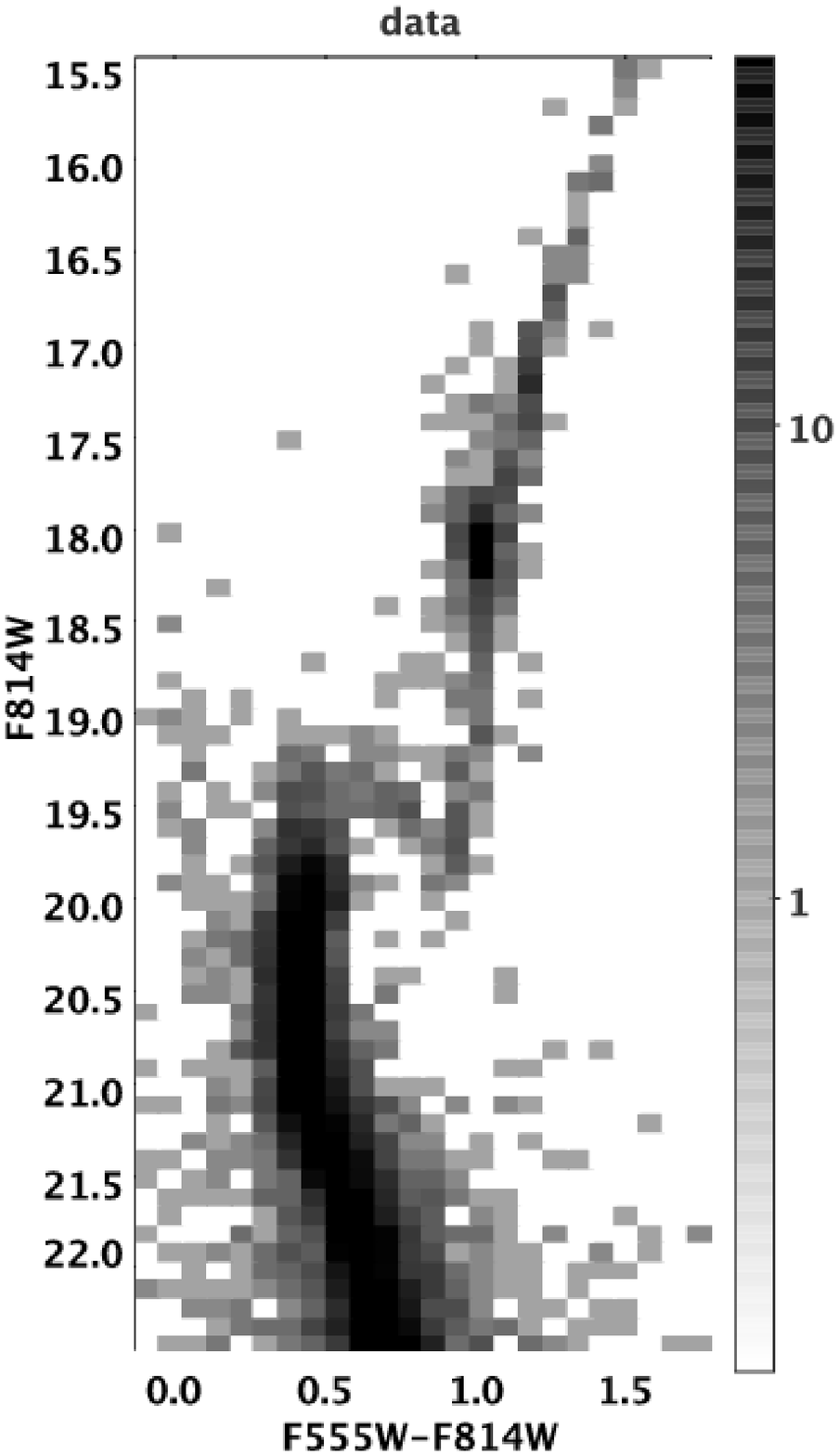}}
\resizebox{0.16\hsize}{!}{\includegraphics[trim=1.0cm 1.0cm 1.0cm 1.0cm]{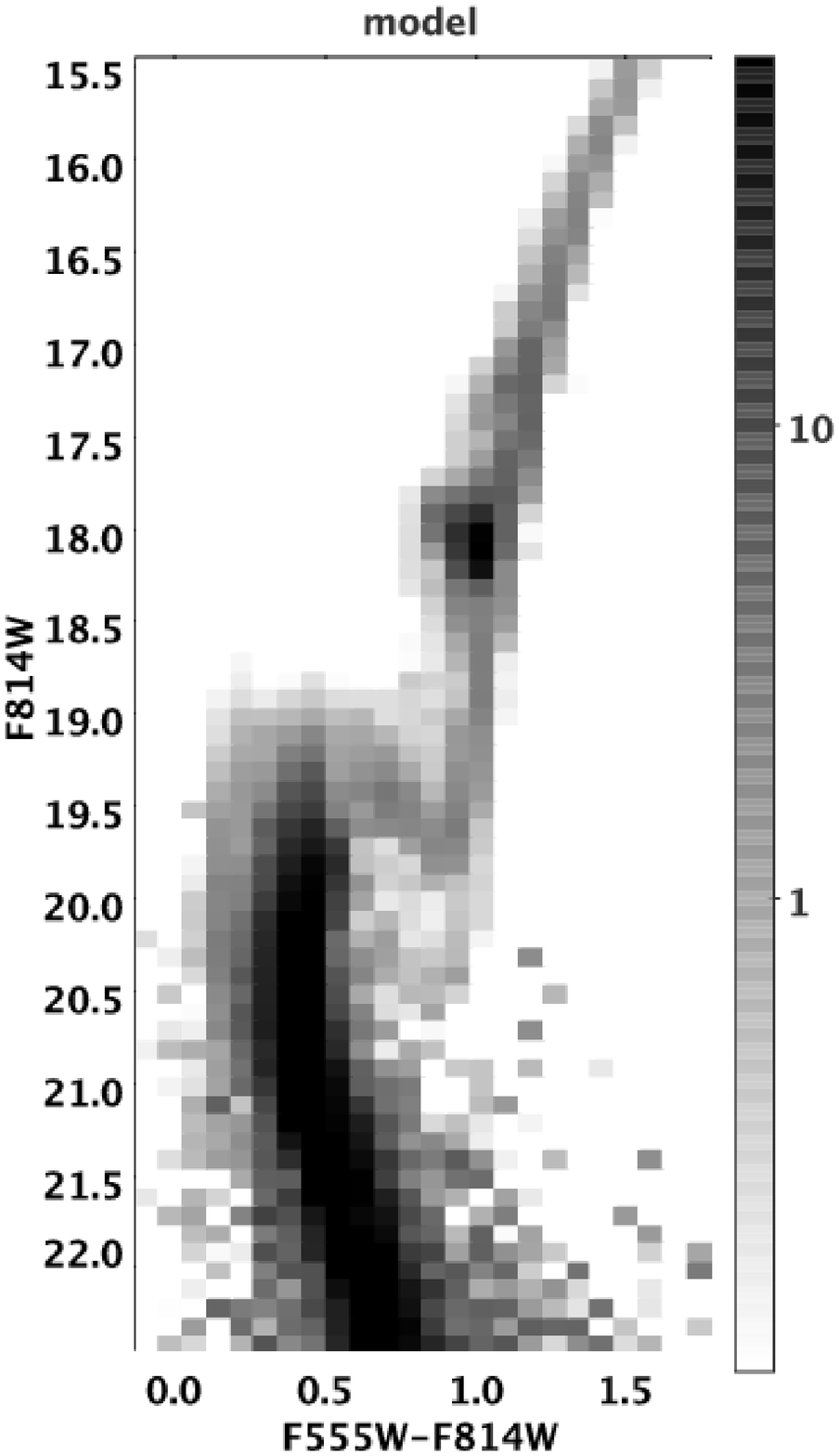}}
\resizebox{0.16\hsize}{!}{\includegraphics[trim=1.0cm 1.0cm 1.0cm 1.0cm]{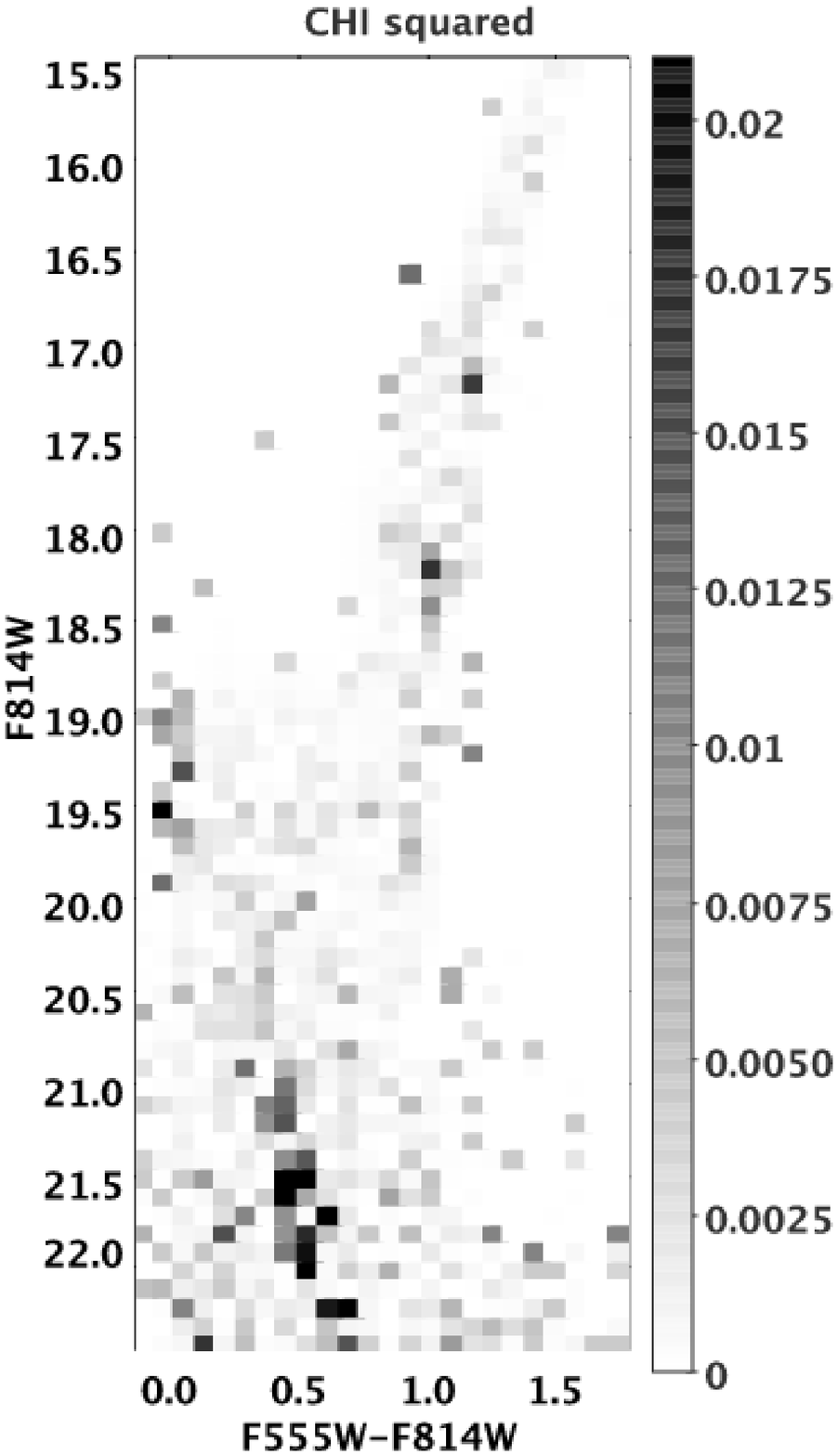}}
}
\end{minipage}
\begin{minipage}{0.96\hsize}
\subfigure[Ring of NGC~1783]{
\resizebox{0.16\hsize}{!}{\includegraphics[trim=1.0cm 1.0cm 1.0cm 1.0cm]{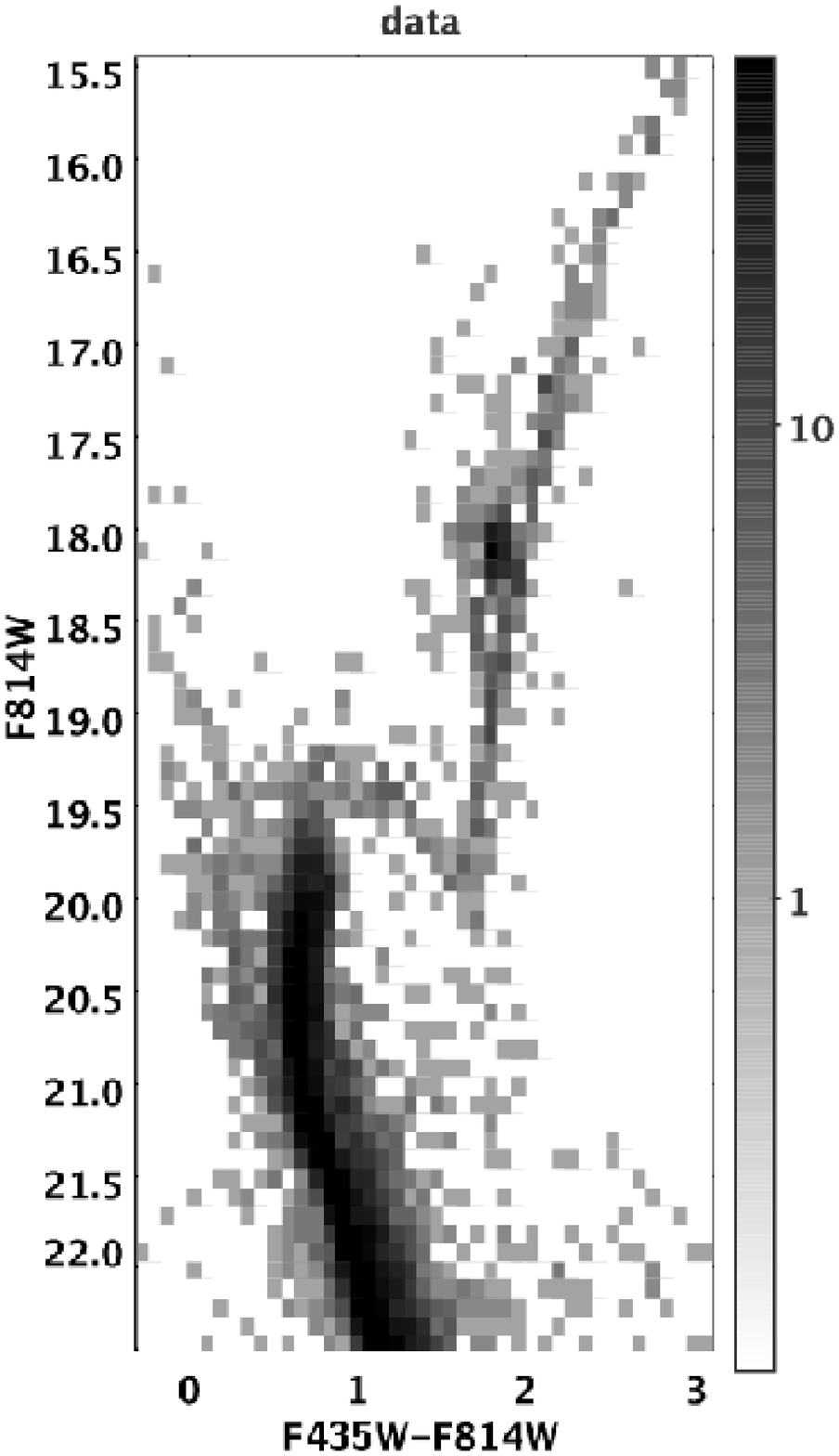}}
\resizebox{0.16\hsize}{!}{\includegraphics[trim=1.0cm 1.0cm 1.0cm 1.0cm]{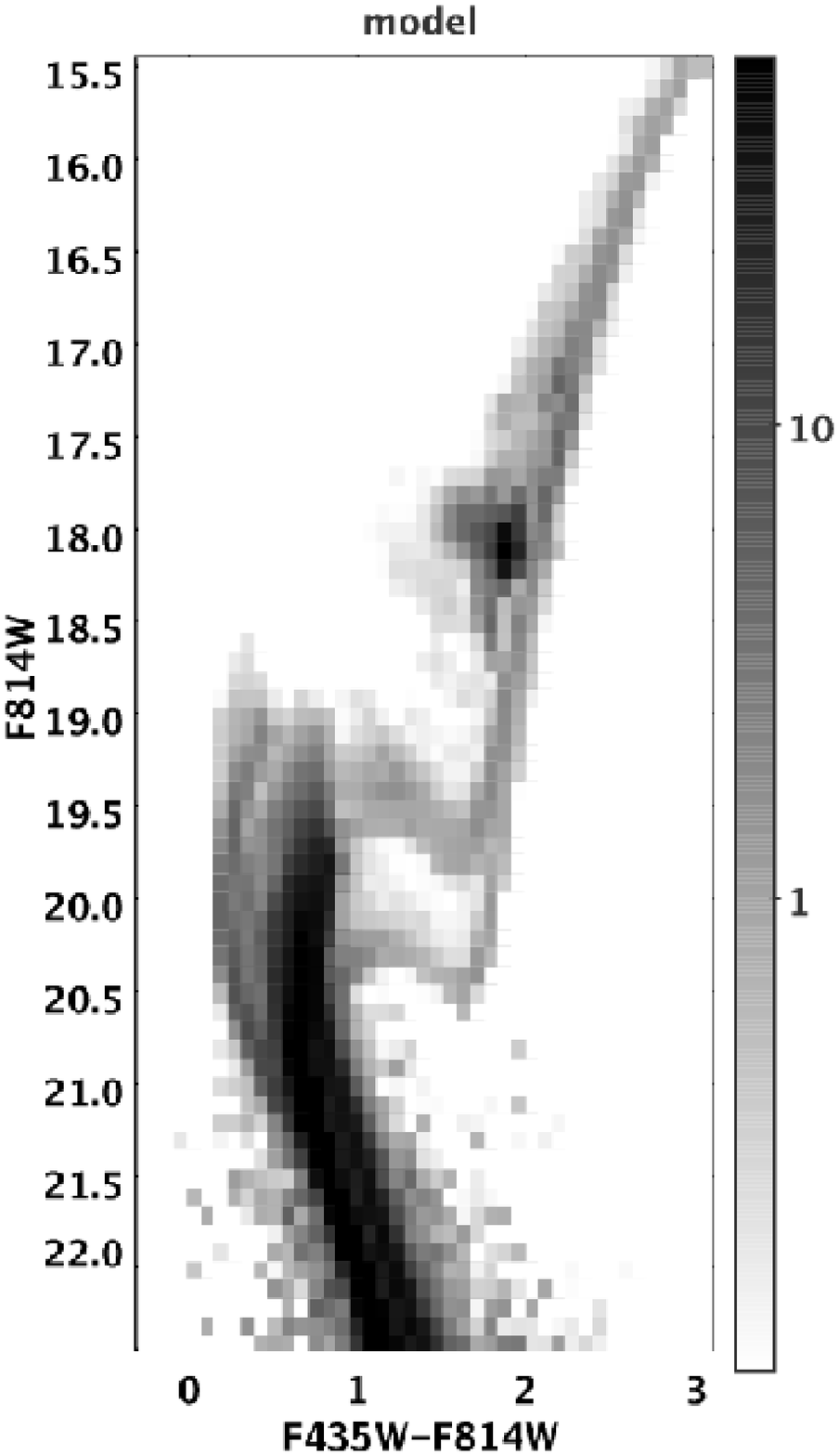}}
\resizebox{0.16\hsize}{!}{\includegraphics[trim=1.0cm 1.0cm 1.0cm 1.0cm]{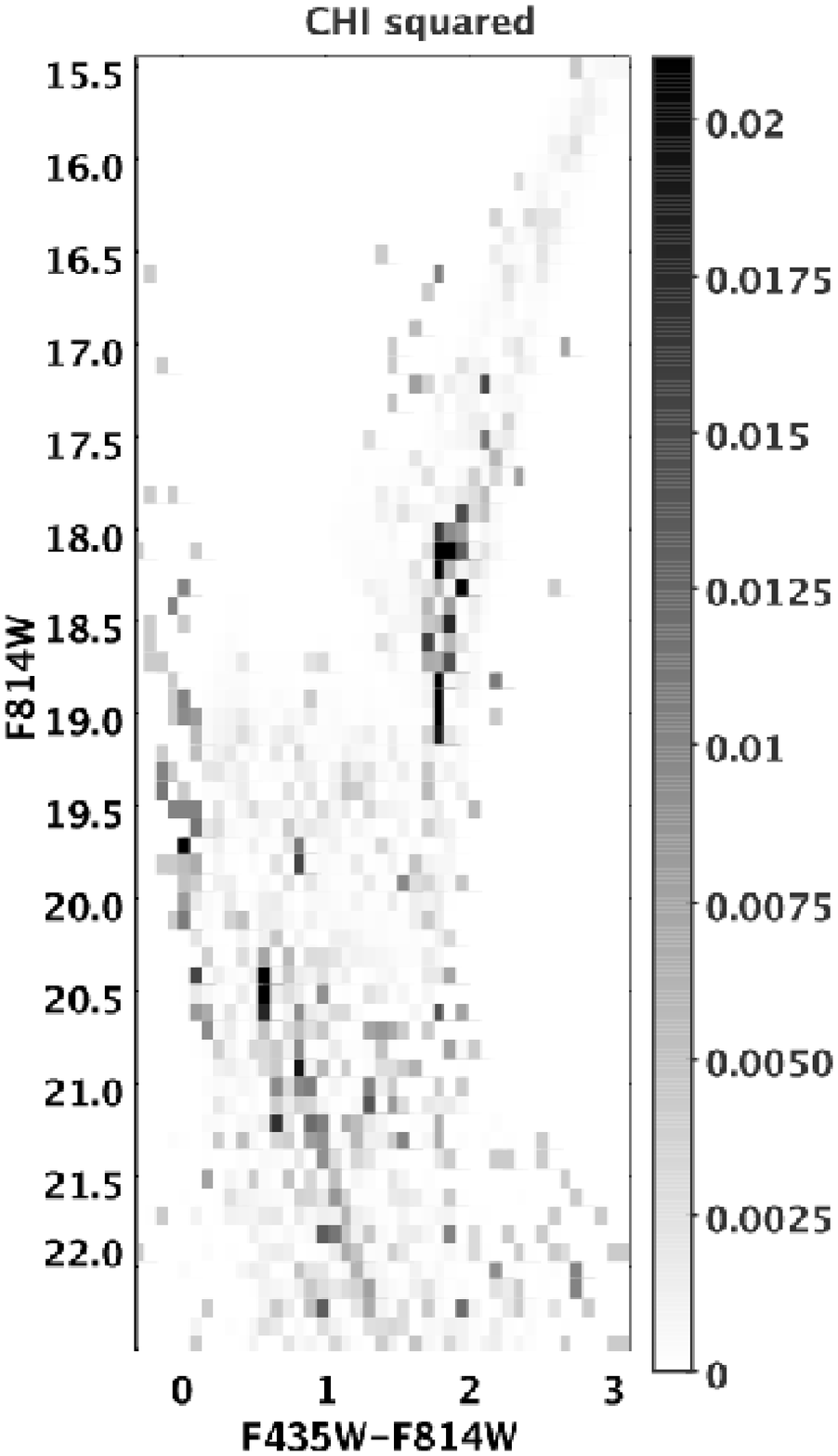}}
\hfill\hspace{0.5cm}
\resizebox{0.16\hsize}{!}{\includegraphics[trim=1.0cm 1.0cm 1.0cm 1.0cm]{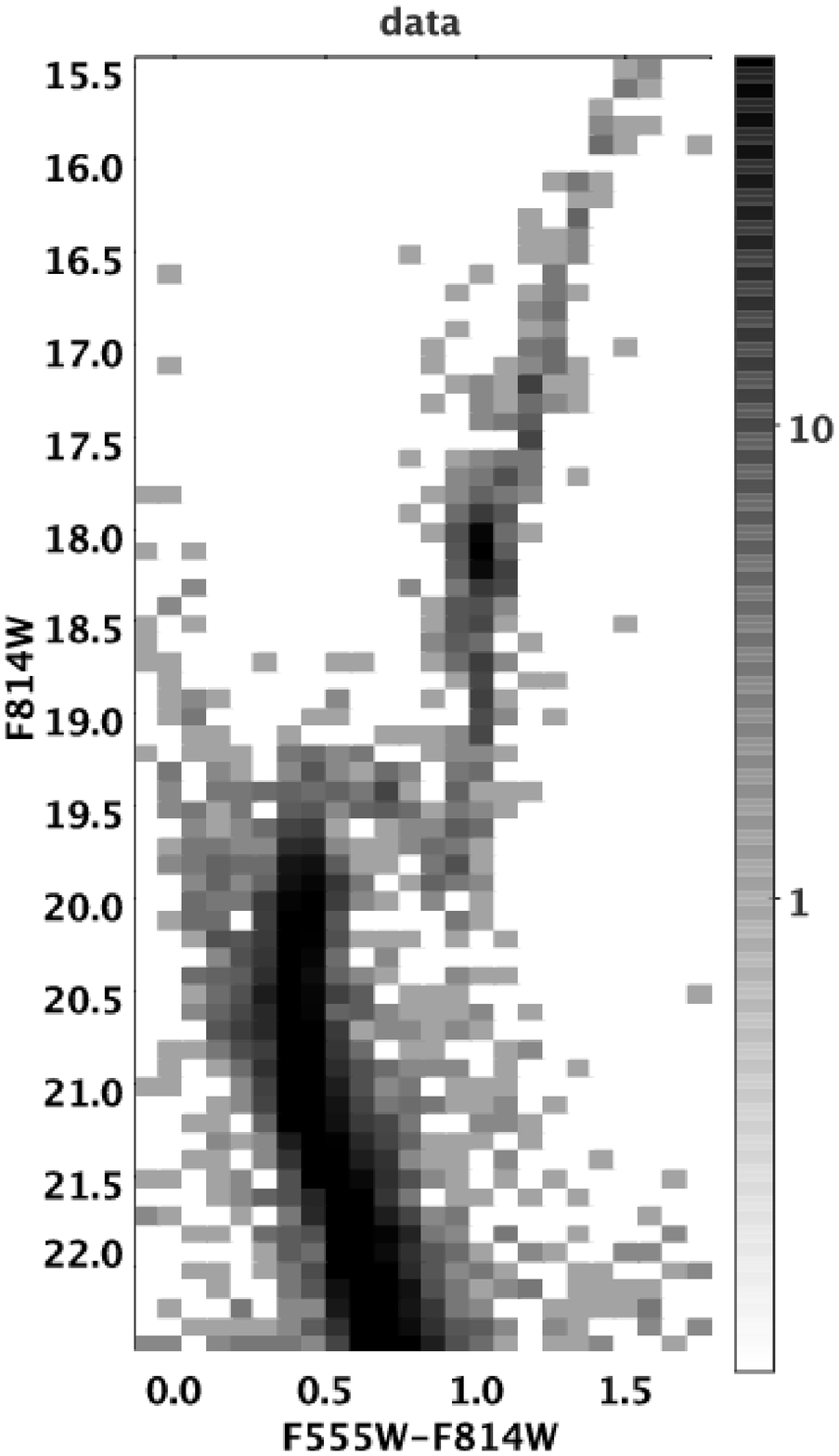}}
\resizebox{0.16\hsize}{!}{\includegraphics[trim=1.0cm 1.0cm 1.0cm 1.0cm]{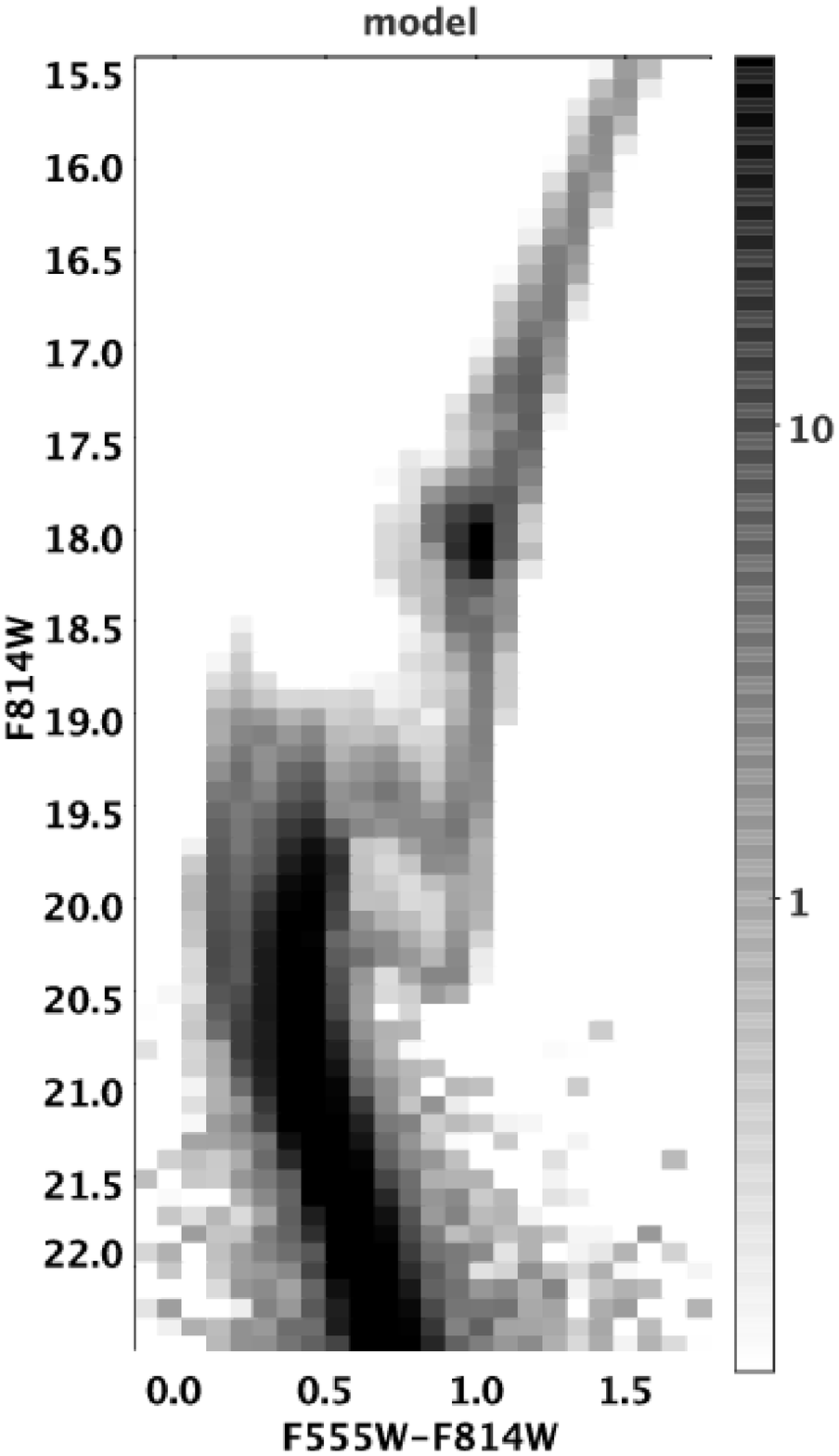}}
\resizebox{0.16\hsize}{!}{\includegraphics[trim=1.0cm 1.0cm 1.0cm 1.0cm]{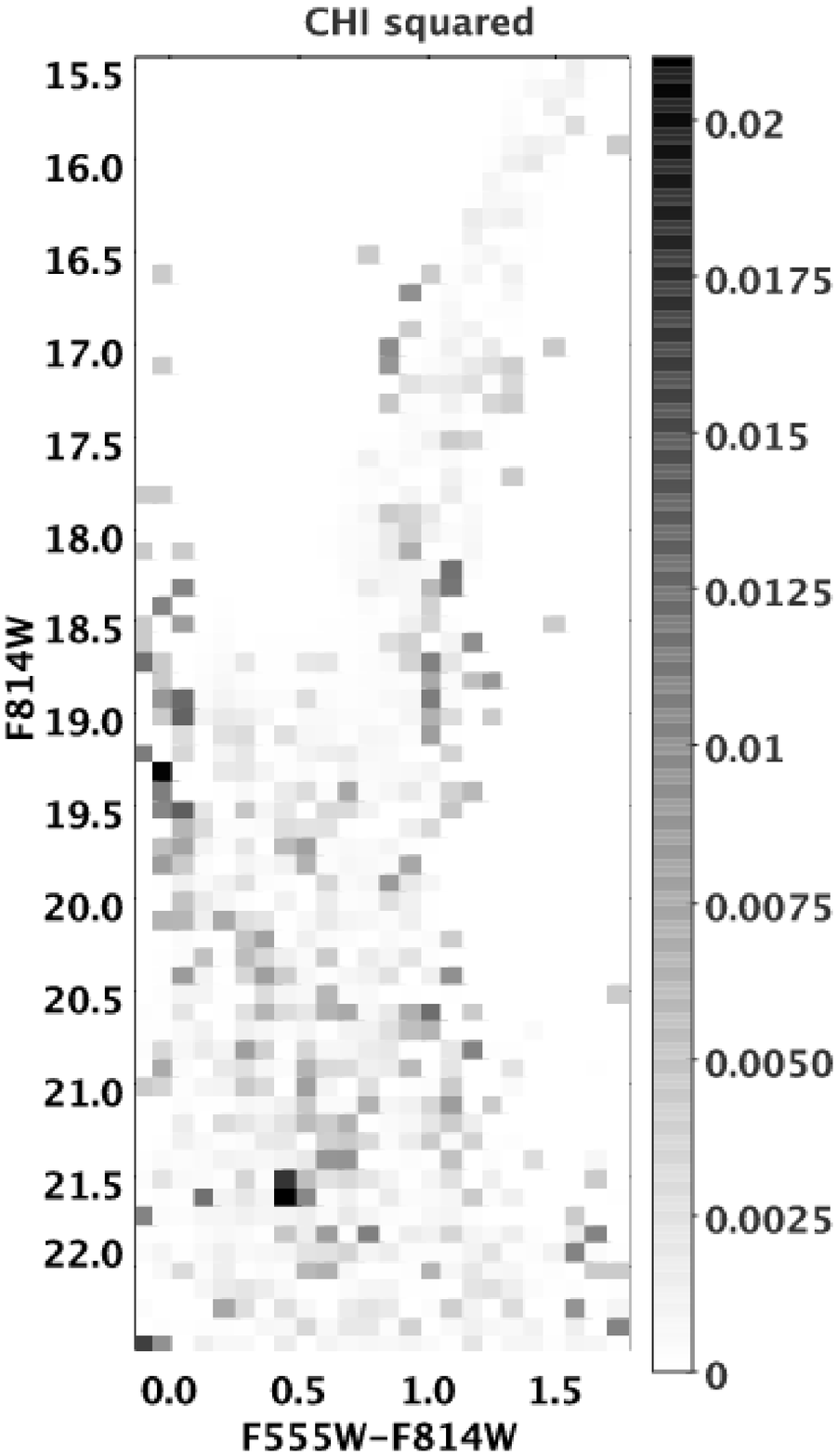}}
}
\end{minipage}
\caption{The same as Fig.~\ref{data_model} but for NGC~1783.}
\label{data_model1783}
\end{figure*}

We proceed in the analysis of NGC~1783 exactly in the same way as for
NGC~1846. Complete maps of $\chisqmin$ for the Centre, as a function
of \dmo, \av\ and metallicity, are presented in the first 7 panels of
Fig.~\ref{chi2_map1783}.  The best solution is found again for
$\mh=-0.49$, with $\dmo=18.57$, $\av=0.22$, and a $\chisqmin=1.39$.
Solutions within the 68\% confidence level span the range of
metallicities between $-0.47$ and $-0.54$.

The metallicity of $\mh=-0.49$ turns out to coincide with the one
previously determined for NGC~1846, which is not unexpected given
their same ages. High resolution spectroscopy by
\citet{Mucciarelli_etal08} instead indicates a value of
$\feh=-0.35\pm0.02$.

Based on Fig.~\ref{chi2_map1783}, we determine $\dmo=18.57\pm0.07$ and
$\av=0.22\pm0.05$ for the cluster Centre (with random errors at the
68~\% significance level). 

\begin{figure*}
\subfigure[SFR$(t)$ for NGC~1783 Centre]{
\resizebox{0.3\hsize}{!}{\includegraphics{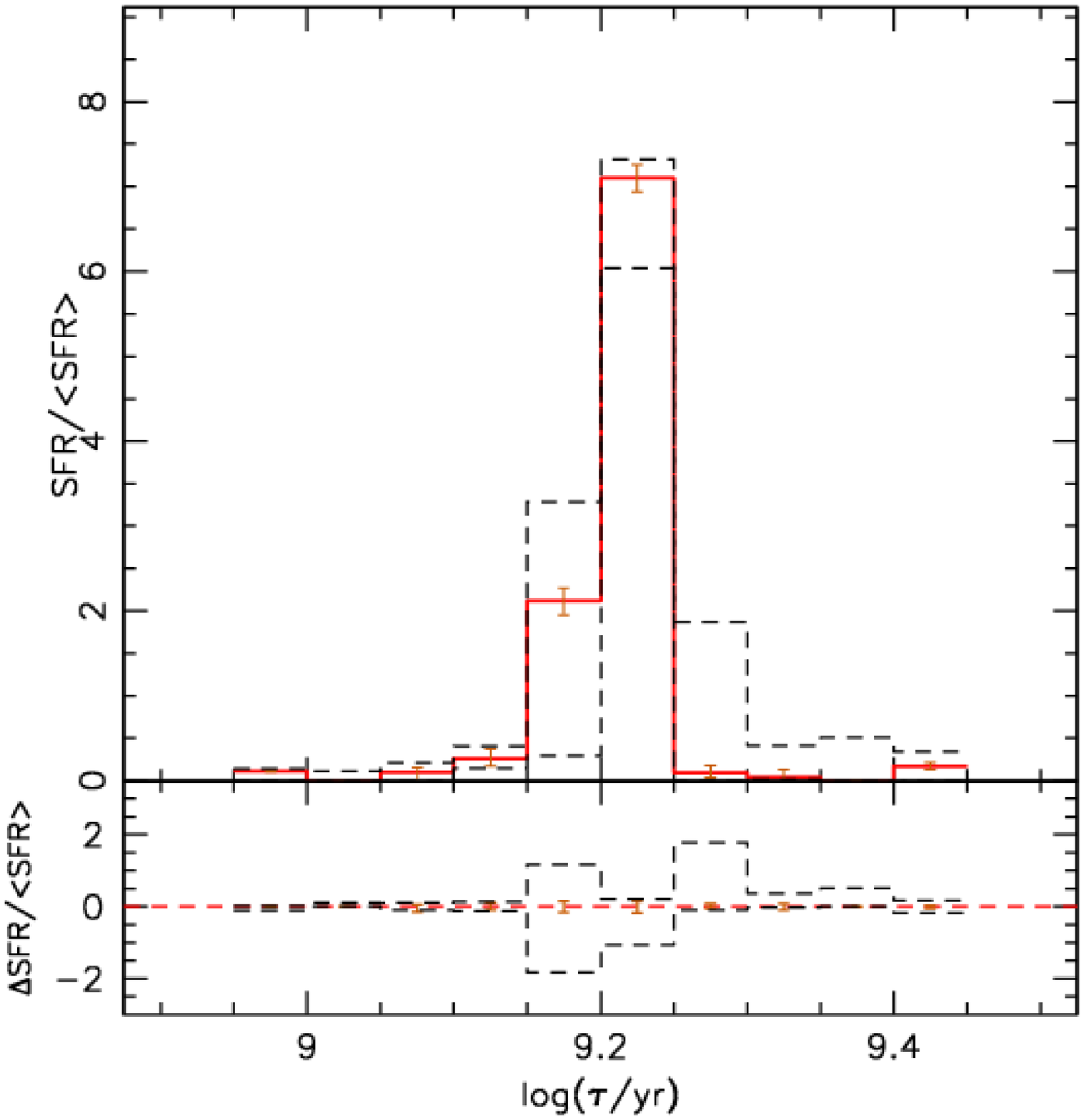}}
\resizebox{0.3\hsize}{!}{\includegraphics{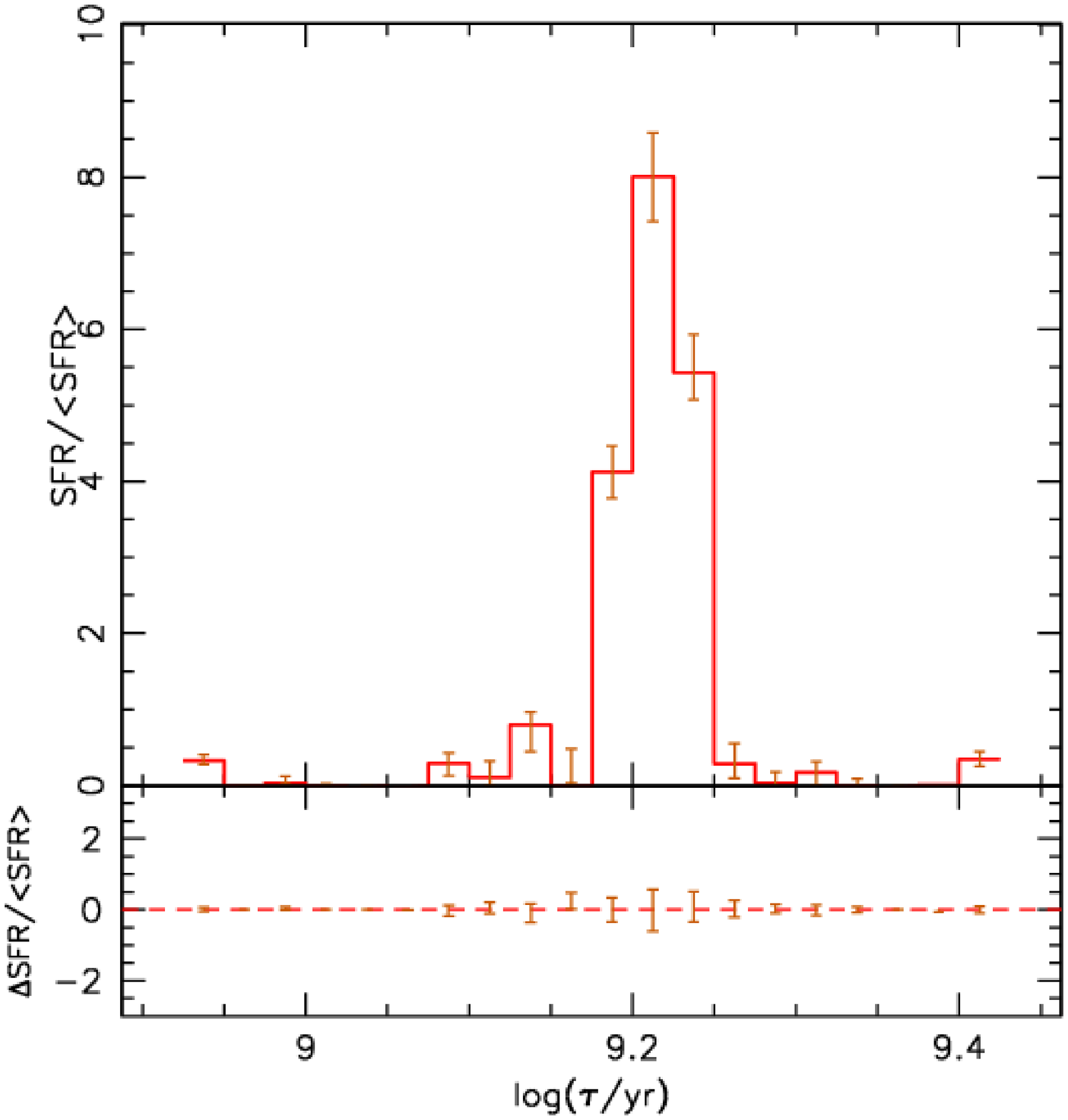}}
\resizebox{0.3\hsize}{!}{\includegraphics{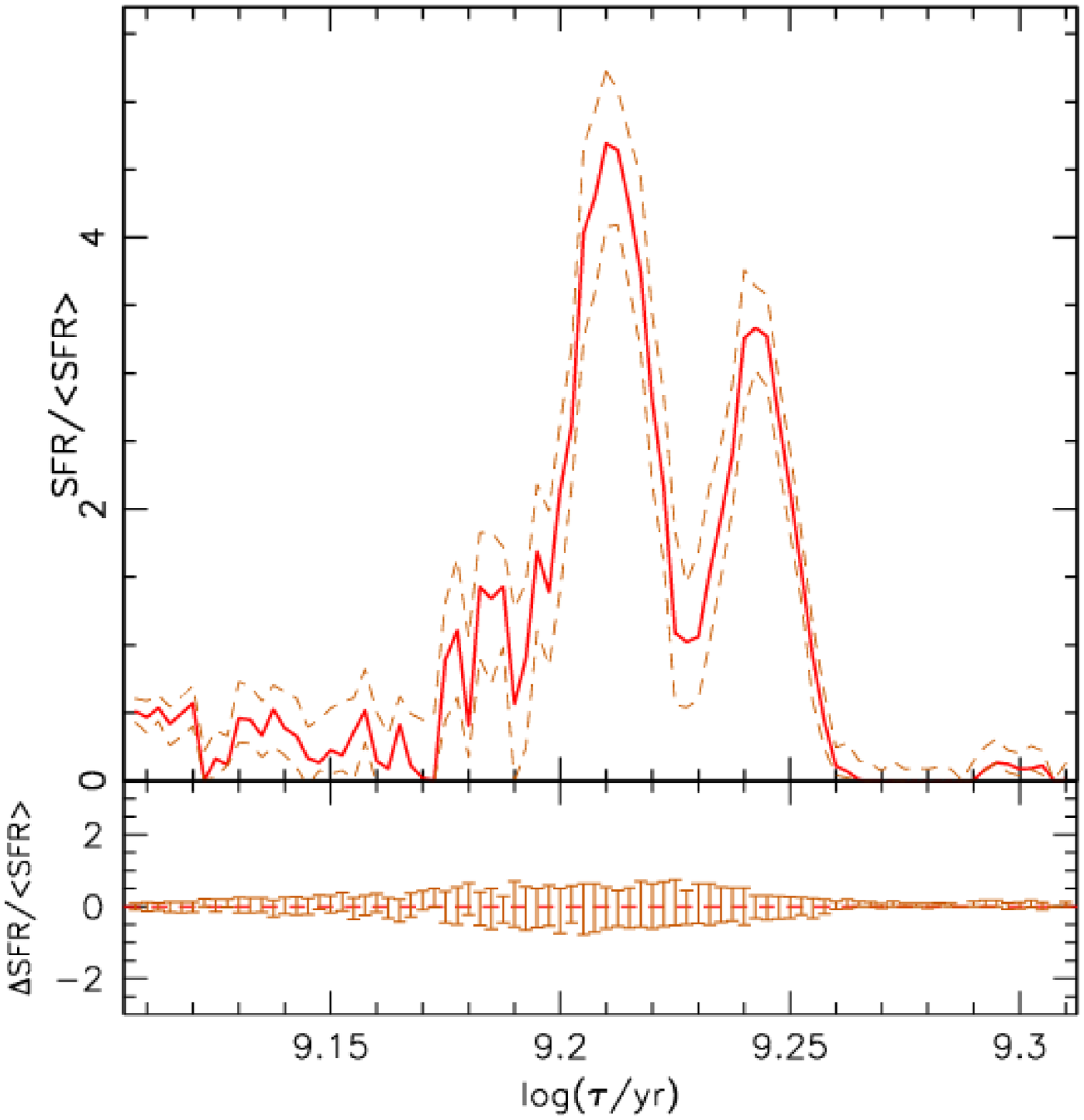}}
}
\subfigure[SFR$(t)$ for NGC~1783 Ring]{
\resizebox{0.3\hsize}{!}{\includegraphics{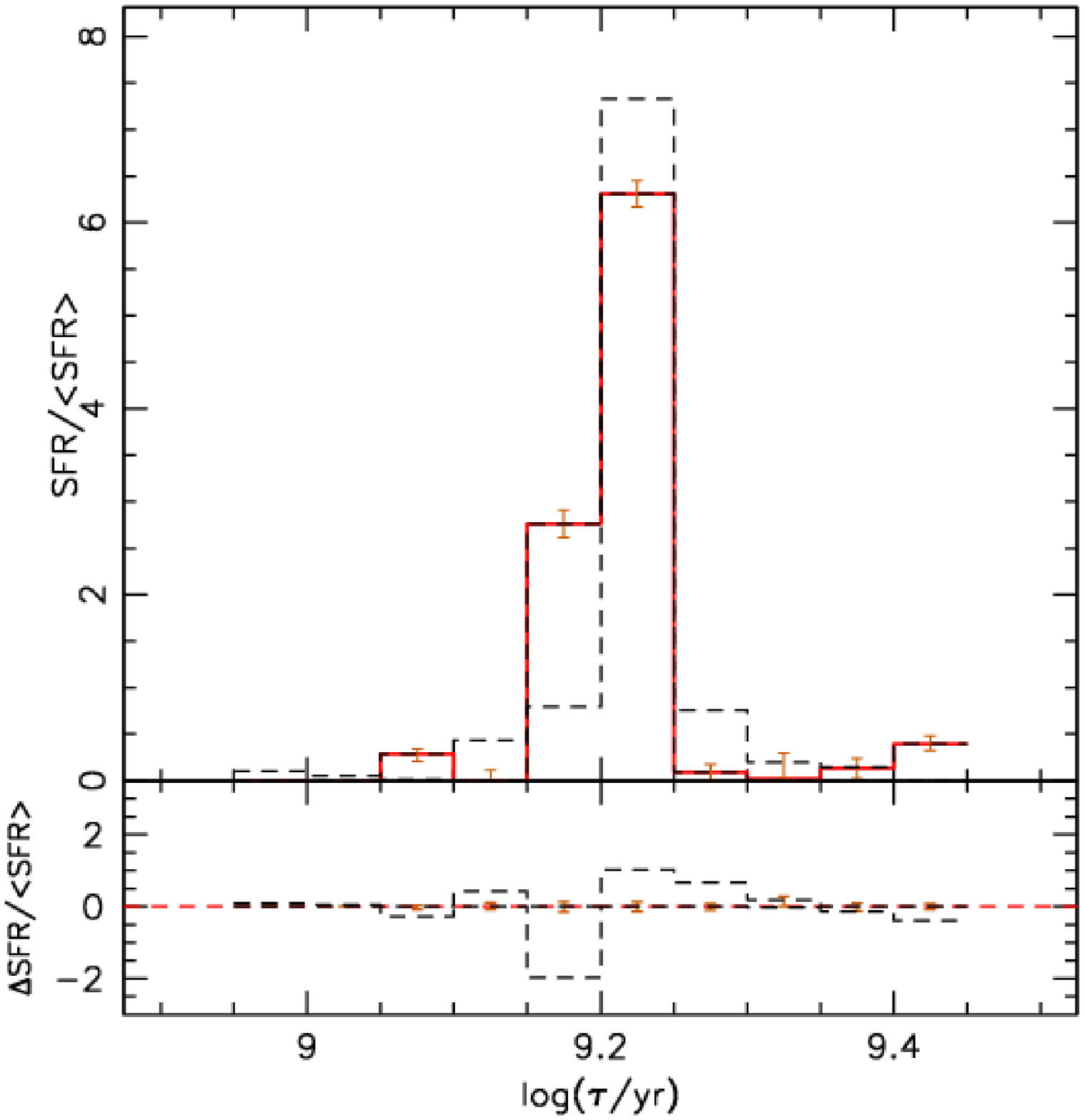}}
\resizebox{0.3\hsize}{!}{\includegraphics{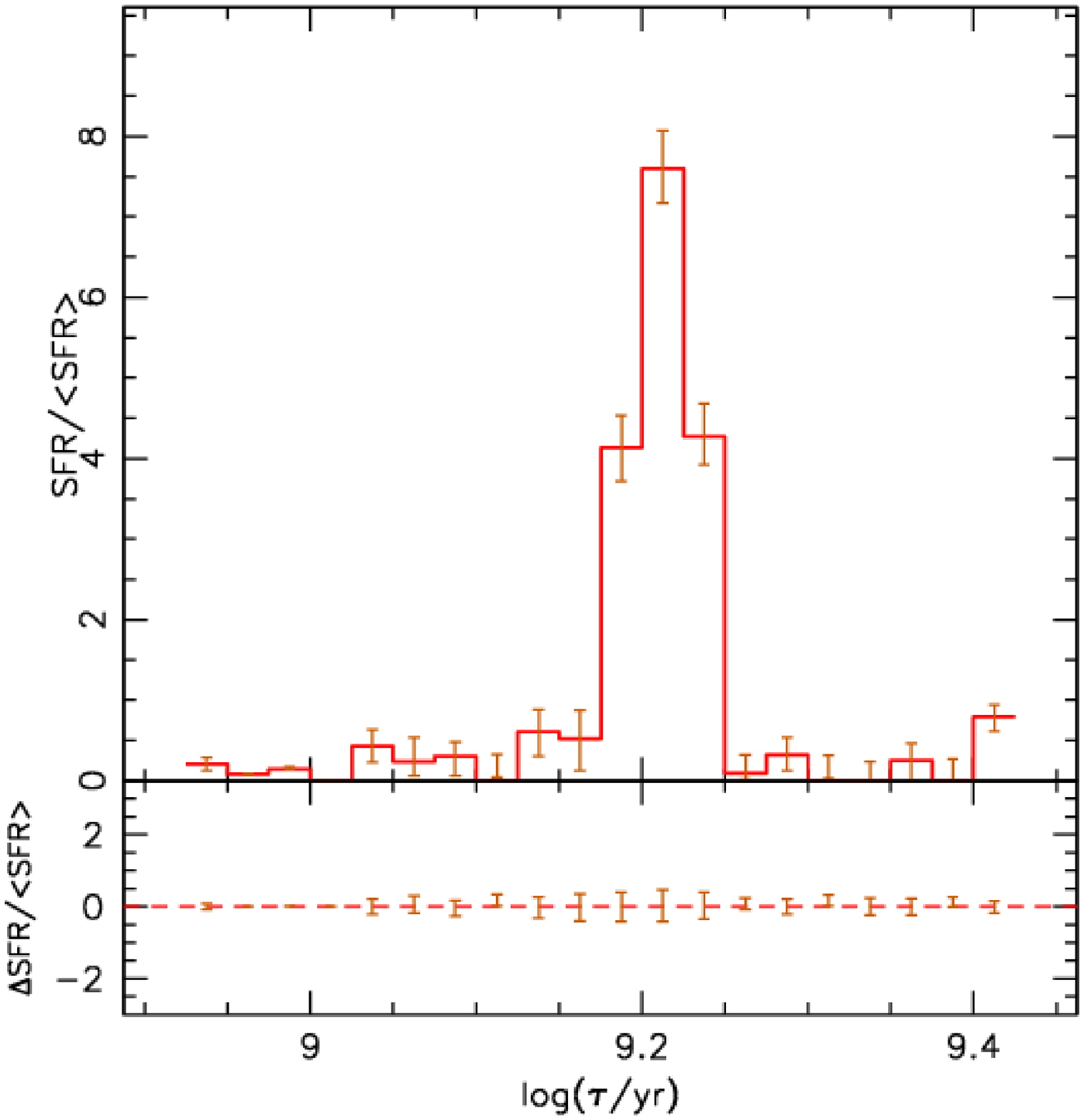}}
\resizebox{0.3\hsize}{!}{\includegraphics{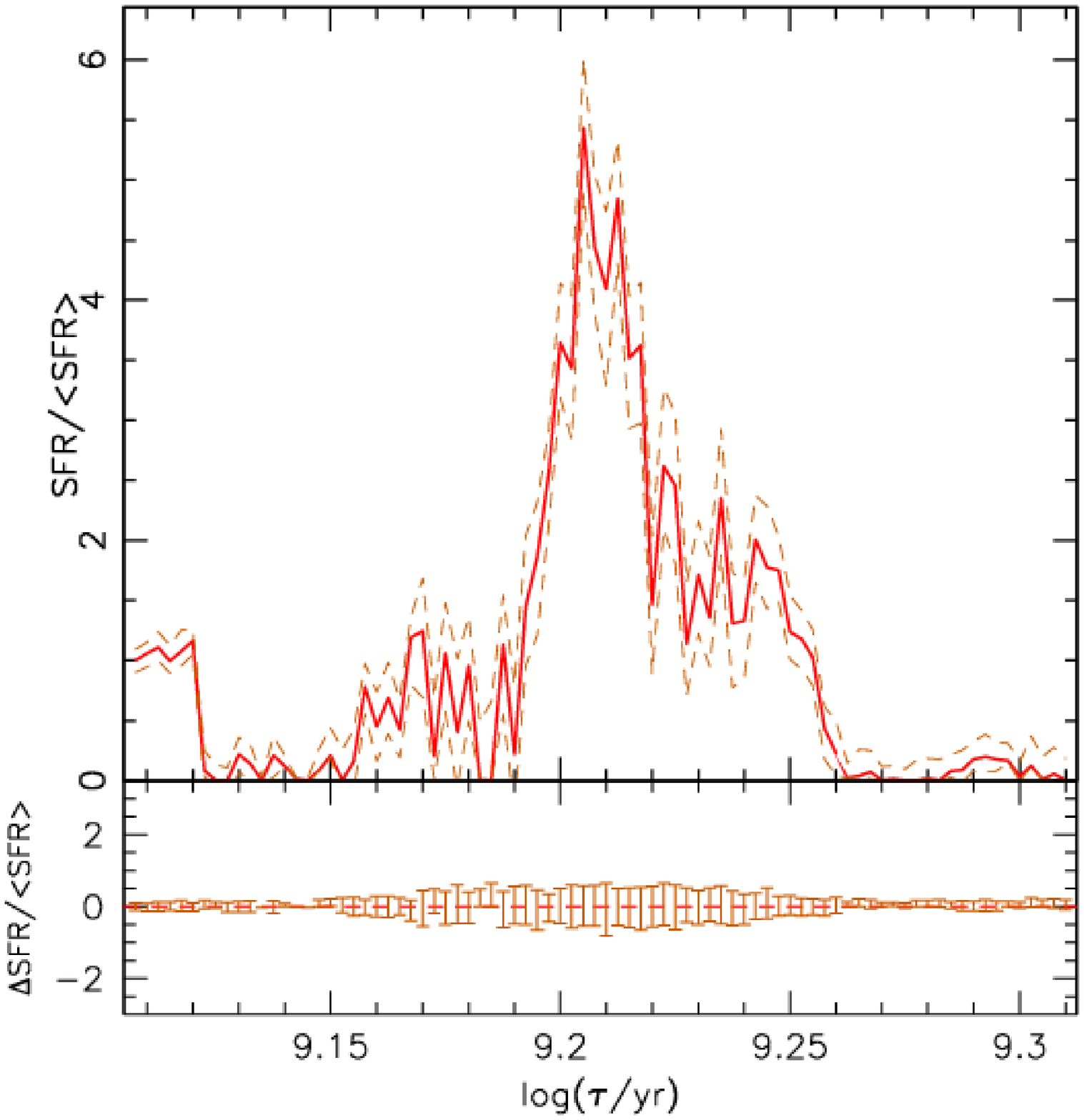}}
}
\caption{The same as Fig.~\ref{sfr1846} but for NGC~1783.}
\label{sfr1783}
\end{figure*}

Using $\mh=-0.49$ for the Ring, we find good solutions in quite a
similar region of the \dmo\ vs. \av\ plane as for the Centre (see last
panel in Fig.~\ref{chi2_map1783}). The overall best solution is at
$\dmo=18.56$, $\av=0.20$, and has a $\chisqmin=1.45$. We however take
the solution at $\dmo=18.57$ as the reference one, which presents a
minimum \chisqmin\ at $\av=0.22$ just as for the Centre. The
best-fitting solutions and map of residuals are also presented in the
Hess diagrams of Fig.~\ref{data_model1783}, while the SFR$(t)$ are
illustrated in the left panels of Fig.~\ref{sfr1783}.

The panels of Fig.~\ref{sfr1783} show the SFR$(t)$ for the cluster
Centre and Ring together with error bars, again at several different
resolutions $\Delta\log t$. These solutions are remarkably similar to
those already found for NGC~1846, in several aspects. Age intervals of
non-null SFR$(t)$ are essentially the same. In particular, also in
NGC~1783 we find indication for a hiatus in the SFR$(t)$ taking place
$\sim\!150$~Myr after the initial burst of star formation.  Comparison
between the middle panels of Fig.~\ref{sfr1783}, however, reveals
something new: the SFR$(t)$ tends to be slightly older in the cluster
Centre than in the Ring. In particular, the Ring seems to present a
less marked peak of star formation at older ages.

\begin{figure*}
\resizebox{0.3\hsize}{!}{\includegraphics{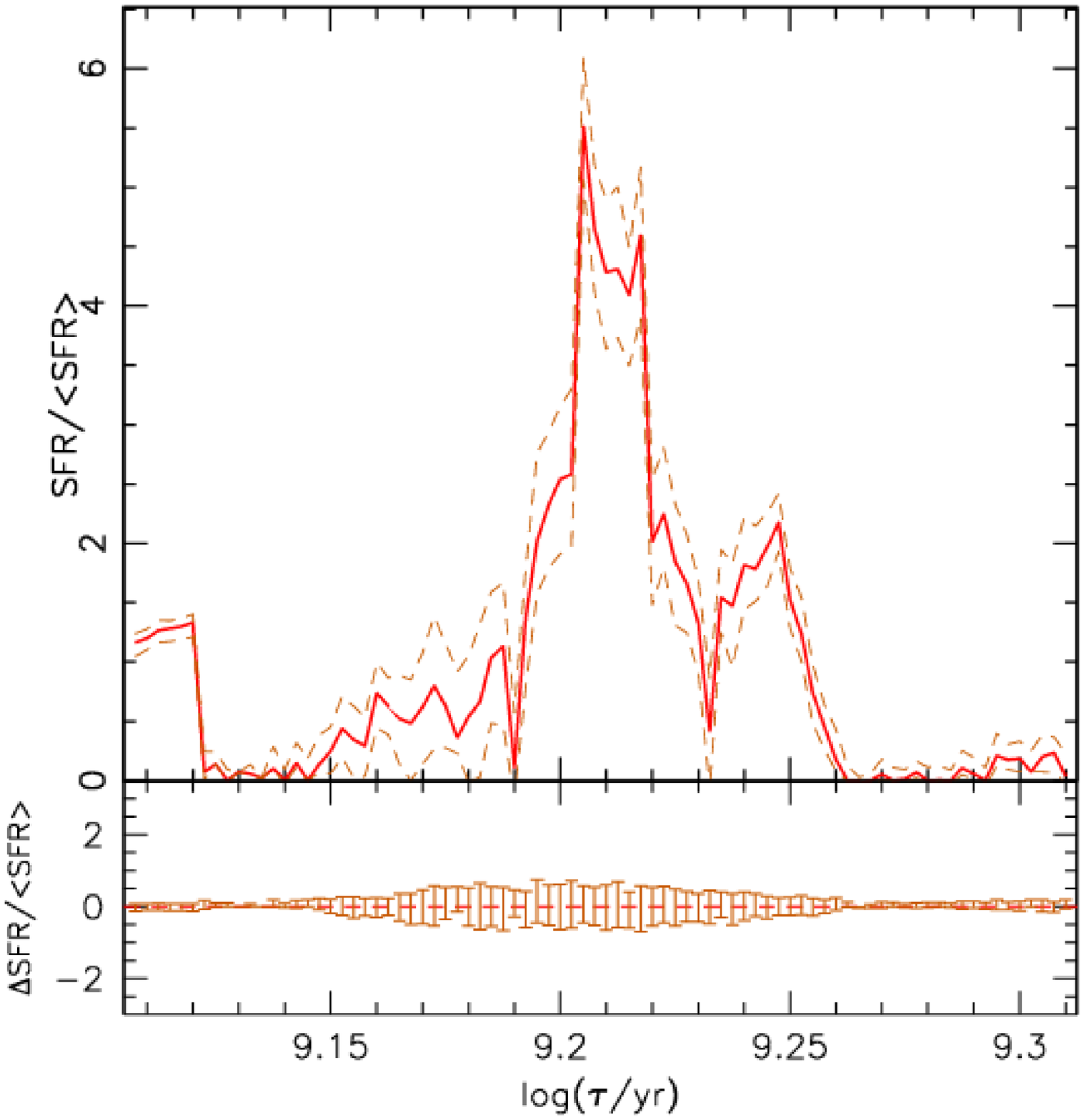}}
\resizebox{0.3\hsize}{!}{\includegraphics{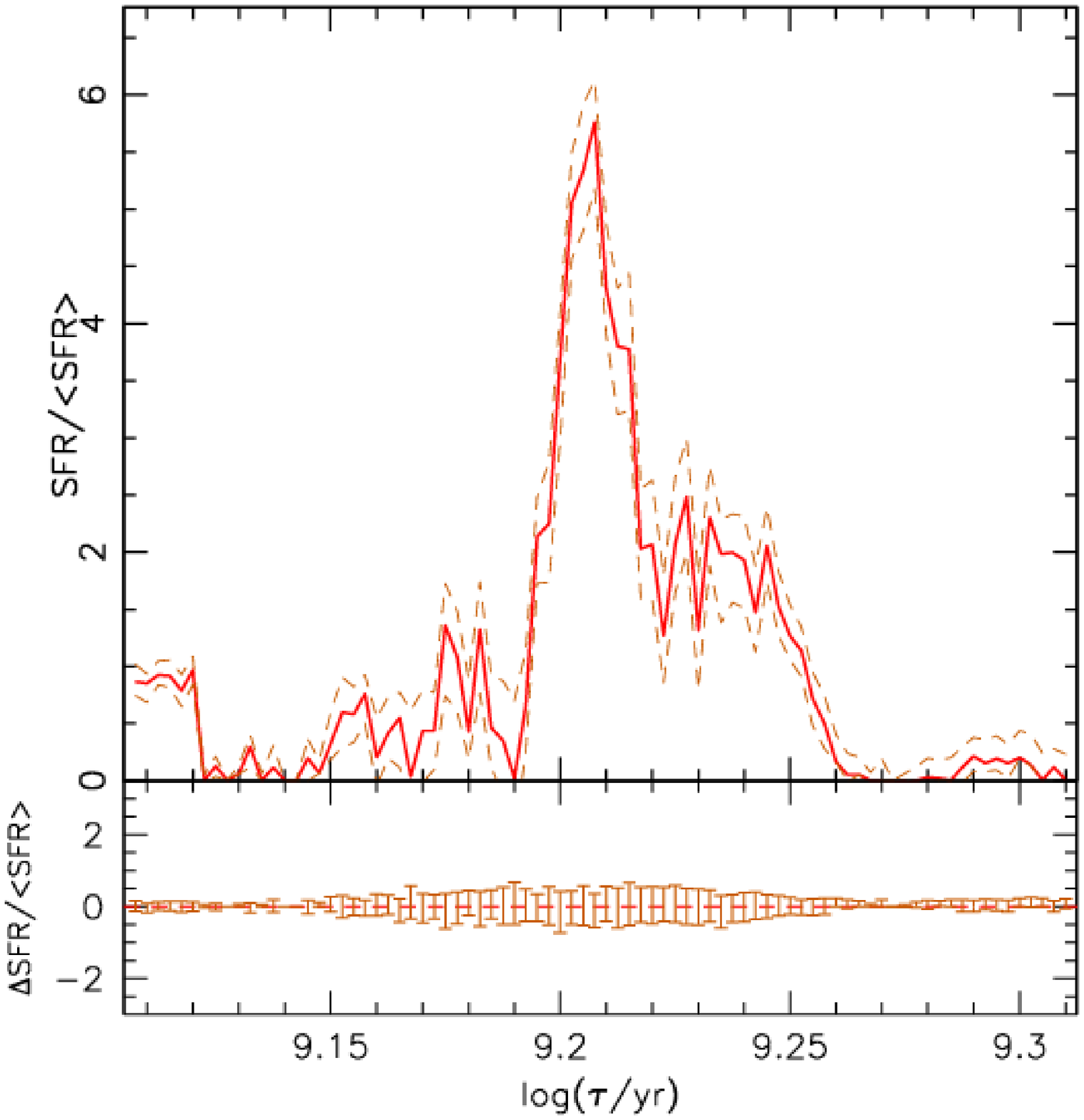}}
\caption{The high-resolution SFH derived for the NGC~1783 Ring, now
  assuming a field contamination decreased/increased by 50~\%
  (left/right panels, respectively), with respect to the values
  expected for the Ring area.}
\label{fig_sfr1783updown}
\end{figure*}

To verify whether this difference could be due to an incorrect account
of the field contamination in the NGC~1783 Ring region, we make an
additional test, illustrated in Fig.~\ref{fig_sfr1783updown}: The
contribution of the field is artificially increased/decreased by
50~\%, with respect to the contribution expected for the Ring area,
and then the SFH recovery is repeated. The results are almost
indistinguishable from those in the bottom-right panel of
Fig.~\ref{sfr1783}, indicating that this is not the factor driving the
different results between Center and Ring in NGC~1783.

\subsubsection{The binary fraction and mass function in NGC~1783}

\begin{figure*}
\resizebox{0.33\hsize}{!}{\includegraphics{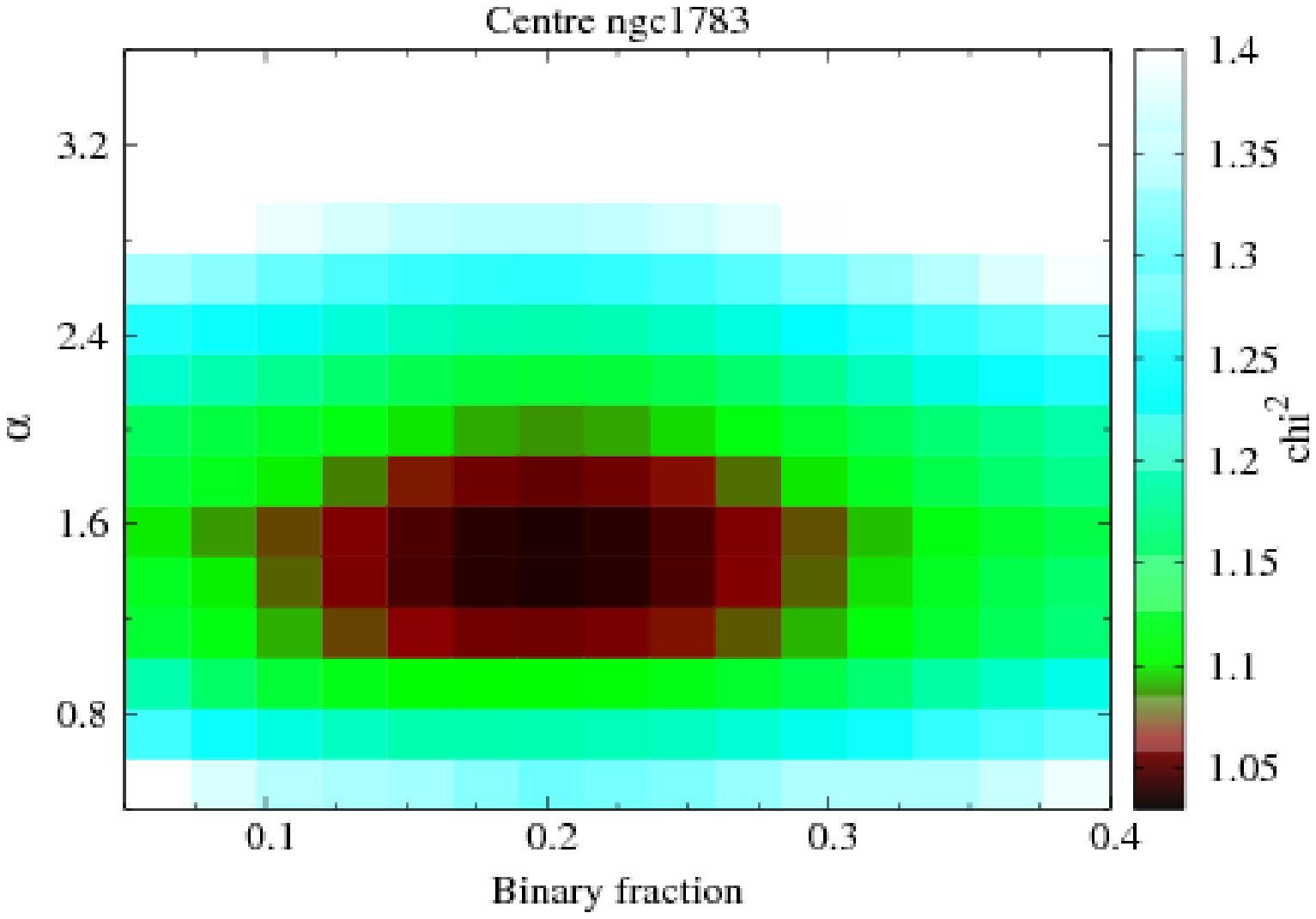}}
\resizebox{0.33\hsize}{!}{\includegraphics{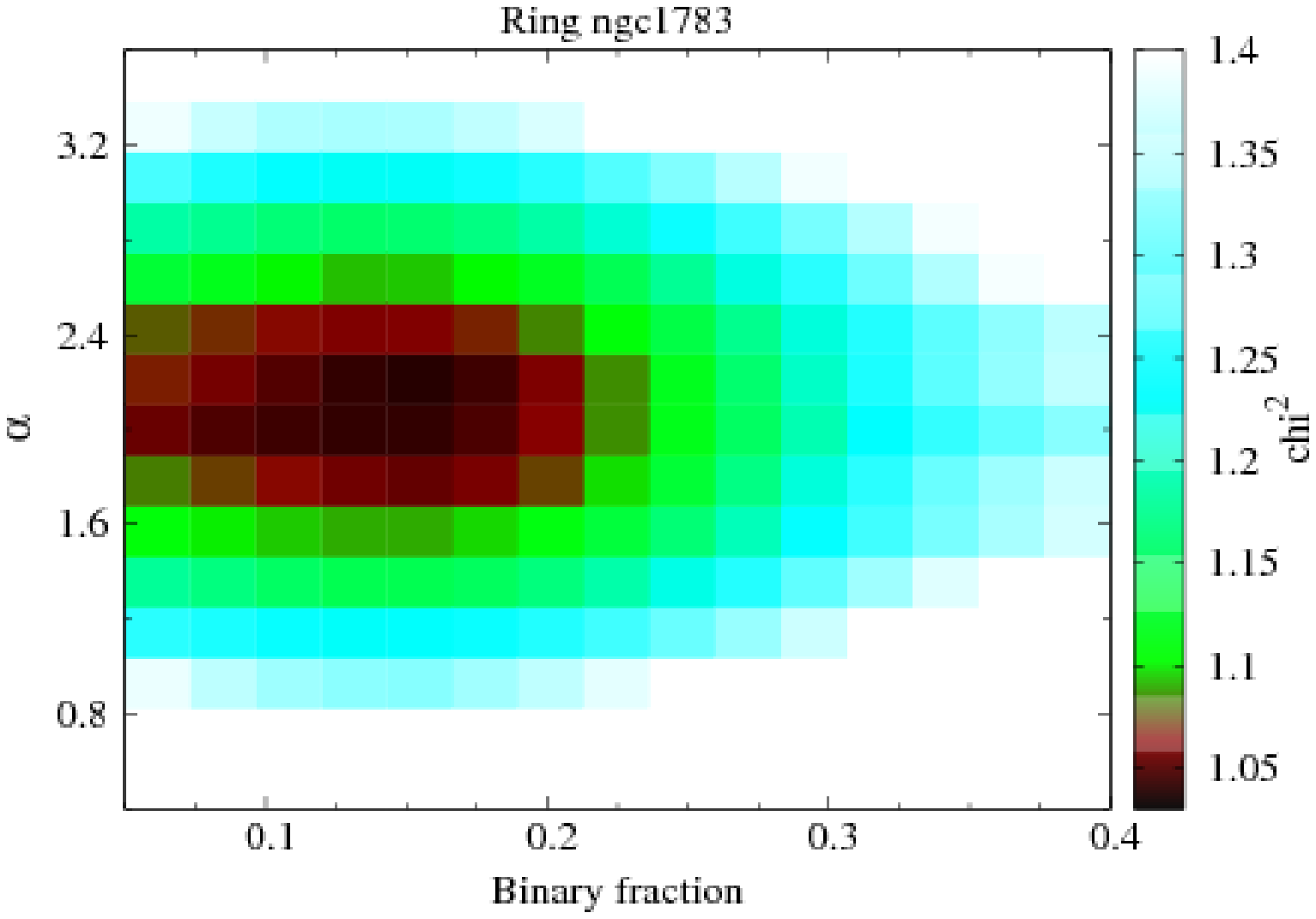}}
\caption{The same as Fig.~\ref{fig_IMF1846}, but for NGC~1783.}
\label{fig_IMF1783}
\end{figure*}

As for NGC~1846, also for NGC~1783 we take the best-fitting solutions
for Centre and Ring and vary its binary fraction $f$ and PDMF slope
$\alpha$. The maps of \chisqmin\ are presented in
Fig.~\ref{fig_IMF1783}. The results this time appear noticeably
different between Centre and Ring: While for the Centre the
best-fitting values are $\alpha=1.3$ and $f=0.20$, for the Ring they
are $\alpha=2.0$ and $f=0.15$. These values of $\alpha$ provide hints
of a possible mass segregation in this cluster.

The interpretation of this finding is not straightforward. Due to CPU
time limitations we did not perform a thorough exploration of the
parameter space, neither the detailed analysis of random and
systematic errors that would be required to define the significance of
the detected PDMF variation. Anyway, we think the results are
interesting enough to be mentioned here, considering the novelty of
determining $\alpha$ in different regions of clusters with MMSTOs.
Since we find a smaller contribution of the older stars in the Ring,
the deficit of less massive stars in the Ring is more likely
associated to the younger population.


\section{Discussion}
\label{conclu}

Our analysis produces a number of results, that we now compare to
previous results for the same clusters, and frame in the present
scenarios for the production of multiple populations in Magellanic
Cloud clusters.

\subsection{Comparing the SFH between NGC~1846 and NGC~1783 Centres}

As easily noticeable by comparing the top right panels of
Figs.~\ref{sfr1846} and \ref{sfr1783}, the timescales for the SFH in
NGC~1846 and NGC~1783 Centres are surprisingly similar. They both seem
to have formed stars for a total period of about 0.3~Gyr. They both
started the star formation activity in a marked burst peaked at
$\log(t/{\rm yr})=9.24$ or 9.25 ($t=1.78$~Gyr). The FWHM of this
marked peak is, in both cases, of 0.02~dex in $\log t$, which is
indistinguishable from the actual resolution of $\Delta\log
t=0.015$~dex of the SFH-recovery method.  In both clusters, this
strong peak is followed by a short hiatus in SFR$(t)$, which lasts
long enough to be detected by the method, that is, at least 0.02~dex
in $\log t$ (or 70~Myr). Then the SFR$(t)$ proceeds up to more recent
times, becoming null at ages of $\log(t/{\rm yr})=9.17$ (1.48~Gyr). In
the following, for obvious reasons we will refer to the marked peak
observed in the SFR$(t)$ as the {\em first generation}, and the
extended period of star formation, after the hiatus, as the {\em
  second generation}.  It should be clear however that more than two
generations may be present, especially in the later periods of star
formation (younger stellar ages).

The presence of separated periods of star formation in these clusters
has been already advanced by other authors, using essentially the same
set of HST images, but different data reductions and methods.
\citet{Mackey_BrobyNielsen2007}, \citet{Mackey_etal08} and
\citet{Milone_etal08} noticed that the MSTO region in NGC~1846 was
clearly split or bimodal, hence suggesting the presence of two main
populations separated in age.  \citet{Goudfrooij_etal09,
  Goudfrooij_etal11a} instead derive a continuous distribution of
stars across the MSTO region, hence concluding that the SFR$(t)$
likely proceeded continuously over a period of $\sim\!300$~Myr. For
NGC~1783, \citet{Mucciarelli_etal07} find no evidence of a broad
turn-off, while \citet{Mackey_etal08} and \citet{Goudfrooij_etal11a}
claimed a MMSTO with a smooth age distribution, whereas
\citet{Milone_etal08} suggests a double MSTO.  Our paper, instead,
would agree more with the interpretation of a split MSTO in both
clusters. Notice that the different results cannot be attributed to
the different data reduction only. The main different between our
analysis and the previous ones is in the fact that we use the
information in the entire CMD, and not only the one across the MSTO,
in order to derive the age distribution.  Our method also attempts to
fit other age-sensitive features like the subgiant branches and red
clumps -- very well drawn in the CMD of both clusters, see
Fig.~\ref{fig_cmd}.

As already mentioned in Sect.~\ref{intro}, the most evident difference
between NGC~1846 and NGC~1783 is in their different radii, while most
of their other parameters are quite similar -- including the total
mass, mean metallicity, and level of contamination by the LMC disk.
Both clusters are among the most massive of their ages in the LMC. It
is not hard to imagine that they formed at about the same age, as a
result of the same sort of dynamical process (or the same large-scale
dynamical event) in the LMC.  All the above-mentioned aspects point to
a common process operating in a similar way in both clusters, from the
onset of star formation up to later stages.

Among the scenarios advanced to explain prolonged star formation in
LMC star clusters, two are particularly worth of being commented here:
First, from dynamical arguments and the inferred escape velocity at an
age of 10~Myr of several clusters exhibiting MMSTOs,
\citet{Goudfrooij_etal11b} suggest that second-generation stars start
being formed from cluster material shed by first-generation stars
featuring slow stellar winds. In our results, the second generation
appears after a period of time to be located between 160 and 220~Myr
after the first generation. This time lag corresponds to the
main-sequence lifetimes of stars with masses between 4.0 and
3.4~\Msun, which are expected to quietly end their nuclear lives as
mass-losing AGB stars with low-velocity winds
\citep[$\sim\!15-20$~km$\,$s$^{-1}$ ;][]{HabingOlofsson}. Thus, our
results for the duration of the SFH hiatus are certainly compatible
with \citet{Goudfrooij_etal11b}'s results.

A somewhat similar scenario has been advanced by
\citet{ConroySpergel11}, who suggest that the supernovae of type II
and the prompt type~Ia from the first-generation stars, first clean up
the interstellar medium of the cluster; the interstellar medium is
then reformed at the cluster centre as intermediate-mass stars start
shedding their envelopes at ages larger than 100~Myr. The accumulated
gas then cools as the Lyman-Werner photon flux from the cluster stars
drops after a few 100~Myr, which allows molecular hydrogen and stars
to form. This starts the second-generation of star formation, after
which no third generation can follow because late SN~Ia from the first
generation clean up the cluster gas again.  Although detailed in the
description of the processes that can allow/interrupt star formation
in clusters, \citet{ConroySpergel11}'s scenario do not make very
precise predictions about the timescales of the different processes
involved. Anyway, it is clear that our findings regarding the SFH,
with the observation of first and second-generation stars, are also in
agreement with their description.  In addition, our observations
indicate that in these $\sim\!10^5$~\Msun\ clusters the final
dissolution of the central interstellar medium by late SN~Ia takes
place after 0.3~Gyr. In the SMC cluster NGC~419, using the same method
we find evidence for a total interval of star formation at least twice
as long \citep{Rubele_etal10}. These numbers may provide important
constraints for a further refinement of the \citet{ConroySpergel11}
scenario.

\subsection{Comparing Centre and Ring SFHs}

Another finding from our methods regards the similarities/differences
between the SFHs at the cluster Centres and Rings. As indicated by the
right panels in Figs.~\ref{sfr1846} and \ref{sfr1783}, we find hints of
a different SFH between Centre and Ring for NGC~1783, in which the
Ring has a much less pronounced first-generation peak. For NGC~1846,
instead, there is no suggestion of a similar effect.

This result is particularly surprising since results from
\citet{Goudfrooij_etal11b}, derived from the same data, seem to
suggest exactly the opposite: after drawing boxes in the CMD for these
and other clusters, they counts stars in the ``upper and lower MSTO
regions'', finding that both kinds of stars present a very similar
radial distribution in NGC~1783, while in NGC~1846 the upper MSTO
stars are clearly more centrally concentrated.

As already commented, the present analysis uses all stars in the CMD,
whereas \citet{Goudfrooij_etal11b} explicitly avoids dealing with
background/field subtraction issues by excluding areas in the CMD
where the background was found to contribute more than 20\% to the
star counts. Their goal is to deal with quite robust star counts as
function of radius. Thus, \citet{Goudfrooij_etal11b} do not use a
large part of the MS, and the top-left extreme of the MSTO, which
instead are included in the present analysis. In addition,
\citet{Goudfrooij_etal11b}'s method is model-independent.

Our approach instead aims to maximize the number statistics by
including all stars irrespective of their origin, and, additionally,
to limit the impact of subjective choices regarding the stars included
in the analysis. This attempt cannot be free of subjective choices,
however. For instance, we had to decide, quite subjectively, the
extent of Centre and Ring regions. Also, we use a particular set of
isochrones, under the implicit assumption that they accurately
describe all relevant phases of stellar evolution. Inaccuracies in
these isochrones would unavoidably appear as systematic errors in all
of our results.

These contrasting results clearly make somewhat uncertain any attempt
to use the radial trend to discuss the scenario for the formation of
second-generation in these clusters. Formally speaking, the results
for NGC~1846 -- no difference between Centre and Ring SFH -- seem just
to indicate that we were not able, with our definition of Centre and
Ring, to detect the expected concentration of the second generation
towards the centre of this cluster. This is not surprising. What is
surprising is the trend derived for NGC~1783, which would suggest
that, in less concentrated clusters like this, second-generation stars
are spatially more spread than first-generation ones. No scenario for
the formation of multiple populations in clusters seem to favour such a
trend. This point clearly deserves more accurate analysis.


As for the determination of the PDMF and binary fractions, the basic
results is that we find hints of mass segregation in NGC~1783, at a
level which is certainly larger than in NGC~1846, but which, however,
may not be statistically significant.  NGC~1783 is also the cluster
with the largest radius, the larger concentration index $c$, and also
the one for which Centre and Ring seem to present a different
SFR$(t)$. It is tempting to suggest a correlation between all these
different aspects, but any conclusion on this is hampered by the
uncertainties in our determinations of $\alpha$ and $f$, in the
determination of $c$ for NGC~1783 \citep{Goudfrooij_etal11a}, and on
the fact that we have analysed just two clusters. It would be very
interesting to extend the same kind of determination to a much larger
sample.

At first sight, our clusters seem to follow the anti-correlation
between the concentration index $c$ and the slope of the global mass
function, which is detected in Galactic globular clusters \citep[GGCs
;][]{DeMarchi_etal07} and interpreted as the result of residual-gas
expulsion in initially mass-segregated star clusters
\citep{Marks_etal08}: indeed, we find the steepest mass function
(larger $\alpha$) in NGC~1783, the cluster with the larger
concentration index. We can also confirm that, after taking into
account the different definitions of $c$ and $\alpha$ (i.e.\ the use
of logarithm for $c$, and the multiplication by $-1$ in the case of
$\alpha$) by different authours, we find values comparable to those
derived by \citet{Glatt_etal11} for 6 SMC intermediate-age and old
clusters; these values place them from $\sim\!2$ to 3~dex bellow the
$\alpha$ vs.\ $c$ relation typical of GGCs. The interpretation of
these trends is not easy, since all these clusters have been analysed
using different methods and heterogeneous data, and, moreover, the
mean \citet{DeMarchi_etal07}'s relation is usually derived for entire
clusters, and not for different cluster regions as in our case.

\subsection{Concluding remarks}

Together with our previous findings for the SMC clusters NGC~419
\citep{Rubele_etal10} and NGC~1751 \citep{Rubele_etal11} using the
same techniques, the present results for NGC~1846 and NGC~1783 confirm
that multiple episodes of star formation provide a good quantitative
description of the observed CMD features. At this point, research in
the field should concentrate on carefully age-dating the different
populations, and in identifying features that can favour/disfavour the
possible scenarios for their formation.  In this sense, the present
work has been partially successful. Without imposing any a priori
limitation or pre-selected shape to the SFH, we have found the
presence of separated first and second-generation episodes of SFH for
both clusters. This observation is compatible with simple schemes
where second-generation SFR$(t)$ can only start after a period of a
few 100~Myr, when the prompt supernovae and the Lyman-Werner flux from
first-generation stars are over \citep{ConroySpergel11}. The stellar
ejecta accumulated to form the second-generation should come partially
from 3.4--4.0~\Msun\ stars of the first generation, and be shed
through slow AGB winds, in accordance with the correlation between the
presence of MMSTOs and $v_{\rm esc}$ at an age with 10~Myr found by
\citet{Goudfrooij_etal11b}.

However, we find an intriguing central concentration of
first-generation stars in the cluster NGC~1783, which is opposite to
what predicted by the above-mentioned scenarios. This cluster has a
large core radius and concentration index, which may be associated
with the observed gradients in the stellar population. The bad
definition of the cluster field does not seem responsible for the
observed trends, because the field contribution is anyway very small.
However, clarification of this issue may requires the analysis of
larger areas around the cluster, and of more clusters of similar age
and different structural parameters. These steps will be pursued in
subsequent papers.

\section*{Acknowledgments}
VKP is grateful to Jay Anderson for sharing his ePSF program.  We
acknowledge the anonymous referee for the many suggestions that
greatly helped us to improve the text.  The data presented in this
paper were obtained from the Multimission Archive at the Space
Telescope Science Institute (MAST). STScI is operated by the
Association of Universities for Research in Astronomy, Inc., under
NASA contract NAS5-26555.
%
%

%
\label{lastpage}
\end{document}